\documentclass[12pt]{amsart}
\usepackage{latexsym,amsmath,amssymb,amscd,eucal}
\usepackage{xypic}
\usepackage{mathrsfs}
\usepackage[all]{xy}

\usepackage{color}

\newcommand{\at}{(\mathbb{C^\times})^n}

\newcommand{\lau}{\mathbb{C}(t_1,\ldots ,t_n)}
\newcommand{\lauq}{\mathbb{C}_{\theta}(t_1,\ldots ,t_n)}
\newcommand{\rs}{\mathbb{C}[\sigma]}
\newcommand{\rsq}{\mathbb{C}_{\theta}[\sigma]}
\newcommand{\rsqq}[1]{\mathbb{C}_{\theta}[#1]}

\newcommand{\newsection}{\setcounter{equation}{0}\section}

\def\appendix#1{\addtocounter{section}{1}\setcounter{equation}{0}
\renewcommand{\thesection}{\Alph{section}}
\section*{Appendix \thesection\protect\indent \parbox[t]{11.715cm} {#1}}
\addcontentsline{toc}{section}{Appendix \thesection\ \ \ #1} }
\newcommand{\eq}{\begin{equation}}
\newcommand{\eqend}{\end{equation}}

\newbox\ncintdbox \newbox\ncinttbox
\setbox0=\hbox{$-$} \setbox2=\hbox{$\displaystyle\int$}
\setbox\ncintdbox=\hbox{\rlap{\hbox to
\wd2{\hskip-.125em\box2\relax \hfil}}\box0\kern.1em}
\setbox0=\hbox{$\vcenter{\hrule width 4pt}$}
\setbox2=\hbox{$\textstyle\int$}
\setbox\ncinttbox=\hbox{\rlap{\hbox to
\wd2{\hskip-.175em\box2\relax \hfil}}\box0\kern.1em}

\newcommand{\Pl}[1]{\Phi_{#1}}

\def\={\ =\ }

\newcommand{\Open}{{\sf Open}}
\newcommand{\module}{{\sf mod}}
\newcommand{\Module}{{\mathscr{M}}}
\newcommand{\coh}{{\sf coh}}
\newcommand{\tor}{{\sf tor}}
\newcommand{\gr}{{\sf gr}}

\newcommand{\complex}{{\mathbb C}} %% complex numbers
\newcommand{\zed}{{\mathbb Z}} %% integers
\newcommand{\nat}{{\mathbb N}} %% naturals
\newcommand{\real}{{\mathbb R}} %% real numbers
\def\alg{{\mathcal A}}
\def\f-alg{{\mathfrak A}}

\def\calg{{\mathcal C}}
\def\galg{{\mathcal G}}
\def\hil{{\mathcal H}}

\def\bun{{\mathcal E}}

\def\sheaf{{\mathcal O}}

\def\Dcal{{\mathcal D}}
\def\Mcal{{\mathcal M}}
\def\Rcal{{\mathcal R}}

\def\Scal{{\mathcal S}}

\def\Vcal{{\mathcal V}}
\def\Jscr{{\mathscr J}}

\def\rank{{\rm rank}}

\def\Pl{{\rm Pl}}
\def\Fun{{\rm Fun}}

\def\P{{\mathbb{P}}}

\def\CP{{\mathbb{CP}}}

\def\U{{\rm U}}
\def\im{{\rm im}}

\def\Ext{{\rm Ext}}
\def\Hom{{\rm Hom}}
\def\End{{\rm End}}
\def\Aut{{\rm Aut}}

\def\ch{{\rm ch}}

\def\Id{{\rm id}}

\newcommand{\Tr}[1]{\:{\rm Tr}\,#1}

\def\e{{\,\rm e}\,}

\def\be{\begin{equation}}
\def\ee{\end{equation}}
\def\bea{\begin{eqnarray}}
\def\eea{\end{eqnarray}}
\def\bd{\begin{displaymath}}
\def\ed{\end{displaymath}}

\def\dd{{\rm d}}

\def\ii{{\,{\rm i}\,}}

\def\calg{{\mathcal C}}
\def\galg{{\mathcal G}}
\def\hil{{\mathcal H}}

\def\bun{{\mathcal E}}

\def\sheaf{{\mathcal O}}

\def\Dcal{{\mathcal D}}
\def\Mcal{{\mathcal M}}
\def\Ncal{{\mathcal N}}
\def\Rcal{{\mathcal R}}

\def\Scal{{\mathcal S}}

\def\Vcal{{\mathcal V}}

\def\Xcal{{\mathcal X}}

\def\Ical{{\mathcal I}}

\newcommand{\Alg}{{\sf Alg}}
\newcommand{\Set}{{\sf Set}}
\newcommand{\Homfun}{{\mathscr{H}}}
\def\Hilb{{\sf Hilb}}

\makeatletter
\newdimen\normalarrayskip              % skip between lines
\newdimen\minarrayskip                 % minimal skip between lines
\normalarrayskip\baselineskip
\minarrayskip\jot
\newif\ifold             \oldtrue            
\def\arraymode{\ifold\relax\else\displaystyle\fi} % mode of array entries
\def\@arrayskip{\ifold\baselineskip\z@\lineskip\z@
     \else
     \baselineskip\minarrayskip\lineskip2\minarrayskip\fi}
\def\@arrayclassz{\ifcase \@lastchclass \@acolampacol \or
\@ampacol \or \or \or \@addamp \or
   \@acolampacol \or \@firstampfalse \@acol \fi
\edef\@preamble{\@preamble
  \ifcase \@chnum
     \hfil$\relax\arraymode\@sharp$\hfil
     \or $\relax\arraymode\@sharp$\hfil
     \or \hfil$\relax\arraymode\@sharp$\fi}}
\def\@array[#1]#2{\setbox\@arstrutbox=\hbox{\vrule
     height\arraystretch \ht\strutbox
     depth\arraystretch \dp\strutbox
     width\z@}\@mkpream{#2}\edef\@preamble{\halign \noexpand\@halignto
\bgroup \tabskip\z@ \@arstrut \@preamble \tabskip\z@ \cr}%
\let\@startpbox\@@startpbox \let\@endpbox\@@endpbox
  \if #1t\vtop \else \if#1b\vbox \else \vcenter \fi\fi
  \bgroup \let\par\relax
  \let\@sharp##\let\protect\relax
  \@arrayskip\@preamble}
\makeatother

\newcommand{\beq}{\begin{eqnarray}}
\newcommand{\eeq}{\end{eqnarray}}

\newcommand{\GL}{{\rm GL}}
\newcommand{\Gr}{\mathbb{G}{\rm r}}
\newcommand{\Fl}{\mathbb{F}{\rm l}}

\newcommand{\fred}{\mathcal{F}}

\def\Sym{{\sf Sym}}

\def\appendix#1{\addtocounter{section}{1}\setcounter{equation}{0}
\renewcommand{\thesection}{\Alph{section}}
\section*{Appendix \thesection. #1}
\addcontentsline{toc}{section}{Appendix \thesection\ \ \ #1} }

\newcommand{\dslash}{\not{\hbox{\kern-2pt $\partial$}}}
\newcommand{\pslash}{\not{\hbox{\kern-2.3pt $p$}}}

 \newtoks\nslashfraction
 \nslashfraction={.13}
 \newcommand{\nslash}[1]{\setbox0\hbox{$ #1 $}
   \setbox0\hbox to \the\nslashfraction\wd0{\hss \box0}/\box0 }

\def\ii{{\,{\rm i}\,}}

\numberwithin{equation}{section}

\newcommand{\lam}[1]{\Lambda^{(#1)}}
\newcommand{\dlam}[1]{\Lambda^{(#1)}\,^{\dagger}}

\begin{document}

\title[Instantons and vortices on noncommutative toric varieties]{Instantons and vortices \\[5pt] on noncommutative toric varieties}

\date{December 2012 \hfill{EMPG--12--25 }}

\author{Lucio S. Cirio}

\address{University of Luxembourg, Mathematical Research Unit,
6, rue Richard Coudenhove-Kalergi, L-1359 Luxembourg, Grand-Duchy of Luxembourg}

\address{Mathematisches Institut, Universit\"at M\"unster,
  Einsteinstrasse 62, 48149 M\"unster, Germany
}

\email{lucio.cirio@uni-muenster.de}

\author{Giovanni Landi}

\address{Dipartimento di Matematica, Universit\`a di
 Trieste, Via A. Valerio 12/1, I-34127 Trieste, Italy}

\address{INFN,
 Sezione di Trieste, Trieste, Italy}

\email{landi@univ.trieste.it}

\author{Richard J. Szabo}

\address{Department of Mathematics, Heriot-Watt University, Colin
  Maclaurin Building, Riccarton, Edinburgh EH14 4AS, U.K.}

\address{Maxwell
  Institute for Mathematical Sciences and Tait Institute, Edinburgh, U.K.}

\email{R.J.Szabo@hw.ac.uk}

\begin{abstract}
We elaborate on the quantization of toric varieties by
combining techniques from toric geometry, isospectral deformations and
noncommutative geometry in braided monoidal categories, and the
construction of instantons thereon by combining methods from
noncommutative algebraic geometry and a quantised
twistor theory. We classify the real structures on a toric
noncommutative deformation of the Klein quadric and use this to derive
a new noncommutative four-sphere which is the unique deformation
compatible with the noncommutative twistor correspondence. We extend
the computation of equivariant instanton partition functions to
noncommutative gauge
theories with both adjoint and fundamental matter fields, finding
agreement with the classical results in all instances. We construct
moduli spaces of noncommutative vortices from the moduli of invariant
instantons, and derive corresponding equivariant partition functions
which also agree with those of the classical limit.
\end{abstract}

\maketitle

\tableofcontents

\parskip 1ex
\linespread{1.1}

\renewcommand{\thefootnote}{\arabic{footnote}}
\setcounter{footnote}{0}

\newsection*{Introduction}

This paper is a continuation of our ongoing work~\cite{CLSI,CLSII}
analysing instanton contributions to gauge theories on noncommutative
deformations of algebraic toric varieties. Part of the motivation
behind this project was to provide a precise notion of noncommutative
toric geometry in which gauge theories could be defined and their
corresponding noncommutative instanton solutions constructed. Drawing on the
machinery of noncommutative algebraic geometry, the instanton moduli
spaces could then be constructed and the corresponding instanton partition
functions explicitly evaluated; the goal was to find in this way a
rigorous analog of the somewhat heuristic calculations in the physics
literature which employ ``local'' noncommutative deformations to
resolve singularities in the instanton moduli spaces and explicitly
compute instanton contributions to the partition functions of
supersymmetric gauge theories on toric varieties. In~\cite{CLSII} we
explicitly carried out this construction for framed instantons on the toric noncommutative
deformation of the projective plane $\CP^2$, and demonstrated that the
\emph{equivariant} instanton partition functions for pure gauge
theories coincide with their classical counterparts on
$\complex^2$, thus providing the desired rigorous derivation in the
context of noncommutative gauge theory. We stress that although the equivariant instanton
counting problems are the same, our instanton moduli spaces are
generically (commutative) deformations of those of the classical case;
in particular, they potentially provide new examples of quantum
integrable systems and could have ramifications in the diverse
relationships that exist
between four-dimensional supersymmetric gauge theories and field
theories in other dimensions, such as two-dimensional conformal field
theory. See~\cite{Szabo} for further physical motivation and context
behind our constructions.

Some of the present article is an overview of the main
results and constructions of~\cite{CLSI,CLSII} with a somewhat more
brief, less technical, and informal style of presentation, which we hope will be more palatable to a
wider audience. We also elaborate and extend many of the points
from~\cite{CLSII}; let us briefly summarise the main new results that
the reader will find in this paper.

We shall give a more precise
treatment of real structures in our functorial quantization approach,
which was glossed over to a certain extent in~\cite{CLSII}. In particular, we
classify the real structures on a toric noncommutative deformation of
the grassmannian $\Gr(2;4)$, and hence show that our new deformation
of $S^4$ found in~\cite{CLSII} is the unique
noncommutative four-sphere which is compatible with the noncommutative
twistor correspondence.

We will also extend the computation of
noncommutative instanton partition functions, presented in~\cite{CLSII}
for the case of pure gauge theories, to gauge theories with matter
fields both in the adjoint and fundamental representations of the
gauge group. In all cases we confirm precise agreement with
the classical results.

Finally, we shall consider the problem of constructing \emph{vortices}
on our noncommutative toric varieties. In the classical case it is well-known that suitable
dimensional reductions of invariant instantons in four dimensions
yield vortices in two dimensions. We will construct noncommutative vortices in our framework via suitable reduction of our instanton
moduli constructions and describe their moduli spaces as smooth
holomorphic submanifolds of the moduli spaces of noncommutative
instantons. We also compute noncommutative vortex partition functions, again including adjoint and fundamental matter
fields as well, and once more find agreement in all cases with
the classical partition functions. We stress that, as in the case of instanton counting,
although the equivariant vortex counting problem is the same as in the
classical limit, the vortex moduli spaces are generically
(commutative) deformations of the classical ones.

The outline of this paper is as follows. In \S1 we summarise our
functorial approach to the quantization of toric varieties, focusing
in particular on noncommutative deformations of projective varieties,
including projective spaces, grassmannians and flag manifolds as
explicit examples. In \S2
we explicitly construct noncommutative instantons in three related
incarnations, and give the construction of instanton bundles based on a
noncommutative version of the twistor transform. In \S3 we construct
smooth moduli spaces for the instanton counting problem on the noncommutative projective plane
$\CP_\theta^2$ and natural bundles over them, and evaluate various
equivariant K-theory partition functions for supersymmetric gauge theories on the complex Moyal plane 
$\complex_\theta^2$. In \S4 we construct moduli spaces for
noncommutative vortices, comparing them with those of non-abelian
vortices in two-dimensional $\Ncal=(2,2)$ supersymmetric gauge
theories, and compute the corresponding noncommutative vortex
partition functions via 
localization in equivariant K-theory.

\subsection*{Acknowledgments}
LSC acknowledges support by the National Research Fund, Luxembourg, AFR project
1164566.
The work of GL was partially supported by the Italian
Project ``Cofin08 -- Noncommutative Geometry, Quantum Groups and
Applications''.
The work of RJS was
supported in part by the Consolidated Grant ST/J000310/1 from the UK
Science and Technology Facilities Council, and by Grant RPG-404 from the Leverhulme Trust.

\newsection{Quantization of toric varieties}

This section summarizes the algebraic constructions which will be used
throughout this paper. We combine constructions from toric geometry
and the theory of isospectral deformations, suitably adapted  from
actions of the real $n$-dimensional torus $\mathbb{T}^n$ on
riemannian (spin) manifolds to actions of the complex torus
$(\mathbb{C^\times})^n$ on algebraic varieties. We use cocycle twist
deformations of Hopf algebras and braided monoidal categories of Hopf
(co)module algebras as a systematic means of deforming the varieties
involved together with geometric objects defined thereon.

\subsection{Hopf cocycle twist deformations\label{twistedsym}}

Let $\hil$ be a Hopf algebra over $\complex$ with coproduct $\Delta:\hil\to\hil\otimes\hil$, counit $\varepsilon:\hil\to\complex$, and antipode $S:\hil\to\hil$. 
We use Sweedler notation $\Delta(h)=h_{(1)}\otimes h_{(2)}$, iterated to $(\Delta \otimes \Id) \circ \Delta(h) = (\Id \otimes \Delta) \circ \Delta(h) = h_{(1)} \otimes h_{(2)} \otimes h_{(3)}$ and so on, with implicit
summation. We consider categories of left/right $\hil$-modules/comodules, with actions and coactions denoted respectively $\triangleright: \hil\otimes V\to V$, $\Delta_V:V\to\hil\otimes V$, 
for the left case: for generic $h\in \hil$ and $v\in V$ we write $\triangleright(h\otimes v)=h\triangleright v$ and $\Delta_V(v) = v^{(-1)}\otimes v^{(0)}$. 
In a similar way, we use $\triangleleft$ and $_V\Delta$ for the right case.
When $V$ has some algebraic structure, for example it is an algebra
and/or a coalgebra, one can ask for a compatibility condition with the $\hil$-module/comodule structure by demanding that
the maps which define the algebraic structure of $V$ (multiplication,
comultiplication, etc.) are morphisms in the category, i.e. they
commute with the $\hil$-action/coaction. When this is the case we have
(left/right) $\hil$-module/comodule algebras/coalgebras $V$. 

We may deform Hopf algebras and their (co)module categories by a  cocycle twist. The construction is due to Drinfel'd, with the dual formulation presented here due to Majid~\cite{Majid1}. Given a Hopf algebra $\hil$, a linear map $F:\hil\otimes\hil\to\complex$ is called a \emph{Drinfel'd twist element} for $\hil$ if it is a convolution-invertible unital $2$-cocycle on $\hil$. These properties ensure that the same coalgebra $\hil$, endowed with multiplication and antipode given for $g,h\in\hil$ by
\begin{align}\label{gammantip}
g\times_{F} h&= F\big(g_{(1)}\otimes h_{(1)}\big)\,
\big(g_{(2)}\cdot h_{(2)}\big)
\,F\,^{-1}\big(g_{(3)}\otimes h_{(3)}\big) \ ,  \\[4pt]
S^{F}(g) &= 
U^{F}\big(g_{(1)}\big)\,S\big(g_{(2)}\big)\,U^{F} 
\,^{-1}\big(g_{(3)}\big) \qquad \mbox{ with} \quad
U^{F}(g) = F\big(g_{(1)}\otimes S(g_{(2)}) \big) \ , \nonumber 
\end{align}
defines a twisted Hopf algebra structure $\hil^F$. If $\hil$ admits a co(quasi)triangular structure 
$\Rcal:\hil\otimes\hil\to\complex$ then $\mathcal{H}^{F}$ is co(quasi)triangular with structure 
$\Rcal_F:\mathcal{H}^{F}\otimes\mathcal{H}^{F}\to\complex$ given by
$\Rcal_{F}(g\otimes h) = F(h_{(1)}\otimes g_{(1)}) \, \Rcal(g_{(2)}\otimes h_{(2)})\, F^{-1}(g_{(3)}\otimes h_{(3)})$. 

A Drinfel'd twist for $\hil$ naturally induces a deformation of the algebraic structure of $\hil$-(co)module (co)algebras. For example if $(A,\mu)$ is a left $\mathcal{H}$-comodule algebra, the deformed product
\begin{equation}
a\star^{F} b := F\big(a^{(-1)}\otimes
b^{(-1)}\big)\, \mu\big(a^{{({0})}}\otimes
b^{{({0})}}\,\big)
\label{dualtwp}\end{equation}
for $a,b\in A$ makes $A^{F}=(A,\star^{F})$ into a left
$\mathcal{H}^{F}$-comodule algebra. 

More generally, Drinfel'd twists naturally affect the whole category of Hopf (co)modules. The induced deformation is encoded in the braiding of the monoidal structure. Let ${}^\hil\Module$ denote the additive category of left $\hil$-comodules. A map $V\xrightarrow{\ \sigma\ }W$ is a morphism of this category if and only if it is $\hil$-coequivariant, i.e. $\Delta_W\circ\sigma = (\Id\otimes\sigma)\circ\Delta_V$. The category ${}^\hil\Module$ has a natural monoidal structure given by the tensor product coaction
$$
\Delta_{V\otimes W}(v\otimes w) = v^{(-1)}\, w^{(-1)}\otimes\big(v^{(0)}\otimes w^{(0)}\big)
$$
for $v\in V$, $w\in W$. When $\hil$ has the trivial cotriangular structure $\Rcal=\varepsilon\otimes\varepsilon$, the category ${}^\hil\Module$ 
{of left $\hil$-comodules} is {(trivially)} braided by the collection
of functorial flip isomorphisms $\Psi_{V,W}(v\otimes w)=w\otimes v$
for each pair of objects $V,W$ of ${}^\hil\Module$, and for all $v\in
V$ and $w\in W$. After deforming $\hil$ via a Drinfel'd twist, the twisted
cotriangular structure now reads $\Rcal_{F}(g\otimes h)= F(h_{(1)}\otimes
g_{(1)}) \, F^{-1}(g_{(2)}\otimes h_{(2)})$ for every $g,h \in \hil$. The monoidal category ${}^{\hil^{F}}\Module$ of left $\hil^{F}$-comodules is then braided by the collection of functorial isomorphisms 
\beq
\Psi^{F}_{V,W}(v\otimes w)=\Rcal_{F}\big(w^{(-1)}\otimes v^{(-1)}\big)\, w^{(0)}\otimes v^{(0)}
\label{PsithetaVW}\eeq
for each pair $V,W$ of left $\hil^F$-comodules with $v\in V$ and
$w\in W$. Since $(\hil^{F},\Rcal_{F})$ is cotriangular the braiding
squares to the identity and makes $ {}^{\hil^F}\Module$ into a
braided monoidal category. 

If $(A,\mu_A)$ and $(B,\mu_B)$ are left $\hil^{F}$-comodule algebras one defines the braided tensor product algebra $A\,{\otimes}_F\,B$ to be the vector space $A\otimes B$ endowed with the product map
$\mu_{A\,{\otimes}_F\,B}:(A\,{\otimes}_F\,B)\otimes(A\,{\otimes}_F\,B) \to A\,{\otimes}_F\,B$ given by
$$
\mu_{A\,{\otimes}_F\,B}=\big(\mu_A\otimes\mu_B\big)\circ \big(\Id\otimes
\Psi^{F}_{B,A} \otimes \Id\big) \ .
$$
The resulting algebra is an object of the category ${}^{\hil^{F}}\Module$ by the tensor
product coaction.

The deformation in passing from $\hil$ to $\hil^{F}$ takes the form of a functorial isomorphism $\mathscr{F}_{F}: {}^{\hil}\Module\to {}^{\hil^{F}}\Module$ of braided monoidal categories. The functor $\mathscr{F}_{F}$ acts as the identity on objects and morphisms of ${}^{\hil}\Module$, but defines a new monoidal structure,
$$
\lambda_{F}\,:\, \mathscr{F}_{F}(V)\otimes \mathscr{F}_{F}(W) ~\longrightarrow~
\mathscr{F}_{F}(V\otimes W) \ , \quad \lambda_{F}(v\otimes w) =
{F}\big(v^{(-1)}\otimes w^{(-1)}\big)\, v^{(0)}\otimes w^{(0)} \ ,
$$
on the category ${}^{\hil_{F}}\Module$. 
This makes $\mathscr{F}_{F}$ into a monoidal functor intertwining the braidings in ${}^{\hil}\Module$ and ${}^{\hil_{F}}\Module$, given respectively by the flip functor $\Psi$ and the functor $\Psi^{F}$ in (\ref{PsithetaVW}).

Since later on we shall also touch upon $*$-structures, we recall
  that a $*$-algebra is an algebra $A$ endowed with an involutive
  anti-linear anti-homomorphism $*:A\rightarrow A$. A Hopf $*$-algebra
  is a Hopf algebra $\hil$ with a $*$-structure such that 
for every $h\in \hil$ one has
$$ 
\Delta( h^*) = \big(h_{(1)}\big)^*\otimes \big(h_{(2)} \big)^* \ , \qquad \varepsilon(h^*) = \overline{\varepsilon(h)} \ .
$$
In general one also requires $(S\circ *)^2 = \Id$; this is however redundant with an invertible antipode since then
$*\circ S^{-1} \circ *$ satisfies all the conditions for the antipode and the latter is unique.  

For a left action of a Hopf $*$-algebra $\hil$ on a $*$-algebra $A$,
with involution denoted $\dag:A\to A$,
one requires compatibility, i.e. that the action be a $*$-action, so that
for $h\in \hil$, $a\in A$, one has
$$
h\triangleright a^\dag = \big(S(h)^*\triangleright a\big)^\dag \ .
$$
Similarly, for a left coaction one requires it to be a $*$-coaction,
i.e. for $a\in A$ one has
$$ 
\Delta_A\big(a^\dag\,\big) = \big(a^{(-1)}\big)^*\otimes \big(a^{(0)}\big)^\dag \ .
$$

When deforming via a Drinfel'd twist $F$ as before, 
if $\hil$ is a Hopf $*$-algebra with invertible antipode (as will
always be the case in the present paper) the only condition for the deformation
$\hil^F$ to be a Hopf $*$-algebra with the same $*$-structure is $(g\times_F h)^* = h^*\times_F g^*$, which forces the Drinfel'd twist $F$ to be \emph{real}, i.e. to satisfy
$$ 
\overline{F(g\otimes h)} = F(h^*\otimes g^*)
$$
for all $g,h \in \mathcal{H}$. If $(A,\mu)$ is a left
$\mathcal{H}$-comodule $*$-algebra, the $*$-structure of $A$ makes
$A^F$ into a $*$-algebra if and only if the Drinfel'd twist is real; in addition, $A^F$ becomes an $\mathcal{H}^F$-comodule $*$-algebra. 

Right analogs of all the structures above are straightforwardly constructed.

\subsection{Cocycle deformations of the complex torus and of $\GL(n)$}\label{se:cdal}

We deform the coordinate algebra of the complex torus $T=
(\complex^\times)^n$ via the cocycle twist procedure. The Pontrjagin
dual group $\widehat{T}=\Hom_\complex(T,\complex^\times)\cong \zed^n$
is the group of characters of the group
$T$. For $p=(p_1,\dots,p_n)\in\zed^n$ and $t=(t_1,\dots,t_n) \in T$, the characters are given by
\beq
t^p:=t_1^{p_1}\cdots t_n^{p_n} \ . 
\label{Tchar}\eeq

The unital algebra $\hil=\alg(T)$ of coordinate functions on the torus
$T$ is the Laurent polynomial algebra $\hil:=\complex(t_1,\dots,t_n)$
of characters. It has Hopf algebra structure
$$
\Delta(t^p)=t^p\otimes t^p \ , \qquad \varepsilon(t^p)=1 \ , \qquad
S(t^p)=t^{-p}
$$
for $p\in \zed^n$, with the coproduct and counit extended as algebra morphisms and the antipode as an anti-algebra morphism. 
The canonical right action of $T$ on itself by group multiplication dualizes to a left $\hil$-coaction
\beq
\Delta_{\alg(T)}\,:\, \alg(T)~\longrightarrow~ \hil\otimes \alg(T) \ , \qquad	
\Delta_{\alg(T)}(u_i^{\pm 1})= t_i^{\pm1} \otimes u_i^{\pm 1}\ , 
\label{DeltaLT}\eeq
where we write $u_i^{\pm 1}$, $i=1,\dots,n$, for the generators of $\alg(T)$ viewed as a left comodule algebra over itself, to distinguish the coordinate algebra $\alg(T)$ from the Hopf algebra~$\hil$. 

To deform the Hopf algebra $\hil$ by a Drinfel'd twist we choose a
complex skew-symmetric $n\times n$ matrix $\theta=(\theta^{ij})$ and
on characters $t^p,t^q\in \hil$ we set
$$
F_\theta(t^p\otimes
t^q)=\exp\Big(\, \frac\ii2 \, \sum_{i,j=1}^n\,  p_i\, \theta^{ij} \,
q_j\, \Big) \ ,
$$ 
and extend it by linearity. Then $F_\theta$ is completely determined by its values on generators
\beq
F_\theta(t_i\otimes t_j)= \exp\big(\mbox{$\frac\ii2$}\,\theta^{ij}\big) =: q_{ij} \ , \qquad i,j=1,\dots,n \ .
\label{Ftheta}
\eeq
One easily checks that $F_{\theta}$ is a Drinfel'd twist element for $\hil$ and that $\hil_{\theta}:=\hil^{F_{\theta}} = \hil$ as a Hopf algebra. 
However, $\hil$ and $\hil_\theta$ differ as cotriangular Hopf
algebras: while $\hil$ has trivial cotriangular structure
$\Rcal=\varepsilon\otimes\varepsilon$, the twisted cotriangular
structure on $\hil_\theta$ is given by
\beq
\Rcal_\theta(t_i\otimes t_j)= F_\theta(t_j\otimes t_i)\,
F_\theta^{-1}(t_i\otimes t_j) =
F_\theta^{-2}(t_i\otimes t_j) =q_{ij}^{-2} \ .
\label{brF}
\eeq
Then $\alg(T)$, regarded as an $\hil$-comodule algebra, acquires by
(\ref{dualtwp}) a deformed noncommutative product
$u_i\star_\theta u_j=F_\theta(t_i\otimes t_j)~ u_i\,u_j =q_{ij}~ u_i\,u_j$ which corresponds to relations
$$
u_i\star_\theta u_j=q_{ij}^2~ u_j\star_\theta u_i \ , \qquad
u_i^{-1}\star_\theta u_j=q_{ij}^{-2} ~u_j\star_\theta u_i^{-1} \qquad \textup{for} \quad  i,j=1,\dots,n \ .  
$$
The algebra $\alg(T_{\theta}):=(\alg(T),\star_{\theta})=\lauq$ is dual
to a noncommutative variety denoted $T_\theta=(\complex_\theta^\times)^n$ and 
called the \emph{noncommutative complex torus}. 
We stress that the deformation affects only the algebra structure of
$\alg(T)$, not the Hopf algebra structure of $\hil$. This means we
deform the complex torus only as a space, not as a group, as in the case of the noncommutative real torus. 

We can also define a noncommutative toric
deformation of the complex general linear group $\GL(n)$ realized
again by a Drinfel'd twist. Consider the Hopf algebra of coordinate
functions $\fred_n:=\Fun(\GL(n))$, with generators $g_{ij}$,
$i,j=1,\ldots ,n$. We identify the generators $t_i\in\hil$ with
$g_{ii}\in\fred_n$ {(dual to the embedding
  $(\complex^\times)^n\hookrightarrow \GL(n)$)}, so that the twist element \eqref{Ftheta} which deforms the complex torus is a Drinfel'd twist for $\fred_n$ too. We have 
$$F_{\theta}(g_{ij}\otimes g_{kl}) =
\exp\big(\mbox{$\frac\ii2$}\,\theta^{ik}\big) \, \delta_{ij}\,
\delta_{kl} = q_{ik}\ \delta_{ij}\, \delta_{kl} \  ,
$$ 
yielding a new Hopf algebra $\fred_n^{\theta}$ with twisted multiplication as in \eqref{gammantip}. Introducing coefficients
\begin{equation}
\label{qcoeff}
Q_{ij\,;\,kl}= q_{ki}\,q_{jl}=q^{-1}_{ik}\,q_{jl} \ , \qquad
Q^2_{ij\,;\,kl} = q^2_{ki}\,q^2_{jl}
\end{equation}
the commutation rules for the generators of $\fred^{\theta}_n$ are computed to be
\begin{equation}
\label{pcommgamma}
g_{ij}\times_{\theta}g_{kl} =
Q^2_{ij\,;\,kl}~g_{kl}\times_{\theta}g_{ij} \ .
\end{equation}
While the coproduct $\Delta(g_{ij})=\sum_k\, g_{ik}\otimes g_{kj}$ and the counit $\varepsilon(g_{ij})=\delta_{ij}$ are left unchanged by the general procedure, direct computations show that the antipode $S(g_{ij})=(g^{-1})_{ij}$ is unaltered as well. 

The noncommutative Hopf algebra $\fred_n^\theta=
(\fred_n,\times_\theta,\Delta,\varepsilon,S)$ is called the
\emph{complex torus deformation quantum group of $\GL(n)$}. It is
dual to a noncommutative variety which we denote by $\GL_\theta(n)$.

\subsection{Quantum determinants and minors\label{Quantumdet}}

The coordinate algebra of $\GL_\theta(n)$ should be properly defined as the Ore localization of the
noncommutative algebra generated by arbitrary matrix units with respect to an invertible and permutable element $\det_\theta$, the determinant element. For a generic $n\times n$ matrix whose entries $g_{ij}$ satisfy commutation relations \eqref{pcommgamma} we adapt the Leibniz formula which expresses the determinant as a linear combination of products $\prod_{i}\,g_{i\,\sigma(i)}$ or $\prod_{i}\,g_{\sigma(i)\,i}$ as the permutation $\sigma$,
weighted by its sign, runs through the symmetric group $S_n$. In order to get a well defined element of $\fred_n^{\theta}$ we put a factor
$Q_{lj\,;\,ki}$ for every pair $g_{ki}\,g_{lj}$ appearing in
$\prod_{i}\,g_{i\,\sigma(i)}$. However, a more succinct expression is
achieved via a $\theta$-deformed Levi--Civita symbol
$\epsilon_\theta$. Since the row and column indices in (\ref{qcoeff})
and (\ref{pcommgamma}) behave differently, we actually need to
introduce two different symbols $\epsilon_{\theta}^{(r)}$, which
refers to row indices, and $\epsilon_{\theta}^{(c)}$, which refers to
column indices; they are given by
\begin{align*}
\epsilon_{\theta}^{i_1\cdots
  i_n\, (c)} &= {\rm sgn}(i_1\cdots
i_n)\,\prod_{k=1}^{n-1}~\prod_{s=1}^{n-k} \,
Q_{s+k\,i_{s+k}\,;\,k\,i_k} \ , \\[4pt]
\epsilon_{\theta}^{j_1\cdots j_n\, (r)} &= {\rm sgn}(j_1\cdots
j_n)\,\prod_{k=1}^{n-1}~\prod_{s=1}^{n-k} \,
Q_{j_{s+k}\,s+k\,;\,j_k\,k} \ .
\end{align*}
They obey the alternating rules
\begin{align}
\epsilon_{\theta}^{j_1\cdots \, j_{\alpha}\cdots j_{\beta} \, \cdots
  j_n \, (c)} &=
- \, q^{-2}_{j_{\alpha}j_{\beta}} ~ \epsilon_{\theta}^{j_1\cdots \,
  j_{\beta}\cdots j_{\alpha} \, \cdots j_n\, (c)} \ , \nonumber \\[4pt]
\epsilon_{\theta}^{i_1\cdots \, i_{\alpha}\cdots i_{\beta} \,
  \cdots i_n \, (r)} &=
- \, q^2_{i_{\alpha}i_{\beta}} ~ \epsilon_{\theta}^{i_1\cdots \,
  i_{\beta}\cdots i_{\alpha} \, \cdots i_n \, (r)} \ ,
\label{qeps}
\end{align}
and the symbol $\epsilon^{(r)}_\theta$ may be thought of as the
inverse of the symbol $\epsilon^{(c)}_\theta$. Using them we may
absorb the $Q$-dependent factors.

The \emph{quantum determinant} is then the element of $\fred_n^\theta$ given by 
\begin{equation}
\label{qdet}
{\det}_{\theta} = \frac{1}{n!} \, \sum_{i_k,j_k=1}^n\, \epsilon_{\theta}^{i_1\cdots i_n\,(r)}\,\epsilon_{\theta}^{j_1\cdots j_n\,(c)}\,g_{i_1j_1}\cdots g_{i_nj_n} \ .
\end{equation}
Note that ${\det}_\theta$ is central in $\fred_n^\theta$ if and only if 
$ \sum_{k=1}^n\,\theta^{ki}=\sum_{k=1}^n\,\theta^{kj}$, modulo $2\pi$, for all $i,j=1,\dots,n$.
On the other hand, the element 
${\det}_{\theta}$ is an $\hil_\theta$-coeigenvector which is left and
right permutable in $\fred_n^\theta$. This means that 
its non-negative integer powers 
form a left and right
denominator set in~$\fred_n^\theta$, and hence we can perform
noncommutative Ore localization with respect to $\det_{\theta}$. 

The determinant element (\ref{qdet}) originates from the braiding $\Psi^{\theta}$
of the category of left $\fred_n^{\theta}$-comodules ${^{\fred_n^{\theta}}\Module}$ given, in the present case,
by \eqref{PsithetaVW} and \eqref{brF}. The $\theta$-deformed exterior algebra of degree $d$ for an object $V$ in ${^{\fred_n^{\theta}}\Module}$ is defined as
\begin{equation}
\label{qest}
\mbox{$\bigwedge^d_{\theta}$}\,V := V^{\otimes d }\,\big/\,\big\langle v_1\otimes v_2 + \Psi^\theta_{V,V}(v_1\otimes v_2) \big\rangle_{v_1,v_2 \in V} \ .
\end{equation}
The braided skew-symmetric algebra $\bigwedge^d_{\theta}\,V$ is spanned by the collection of minors of order $d\leq n$ in elements of $V$ when $n$ is the number of generators of $V$. Indeed consider two multi-indices $I=(i_1 \cdots i_d)$ and $J=(j_1 \cdots j_d)$ labelling the rows and columns of a given minor, and define the determinant $\Lambda^{IJ}$ of this sub-matrix as
\begin{equation}
\label{dets}
\Lambda^{IJ} = \frac{1}{d!} \,\sum_{i_k\in I,j_k\in J} \, \epsilon_{\theta}^{i_1\cdots
  i_d\,(r)}\,\epsilon_{\theta}^{j_1\cdots j_d\,(c)}\,g_{i_1j_1}\cdots
g_{i_dj_d} 
\end{equation}
where the symbols $\epsilon_{\theta}$ satisfy alternating rules derived from (\ref{qeps}). Here the $\fred_n^\theta$-comodule structure of $\GL(n)$ is induced from the $\fred_n^\theta$-comodule structure of $V$ and of its
dual $V^*$. When this $\fred_n^\theta$-comodule structure induces the noncommutative product (\ref{pcommgamma}) among the entries of
elements of $\GL(n)$, the alternating properties of the deformed Levi--Civita symbols coincide with those of (\ref{qeps}).

For later use we introduce a deformation of the Laplace expansion of the determinant in terms of lower order minors. If $I$ is a row multi-index, $J$ a column multi-index with $|I|=|J|=d$ we write $I^{\alpha}=I\setminus \{i_{\alpha}\}= (i^{\alpha}_1, \dots, i^{\alpha}_{d-1})$  and $J^{\alpha}=J\setminus\{j_{\alpha}\}=(j^{\alpha}_1, \dots, j^{\alpha}_{d-1})$ for $\alpha\in (1,\ldots ,d)$. 
By taking into account the appropriate $Q$-factors, the noncommutative
version of the classical expansion of the determinant $\Lambda^{IJ}$ with respect to the $k$-th row is 
\begin{equation}
\label{nlapkr}
\Lambda^{IJ} = \sum_{\alpha=1}^d \ \prod_{\beta =1}^{d-1}\,
(-1)^{k+\alpha} \, Q_{i^k_{\beta}j^{\alpha}_{\beta}\, ;\, k\alpha} \,
g_{k\alpha}\, \Lambda^{I^k J^{\alpha}} = \sum_{\alpha=1}^d \
\prod_{\beta =1}^{d-1} \, (-1)^{k+\alpha}\, Q_{k\alpha\, ;\,
  i^k_{\beta}j^{\alpha}_{\beta}}\, \Lambda^{I^k J^{\alpha}} \, g_{k\alpha} \ .
\end{equation}
Similarly, if we expand with respect to the $k$-th column we get 
\begin{equation}
\label{nlapkc}
\Lambda^{IJ} = \sum_{\alpha=1}^d \ \prod_{\beta =1}^{d-1}\,
(-1)^{k+\alpha} \, Q_{i^k_{\beta}j^{\alpha}_{\beta}\,;\, \alpha k} \,
g_{\alpha k}\, \Lambda^{I^{\alpha} J^k} = \sum_{\alpha=1}^d \
\prod_{\beta =1}^{d-1}\, (-1)^{k+\alpha} \, Q_{\alpha k\, ;\,
  i^k_{\beta}j^{\alpha}_{\beta}} \, \Lambda^{I^{\alpha} J^k} \, g_{\alpha k}\ .
\end{equation}

We will also need the commutation rules between any two $d\times d$ and $d'\times d'$ minors $\Lambda^{IJ}$ and $\Lambda^{I'J'}$, when $\fred_n^\theta$ is regarded as the coordinate algebra $\alg(\GL_\theta(n))$ of the noncommutative variety
$\GL_\theta(n)$. A direct computation gives
\begin{equation}
\label{ncmin}
\Lambda^{IJ} \, \Lambda^{I'J'} = R^2_{IJ\,;\,I'J'} ~\Lambda^{I'J'} \, \Lambda^{IJ} 
\end{equation}
where we introduced the commutation coefficient
\beq
\label{rcoeff}
R_{IJ\,;\,I'J'}= \prod_{\alpha=1}^d~\prod_{\alpha'=1}^{d'}\,
Q_{i_{\alpha}j_{\alpha}\,;\,i'_{\alpha'}j'_{\alpha'}} \ .
\eeq
In particular, this shows that the minors of order $d$ generate a subalgebra.

Finally, consider a pair of multi-indices differing only by transposition of two columns, $J=(j_1\cdots j_{\alpha}\cdots j_{\beta}\cdots j_d)$ and $J^{t_{\alpha\beta}}=(j_1\cdots
j_{\beta}\cdots j_{\alpha}\cdots j_d)$. From (\ref{dets}) one obtains
\begin{equation}
\label{qalt}
\Lambda^{IJ} = (-1)^{\beta-\alpha} \, \Lambda^{IJ^{t_{\alpha\beta}}} \ ,
\end{equation}
which are alternating relations that can be further generalized to arbitrary permutations. 

\subsection{Noncommutative deformations of toric varieties}\label{ncdtv}

A \emph{toric variety} $X$ of dimension $n$ is an irreducible
algebraic variety over $\complex$ which contains the complex torus $T=
\at$ as a Zariski open subset and the regular action of $T$ on itself
extends to an action on the whole of $X$. Basic examples are the
complex torus $T$ itself, the affine planes $\mathbb{C}^n$, the projective spaces $\mathbb{CP}^n$, and the weighted projective spaces $\mathbb{CP}^n[a_0,a_1,\ldots ,a_n]$. 

One of the main interests in studying toric varieties is that their
geometry can be encoded by combinatorial data, a fan, that describes
the way in which $T$ acts on $X$. Various properties of a toric
variety, such as smoothness and compactness, may be formulated
entirely in terms of the fan structure. We will quantize toric
varieties by using the same fan description and deforming the coordinate algebra of every cone of the fan.

Recall that a rational polyhedral cone $\sigma\subset \real^n$
is a cone $\sigma=\real_{\geq0} v_1+ \cdots +\real_{\geq0} v_s$ generated by finitely
many elements $v_1,\dots,v_s\in \zed^n$. It is {strongly convex} if it does not contain any real line, $\sigma \cap (-\sigma) = 0$.  For every strongly convex rational polyhedral cone $\sigma\subset \real^n$  
we define the dual cone 
$\sigma^{\vee} := \{ m\in \real^n~|~ m\cdot u \geq 0 \quad \forall u\in\sigma \}$.
The set $\sigma^\vee\cap \zed^n$ is a finitely generated semigroup
under addition, with generators $(m_1,\ldots ,m_l)$ where $l\geq n$.
To each $m_a=((m_a)_1,\dots, (m_a)_n)$ one associates a Laurent monomial
in $\lau$ by the assignment $m_a \mapsto t^{m_a}=t_1^{(m_a)_1}\cdots t_n^{(m_a)_n}$. 
The product between any two such elements is via the
corresponding sum of characters, $t^{m_a}\cdot t^{m_b} :=
t^{m_a+m_b}$. Thus the generators of $\sigma^{\vee}\cap \zed^n$ span a
subalgebra of $\lau$ denoted $\rs$, the coordinate
algebra of a normal affine toric variety $U[\sigma]=\textup{Spec}(\rs)$. 
Note that the inclusion $0\hookrightarrow \sigma$ induces an embedding of the torus
$T=U[0]$ as a dense open subset of $U[\sigma]$. 

Our noncommutative
deformation is induced by the cocycle deformation of this
complex torus. It starts with the same complex skew-symmetric matrix $\theta$ as in \eqref{Ftheta}. The algebra $\rsq$ is defined to be the subalgebra of $\lauq$ generated by $\{t^{m_a}\}$ with product
$$
t^{m_a}\star_{\theta}t^{m_b}:=\exp\Big(\, \frac\ii2 \
\sum_{i,j=1}^n \, (m_a)_i\, \theta^{ij}\,(m_b)_j\, \Big)~ t^{m_a+m_b} \ .
$$
This corresponds to a deformation of the algebra generated by the
characters, but not of their group structure. It is a finitely-generated
$\hil_\theta$-comodule subalgebra of the algebra $\alg(T_\theta)$ of the
noncommutative torus. The noncommutative
affine variety corresponding to the algebra $\rsq$ is denoted $U_\theta[\sigma]$. It is a multi-parameter deformation of $U[\sigma]$.

The variety $U[\sigma]$ may also be described as an embedding in the complex plane $\complex^l$. If $\sigma^{\vee}\cap \zed^n$ has $l$ generators, consider the polynomial algebra $\mathbb{C}[x_1,\ldots ,x_l]$ (one variable $x_a$ for each $m_a$). Since the generators $m_a$ are $l$ rational vectors in $\real^n$, there are $l-n$ linear relations among them. Then we may quotient the algebra $\mathbb{C}[x_1,\ldots ,x_l]$ by the ideal generated by the $l-n$ relations for the vectors $m_a$, realized as multiplicative relations for the variables $x_a$. 
If $R[m_a]\subset\mathbb{C}[x_1, \ldots ,x_l]$ denotes the subspace
generated by these relations, then $U[\sigma]$  
is realized as the spectrum of the quotient algebra $\rs=\mathbb{C}[x_1, \ldots , x_l]/\langle R[m_a]\rangle$. An analogous realization is possible for noncommutative affine toric
varieties. The noncommutative deformation of the polynomial algebra $\mathbb{C}[x_1, \ldots, x_l]$
is obtained from the multiplicative relations between the Laurent monomials
$t^{m_a}$. These relations become
\begin{equation}
\label{nclm}
t^{m_a}\star_{\check\theta}t^{m_b}:=\check q_{ab}~t^{m_a+m_b} \ ,
\end{equation}
where we denote $\check\theta_{ab}:= \sum_{i,j}\, (m_a)_i\,\theta^{ij}\,(m_b)_j$ and $\check q_{ab}=\exp(\frac\ii2\,\check\theta_{ab})$ with $a,b=1,\ldots ,l$, and $i,j=1,\ldots ,n$.
Then the generators of the algebra of the affine variety obey
\begin{equation}
\label{ncav}
x_a\star_{\check\theta} x_b = \big(\check q_{ab}\big)^2~ x_b \star_{\check\theta} x_a \ .
\end{equation}
These relations define the $l$-dimensional complex Moyal
plane {$\complex_{\check\theta}^l$} with coordinate algebra~$\mathbb{C}_{\check\theta}[x_1,\ldots ,x_l]$. The $l-n$ linear relations among the generators of the dual cone $\{m_a\}$ are now expressed in the character algebra. They can always be brought to the form
$$
\sum_{a=1}^l\, (p_{s,a}-r_{s,a}) ~m_a=0 \qquad \mbox{for} \quad
s=1,\dots,l-n \ , 
$$
with non-negative integer coefficients $p_{s,a}$ and $r_{s,a}$. 
Then for each $s$, one obtains from (\ref{nclm}) the additional relation
\begin{equation}
\label{excon}
x_1^{p_{s,1}}\star_{\check\theta}\cdots\star_{\check\theta}x_l^{p_{s,l}}= \Big(\, \prod_{1\leq a<b\leq l}\,\big(\check q_{ab}\big)^{p_{s,a}\,p_{s,b}-r_{s,a}\,r_{s,b}}\,\Big)~ x_1^{r_{s,1}}\star_{\check\theta}\cdots\star_{\check\theta}x_l^{r_{s,l}} \ . 
\end{equation}
The subspace of relations (\ref{excon}) is denoted $R_{\check\theta}[m_a]$. It is a multi-parameter deformation of the subspace $R[m_a]$, which generates a two-sided ideal in $\mathbb{C}_{\check\theta}[x_1,\ldots ,x_l]$. 
{The quotient algebra $\mathbb{C}_{\check\theta}[x_1,\ldots
  ,x_l]/\langle R_{\check\theta}[m_a] \rangle$ is an alternative realization of the coordinate algebra $\rsq$ of the noncommutative affine variety $U_{\theta}[\sigma]$.}

We obtain generic toric varieties by gluing together affine toric
varieties. For this, we recall the notion of a
fan. Given a cone $\sigma\subset \real^n$, a {face}
$\tau\subset\sigma$ is a subset of the form $\tau=\sigma\cap m^\perp$
for some $m\in\sigma^\vee$, where $m^\perp:=\{u\in \real^n~|~m\cdot
u=0\}$. A {fan} $\Sigma\subset \real^n$ is a non-empty finite
collection of strongly convex rational polyhedral cones in $\real^n$
satisfying two conditions: firstly, if $\sigma\in\Sigma$ and $\tau$ is
a face of $\sigma$, then $\tau\in\Sigma$; secondly, if $\sigma, \sigma'
\in \Sigma$, then the intersection $\sigma\cap\sigma'$ is a face of both
$\sigma$ and $\sigma'$. To a fan $\Sigma$ in $\real^n$ we associate a toric variety $X=X[\Sigma]$. The cones 
$\sigma\in\Sigma$ correspond to the open affine subvarieties
$U[\sigma]\subset X[\Sigma]$, and $U[\sigma]$ and $U[\sigma'\,]$ are
glued together along their common open subset $U[\sigma\cap\sigma'\,
]=U[\sigma]\cap U[\sigma'\, ]$. 

In a similar way, we obtain generic noncommutative toric varieties
$X_\theta[\Sigma]$ by gluing together noncommutative affine toric
varieties. If $\sigma$ and $\sigma'$ are two cones in the fan $\Sigma$
which intersect along the face $\tau=\sigma\cap\sigma'$, then there are canonical morphisms between the associated noncommutative algebras $\rsqq{\sigma} \rightarrow \rsqq{\tau}$ and $\rsqq{\sigma'\,} \rightarrow \rsqq{\tau}$ induced by the inclusions
$\tau\hookrightarrow\sigma$ and $\tau\hookrightarrow\sigma'$. The images of these morphisms in $\rsqq{\tau}$ are related by an algebra automorphism in the category
${}^{\hil_\theta}\Module$ playing the role of a ``coordinate transition function'' between $U_{\theta}[\sigma]$ and $U_{\theta}[\sigma'\,]$. See~\cite[\S 3.1]{CLSI} for the explicit description in the general case, and \S\ref{NCprojvar} below for the example of the noncommutative projective plane. 

Since each algebra $\complex_\theta[\sigma]$ for $\sigma\in\Sigma$ is a subalgebra of $\lauq$, the intersection of the algebras $\complex_\theta[\sigma]$ is
well-defined, and the ``algebra of coordinate functions'' $\alg\big(X_\theta[\Sigma]\big)$ on $X_\theta[\Sigma]$ can be represented via the exact sequence
\beq
0~\longrightarrow~\alg\big(X_\theta[\Sigma]\big)~\longrightarrow~ \prod_{\sigma\in\Sigma}\,\complex_\theta[\sigma]~\longrightarrow~
\prod_{\sigma,\sigma'\in\Sigma}\,\complex_\theta[\sigma\cap\sigma'\,] \ ,
\label{XSigmaalg}\eeq
with the gluing automorphisms discussed above. The sequence
(\ref{XSigmaalg}) is an exact sequence in the category
${}^{\hil_\theta}\Module$, and hence $\alg\big(X_\theta[\Sigma]\big)$ is a left
$\hil_\theta$-comodule algebra.

\subsection{Noncommutative projective varieties\label{NCprojvar}}

We specialize to noncommutative projective spaces
$\complex\P_\theta^n$. 
These varieties admit also a
``global'' description via noncommutative homogeneous coordinate
algebras which, after Ore localization, reduces to the local
description provided by the noncommutative affine varieties
$U_\theta[\sigma]$. Moreover, they are used to define 
noncommutative deformations of projective varieties, such as grassmannians, via restriction. 

\subsubsection*{Noncommutative projective plane}

Consider lattice vectors $v_1=(1,0)$, $v_2=(0,1)$ and $v_3=(-1,-1)$ in
$\zed^2$. The fan $\Sigma_{\complex\P^2}$ of $\complex\P^2$ has one-dimensional cones $\tau_i=\real_{\geq0}v_i$, while the three maximal cones are generated by pairs of these vectors as
$
\sigma_i=\real_{\geq0}v_{i+1}+\real_{\geq0}v_{i+2}
$
(with indices $i=1,2,3$ read mod~3) with
$\sigma_i\cap\sigma_{i+1}=\tau_{i+2}$ and $\sigma_i\cap\sigma_j=0$
otherwise. The zero cone is the triple overlap $\sigma_1\cap\sigma_2\cap\sigma_3=0$. Pictorially
$$
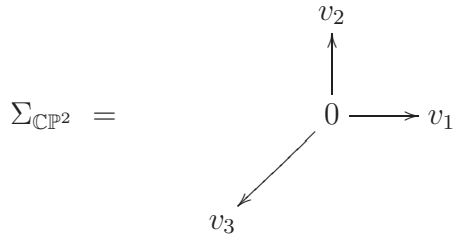
\xymatrix{
 & & {v_2} & \\
\Sigma_{\complex\P^2} \ = \ 
 & & {0} \ar[dl] \ar[u] \ar[r] &{ v_1} \\
 & { v_3} & &
}
$$
The corresponding open affine
subvarieties $U[\sigma_i]$ generate an open cover of $X[\Sigma_{\complex\P^2}]=\complex\P^2$. 

For the noncommutative deformation we write out the
relations among the generators of the subalgebras
$\complex_\theta[\sigma_i]\subset\complex_\theta(t_1,t_2)$. Each dual
cone $\sigma_i^\vee$ is strongly convex and hence the generators of
$\sigma_i^\vee\cap \zed^2$ are independent. The semigroup
$\sigma_3^\vee\cap \zed^2$ is generated by $m_1=(1,0)$ and $m_2=(0,1)$, so
that $\check\theta=\theta:= \theta^{12}$ for the deformation matrix, and the algebra $\complex_\theta[\sigma_3]=\complex_\theta[x_1,x_2]$ is generated by $x_a=t^{m_a}=t_a$, $a=1,2$, with the relation
  \beq x_1\,x_2=q^2~x_2\,x_1
  \label{x1x2basic}
  \eeq
where $q:=q_{12}$.
The other two cones $\sigma_i$ for $i=1,2$ are similarly treated and,
after suitable redefinitions of the generators, in each case one finds
$\check\theta=\theta$ and that $\complex_\theta[\sigma_i]$ is
generated by elements $x_1,x_2$ satisfying the relations (\ref{x1x2basic}). All
three varieties $U_\theta[\sigma_i]\cong\complex_\theta^2$ are thus
copies of the two-dimensional complex Moyal plane.

To glue the noncommutative affine toric varieties together, consider for example the face $\tau_1=\sigma_3\cap\sigma_1$. The semigroup $\tau_1^\vee\cap \zed^2$ is generated by $m_1=(1,0)$, $m_2=(0,1)$ and $m_3=-m_2$. The generators of the subalgebra
$\complex_\theta[\tau_1]=\complex_\theta[t_1,t_2,t_2^{-1}]$ are the elements $y_1=t_1$, $y_2=t_2$ and $y_3=t_2^{-1}$ with the relations
\beq
y_1\,y_2=q^2~y_2\,y_1 \ , \qquad
y_1\,y_3=q^{-2}~y_3\,y_1 \ , \qquad
y_2\,y_3=1=y_3\,y_2 \ ,
\label{y1y2y3basic}\eeq
which we may identify as the algebra dual to a noncommutative
projective line $\complex\P_\theta^1$. The algebra morphisms
$\complex_\theta[\sigma_1]\to\complex_\theta[\tau_1]$ and
$\complex_\theta[\sigma_3]\to\complex_\theta[\tau_1]$ are both natural
inclusions of subalgebras, and in this manner there is a natural algebra automorphism $\complex_\theta[\sigma_1]\to\complex_\theta[\sigma_3]$. The other
faces are similarly treated, and thus the noncommutative toric geometry of $\complex\P^2_\theta=X_\theta[\Sigma_{\complex\P^2}]$ can be assembled into a
diagram of gluing morphisms
$$
\xymatrix{
 & \complex_\theta\big[t_1^{-1}\,,\,(t_1\,t_2^{-1})\,,\,
 (t_1\,t_2^{-1})^{-1} \big] \ar[d]&
 \\ \complex_\theta\big[t_1^{-1}\,,\,t_1^{-1}\,t_2\big]~
\ar[ur]\ar[r]\ar[d]& ~
\complex_\theta(t_1,t_2) ~ & ~ \ar[l]\ar[d]\ar[ul] 
\complex_\theta\big[t_1\,t_2^{-1}\,,\,t_2^{-1}\big] \\
\complex_\theta\big[t_1\,,\,t_1^{-1}\,,\,t_2\big] ~ \ar[ur] & ~ 
\complex_\theta[t_1,t_2] ~ \ar[l]\ar[r] \ar[u]& ~ \complex_\theta
\big[t_1\,,\,t_2\,,\,t_2^{-1}\big]\ar[ul]
}
$$
The noncommutative affine variety associated to the zero cone is the spectrum of the full deformed character algebra
$\complex_\theta[0]=\complex_\theta(t_1,t_2)$, corresponding to the open embedding of the noncommutative complex torus {into $\complex\P^2_\theta$}.

This construction generalizes straightforwardly to higher-dimensional projective spaces $\complex\P^n$, $n>2$,
leading to noncommutative projective spaces $\complex\P_\theta^n$. 

\subsubsection*{Homogeneous coordinate algebras\label{Homcoordalg}} 

We now describe a homogeneous coordinate algebra for the noncommutative projective spaces
$\complex\P_\theta^n$, recovering a local description by noncommutative Ore localization. 
For this, consider the embedding $(\complex_\theta^\times)^n\hookrightarrow (\complex_{\tilde\theta}^\times)^{n +1}$ with
$$
\tilde\theta=\begin{pmatrix}
\theta & 0 \\
0      & 0
\end{pmatrix} \ .
$$
The corresponding complex Moyal plane $\complex_{\tilde\theta}^{n+1}$ is defined by the graded polynomial algebra $\complex_{\tilde\theta}[w_1,\dots,w_{n+1}]$ in $n+1$ generators $w_i$, $i=1,\dots,n+1$, of degree~$1$ with quadratic relations
\bea
w_{n+1}\,w_i&=&w_i\,w_{n+1} \ , \qquad i=1,\dots,n \ , 
\nonumber \\[4pt]
w_i\,w_j&=&q_{ij}^2~w_j\,w_i \ , \qquad i,j=1,\dots,n \ .
\label{wquadrels}\eea
This graded algebra is called the homogeneous coordinate algebra $\alg=\alg(\complex\P_\theta^n)$ of the noncommutative variety $\complex\P_\theta^n$. 
The grading on $\alg$ is by the usual polynomial degree. There is a
natural coaction $_{\alg}\Delta:\alg\to\alg\otimes\fred^\theta_{n}$ with
$_{\alg}\Delta(w_i)=w_i\otimes g_{ii}$ for $i=1,\ldots, n$, and
$_{\alg}\Delta(w_{n+1})=w_{n+1}\otimes 1$, defined via the embedding
$(\complex_\theta^\times)^n\hookrightarrow(\complex_{\tilde\theta}^\times)^{n+1}$. This
makes $\alg$ into an object of the category ${}^{\hil_\theta}\Module$.

It is straightforward to verify that each monomial $w_i$ generates a left (and right) denominator set in $\alg$. Let $\alg{[w^{-1}_i]}$ be the left Ore localization of $\alg$ with respect to $w_i$. Since $w_i$ is homogeneous of degree~$1$, the algebra $\alg[w_i^{-1}]$ is also $\nat_0$-graded. Elements of degree~0 in $\alg{[w^{-1}_i]}$ form a subalgebra which we denote by $\alg{[w^{-1}_i]}_0$. Noncommutative affine subvarieties associated to maximal cones can be recovered as the Ore localizations of the homogeneous coordinate algebra, since for each maximal cone $\sigma_i$ there is a natural isomorphism of noncommutative algebras
$\complex_\theta[\sigma_i]\cong\alg\big[w^{-1}_i\big]_0$ in the
category ${}^{\hil_\theta}\Module$~\cite[Thm. 5.4]{CLSI}. Note,
however, that the surjective map $\alg\to\alg\big[w^{-1}_i\big]_0$ is
in general an algebra homomorphism only for $i=n+1$ corresponding to
the central generator $w_{n+1}$. In the
classical case $\theta=0$, the category of coherent sheaves on the
projective space $\complex\P^n$ is equivalent to the category of
finitely-generated $\alg$-modules~\cite[Cor.~2.4]{AKO}; in
\S\ref{NCsheaves} we show that this equivalence extends to the
noncommutative case $\theta\neq0$ as well.

We use noncommutative homogeneous coordinate algebras to introduce {\it noncommutative projective varieties}.
If $I\subset\alg$ is a graded two-sided ideal generated by a set of homogeneous polynomials $f_1,\dots,f_m\in\complex_{\tilde\theta}[w_1,\dots,w_{n+1}]$, then the quotient algebra $\alg_I:=\alg/I$ is the coordinate algebra of a noncommutative projective variety. The projection
$\pi_I:\alg\to\alg_I$ can be regarded as the dual of a closed embedding given by  $X_\theta(I)\hookrightarrow\complex\P_\theta^n$, identified with the common zero locus in $\complex_{\tilde\theta}^{n+1}$ given by the set of relations $\{f_1=0,\dots,f_m=0\}$. Its homogeneous coordinate algebra
$\pi_I(\complex_{\tilde\theta}[w_1,\dots,w_{n+1}])$ has relations (\ref{wquadrels}) and $f_1=0,\dots,f_m=0$. It is also graded,
with $(\alg_I)_0=\complex$ and $\dim_\complex(\alg_I)_k<\infty$ for all $k\geq0$. The $\hil_\theta$-coaction on $\alg$ naturally restricts to $\alg_I$. 

\subsubsection*{Noncommutative grassmannians and flag varieties\label{ncgrass}}

A prominent example of noncommutative projective varieties in the
following will be noncommutative Grassmann varieties {$\Gr_\theta(d;n)$} associated to an $\hil_\theta$-comodule $V$ of dimension~$n>d$. 
An element of the homogeneous coordinate algebra of $\Gr_\theta(d;n)$
is defined as an element in $\mathbb{P}(\bigwedge_{\theta}^dV)$,
obtained by taking the $\theta$-deformed exterior product of $d$ rows
of a matrix in $\alg(\GL_{\theta}(n))$ (and quotienting by the
appropriate equivalence relation). We take a noncommutative $d\times
n$ matrix representing an element of $\alg(\Gr_{\theta}(d;n))$ and
send it into the ${n\choose d}$-tuple of its minors; we label each of
them $\Lambda^{J}$ by ordered $d$-multi-indices $J=(j_1\cdots j_d)$,
$1\leq j_\alpha\leq n$. Then we provide noncommutative relations
between minors, seen now as homogeneous coordinates in
$\alg(\mathbb{CP}_{\Theta}^{N})$ {with $N={n\choose d}-1$}, together with noncommutative Pl\"ucker relations between them.

From (\ref{ncmin}) with $|J|=|J'\,|=d$ representing two different minors we have
\begin{equation}
\label{ncrelpr}
\Lambda^{J}\,\Lambda^{J'} = \Big(~\prod_{\alpha,\beta =1}^d
\,q^2_{j_{\alpha}j'_{\beta}}~ \Big)~ \Lambda^{J'}\,\Lambda^{J} \ .
\end{equation}
With respect to the noncommutative algebraic torus
$T_\theta=(\complex^\times_\theta)^n$, the commutation relations
(\ref{ncrelpr}) come
from regarding the coordinate algebra $\alg(\Gr_{\theta}(d;n))$
naturally as an object of the category ${}^{\hil_\theta}\Module$ by the left coaction
\beq
\Delta_{\alg(\Gr_{\theta}(d;n))} \,:\, \alg\big(\Gr_{\theta}(d;n)\big)
~\longrightarrow~ \hil_\theta\otimes \alg\big(\Gr_{\theta}(d;n)\big)
\nonumber 
\eeq
given on generators by
\beq
\Delta_{\alg(\Gr_{\theta}(d;n))}\big(\Lambda^J \big)=t_J\otimes
\Lambda^J \ , \qquad t_J:=t_{j_1}\cdots t_{j_d} \ .
\label{grasstoric}\eeq
This implies that the $N\times N$ noncommutativity matrix $\Theta$ of the projective space containing the embedding of $\Gr_{\theta}(d;n)$
is completely determined (mod~$2\pi$) from the $n\times n$
noncommutativity matrix $\theta$ of the grassmannian. Necessary and
sufficient conditions for the existence of the embedding of
$\Gr_{\theta}(d;n)$ into $\alg(\mathbb{CP}_{\Theta}^{N})$ are thus
\begin{equation}
\label{Theta}
\Theta^{JJ'} = \sum_{\alpha,\beta =1}^d
\,\theta^{j_{\alpha}j'_{\beta}} \ .
\end{equation}
Noncommutative Pl\"ucker relations generate an ideal in the homogeneous coordinate algebra $\alg(\complex\P^N_{\Theta})$ of the projective space,
and we will \emph{define} the noncommutative quotient algebra to be the homogeneous coordinate algebra $\alg(\Gr_\theta(d;n))$ of the (embedding
of the) noncommutative grassmannian. From the noncommutative Laplace expansions \eqref{nlapkr} and \eqref{nlapkc} we see that noncommutative minors of order $d$ in $\GL_{\theta}(n)$ obey \textup{Pl\"ucker relations}
\begin{equation}\label{ncysr}
\sum_{\gamma =1}^{d+1} \ \prod_{\alpha =1}^d \ \prod_{\beta =1}^{d-1}
\; (-1)^{\gamma +1} \; q_{i_{\gamma}i^{\gamma}_{\alpha}} \,
q_{i_{\gamma}j_{\beta}} \, \Lambda^{I^{\gamma}} \, \Lambda^{i_{\gamma}\cup J} = 0 \ ,
\end{equation}
for every choice of multi-indices $I$ and $J$ with $|I|=d+1$ and
$|J|=d-1$.

The generalization to flag varieties is straightforward. Consider an increasing chain of nested vector subspaces of $V$,
$$
0=V_0~\varsubsetneq~ V_1~\varsubsetneq~ V_2~\varsubsetneq~ \cdots~
\varsubsetneq~ V_{r+1}=V \ , 
$$
such that $\gamma_i=\dim_\complex (V_i)-\dim_\complex (V_{i-1})$ for $i=1,\dots,r+1$ and $\gamma=(\gamma_1,\ldots ,\gamma_{r+1})$ with $1\leq r\leq n-1$ is a partition of $n$. The corresponding flag variety $\Fl(\gamma;n)$ is the moduli space of chains (or ``flags'') associated to the sequence $\gamma=(\gamma_1,\ldots ,\gamma_{r+1})$. 
The Pl\"ucker embedding of flag varieties into projective spaces is similar to the case of grassmannians. Set $d_i=\sum_{a\leq i}\,\gamma_a=\dim_\complex(V_i)$ for $i=1,\dots,r+1$. A point in $\Fl(\gamma;n)$ is represented by an equivalence class $[A]$ in $\GL(n)$ with respect to the equivalence relation given by the action of block diagonal matrices whose blocks are invertible transformations in each nested subspace $V_i$. For each $i$ there is a Pl\"ucker map $\Pl_i:\Fl(\gamma;n)\rightarrow {\mathbb{CP}^{N_i}} $, {with $N_i={n\choose d_i}-1$}, whose image is the ${n\choose d_i}$-tuple of all minors of $A$ obtained from the first $d_i$ rows. Hence each minor is labelled by a multi-index representing the $d_i$ columns involved while the rows are always given by the standard ordered multi-index
$(1\,2\cdots d_i)$. Assembling all of these maps together we get a
Pl\"ucker embedding 
\begin{equation}
\label{plufl}
\Pl\,:\,\Fl(\gamma;n)~\longrightarrow
 ~\mathbb{CP}(\gamma;n):= 
 \mathbb{CP}^{N_1}\times \cdots \times \mathbb{CP}^{N_r} \ ,
\end{equation}
where the last factor for $i=r+1$ gives a trivial contribution since
{$N_{r+1}={n\choose n}-1=0$}. The image of the Pl\"ucker map $\Pl$ in
$\mathbb{CP}(\gamma;n)$ is described by a set of quadratic equations
called Young symmetry relations. In the noncommutative setting we
follow the same idea. In addition to noncommutative Pl\"ucker relations \eqref{ncysr} we now have noncommutative relations (\ref{ncmin}) between minors of different size, i.e. with multi-indices of different lengths $|I|=|J|=d$ and $|I'\,|=|J'\,|=d'$, and ``genuine'' noncommutative Young symmetry
relations among minors of different order. 
Again from the Laplace expansion of the minors in \eqref{nlapkr} and
\eqref{nlapkc} one shows that noncommutative minors of order $d$ and $d'$ in $\GL_{\theta}(n)$ obey \textup{Young symmetry relations}
\begin{equation}
\label{ncplfl}
\sum_{\gamma =1}^{d+1} \ \prod_{\mu =1}^d \ \prod_{\nu =1}^{d' -1} \, (-1)^{\gamma +1} \, q_{i_{\gamma}i^{\gamma}_{\mu}} \, 
q_{i_{\gamma}j_{\nu}} \, \Lambda^{I^{\gamma}} \, \Lambda^{i_{\gamma}\cup J} = 0 \ ,
\end{equation}
for every choice of multi-indices $I,J$ with $|I|=d+1$ and $|J|=d' -1$. The coordinate algebra of the noncommutative flag variety $\Fl_{\theta}(\gamma;n)=\Fl_\theta(d_1,\dots,d_r;n)$ is the quotient of the homogeneous coordinate algebra of $\mathbb{CP}_{\Theta}(\gamma;n)$ by the ideal generated by the noncommutative Young symmetry relations \eqref{ncplfl}.

\newsection{Noncommutative instantons}

In this section we spell out our explicit construction of
noncommutative instantons in three related guises: in the algebro-geometric form of framed sheaves on the
noncommutative projective plane $\CP_\theta^2$, in the linear algebraic form of a braided ADHM
parametrization, and in the differential geometric form of
anti-selfdual connections on a noncommutative four-sphere
$S_\theta^4$. The construction of instanton bundles is based on a
noncommutative version of the twistor transform. Via a classification of the real structures on the
grassmannian $\Gr_\theta(2;4)$ we prove here that our particular deformation of $S^4$ is in fact the unique noncommutative four-sphere which is compatible with the noncommutative twistor correspondence.

\subsection{Sheaves on noncommutative toric varieties\label{NCsheaves}}

A sheaf theory on noncommutative toric varieties is straightforwardly developed. The idea is that the ``topology''
of the noncommutative variety $X_\theta=X_\theta[\Sigma]$ is given by
the cones in the fan $\Sigma$; they form the category $\Open(X_\theta)$ of toric open sets. Then
$\Open(X_\theta)$ always contains a sufficiently fine open cover,
i.e. sets of inclusions $(\sigma_i\hookrightarrow\sigma)_{i\in I}$ of
cones of $\Sigma$ such that $\sigma=\bigcup_{i\in I}\,\sigma_i$. The category $\Open(X_\theta)$ with the data of coverings forms a
Grothendieck topology on $X_\theta$.

The assignment $\sigma\mapsto\complex_\theta[\sigma]$ of the
noncommutative algebra $\complex_\theta[\sigma]$ to every cone
$\sigma\in\Sigma$ defines a sheaf of algebras
on $\Open(X_\theta)$ \cite[Prop. 4.2]{CLSI}; it is
regarded as the {structure sheaf} $\sheaf_{X_\theta}$ of the noncommutative toric variety $X_\theta$.
Let $\module(X_\theta)$ be the category of sheaves of right $\sheaf_{X_\theta}$-modules on $\Open(X_\theta)$. If $\Sigma$
consists of a single cone $\sigma$,
i.e. $X_\theta[\Sigma]=U_\theta[\sigma]$ is an affine variety with
coordinate algebra 
$\complex_\theta[\sigma]$, then 
\beq
\module\big(U_\theta[\sigma]\big)
\cong \module\big(\complex_\theta[\sigma]\big)
\label{moduleiso}\eeq
coincides with the category of right $\complex_\theta[\sigma]$-modules. We denote by $\widetilde{M}$ the
sheaf associated to a module $M$ under the isomorphism (\ref{moduleiso}). A sheaf of right $\sheaf_{X_\theta}$-modules is
{coherent} if its restriction to each affine open set
$U_\theta[\sigma]$ is of the form $\widetilde{M}$ for some finitely-generated right
$\complex_\theta[\sigma]$-module $M$. Coherent sheaves of right
$\sheaf_{X_\theta}$-modules form a category denoted $\coh(X_\theta)$. There are restriction functors
\beq
j_\sigma^\bullet\,:\,\coh(X_\theta)~\longrightarrow~\module\big(\complex_\theta[\sigma]\big)
\label{restrfunct}\eeq
for each open inclusion $j_\sigma:U[\sigma]\hookrightarrow X[\Sigma]$. Let $\tor(\sigma)$ be the full Serre subcategory of $\coh(X_\theta)$ generated by objects $E$ such that $j_\sigma^\bullet(E)=0$. In~\cite[Prop.~4.3]{ingalls} it is proven that the restriction functor (\ref{restrfunct}) is exact, and there is a natural equivalence of categories
\beq
\label{equiv-cat}
\coh(X_\theta)\,\big/\,\tor(\sigma)~\cong~\module\big(\complex_\theta[\sigma]\big) \ .
\eeq
We use \eqref{equiv-cat} to reduce geometric problems in the category $\coh(X_\theta)$ to algebraic problems in the algebra $\complex_\theta[\sigma]$ via the localization functors $j_\sigma^\bullet$. 
A coherent sheaf $\bun\in\coh(X_\theta)$ is {locally free} or a {bundle} if each $\bun_\sigma:=j_\sigma^\bullet(\bun)$, $\sigma\in\Sigma$,
  corresponds to a free module $\complex_\theta[\sigma]^{\oplus r}$
  for some integer $r\in\nat$ called the {rank} of $\bun$. A coherent
  sheaf $E\in\coh(X_\theta)$ is {torsion free} if each $E_\sigma$,
  $\sigma\in\Sigma$, has no finite-dimensional submodules, or
  equivalently if it admits an embedding $E\hookrightarrow\bun$ into a
  locally free sheaf $\bun$; the {rank} of $E$ is the rank of $\bun$ minus the rank
  of $\bun/E$.
For any pair of sheaves $E,F\in\coh(X_\theta)$, let $\Ext^p(E,F)$ be
the $p$-th derived functor of the functor
$\Hom(E,F)=\Hom_{\coh(X_\theta)}(E,F)$. For a sheaf
$E\in\coh(X_\theta)$, we define
$
H^p(X_\theta,E):=\Ext^p(\sheaf_{X_\theta},E) $.

We will mostly work on projective varieties. Here many algebraic and
geometric properties are inherited from $\mathbb{CP}^n_{\theta}$ and
its homogeneous coordinate algebra $\alg=\alg(\mathbb{CP}^n_{\theta})$
by restriction. The algebra $\alg$ has nice ``smoothness''
properties which enable straightforward extensions of many classical
constructs within the realm of noncommutative algebraic geometry.
In
particular, $\alg$ is an Artin--Schelter regular algebra of global
homological dimension $n+1$.
Let $\gr(\alg)$ be the category of finitely-generated graded right
$\alg$-modules $M=\bigoplus_{k\geq0}\,M_k$ and degree zero morphisms, and let $\tor(\alg)$ be the full Serre subcategory of $\gr(\alg)$
consisting of finite-dimensional graded $\alg$-modules $M$,
i.e. $M_k=0$ for $k\gg0$. 
From (\ref{equiv-cat}) it follows that the
category of coherent sheaves on $\Open(\CP_\theta^n)$ can be
identified with the abelian quotient category $\coh(\CP_\theta^n):=
\gr(\alg)/\tor(\alg)$. Let $\pi:\gr(\alg)\to\coh(\CP_\theta^n)$ be the
canonical projection functor; it has a natural right adjoint functor
$\Gamma:\coh(\CP_\theta^n)\to \gr(\alg)$. Under this correspondence, the structure
sheaf $\sheaf_{\CP_\theta^n}$ is the image $\pi(\alg)$ of the homogeneous coordinate algebra itself, regarded as a free right
$\alg$-module of rank one. If $E=\pi(M)$ where $M\in\gr(\alg)$ is a graded right $\alg$-module, then $M[w_i^{-1}]_0=
\big(M\otimes_\alg\alg[w_i^{-1}] \big)_0$ is a right
$\complex_\theta[\sigma_i]$-module for each $i=1,\dots,n+1$. 
For $k\in\zed$, we let $E(k)$ be the sheaf corresponding to the module $M$ with its grading shifted 
by $k$ units.

\subsubsection*{Tautological bundle on $\Gr_\theta(d;n)$}

An important example of a locally free sheaf in the following is the
tautological bundle $\Scal_{\theta}$ on $\Open(\Gr_{\theta}(d;n))$,
which further admits a straightforward extension to noncommutative
flag varieties.  The bundle $\Scal_{\theta}$ is the subsheaf of elements of the free module $(f_1(\Lambda),\ldots ,f_n(\Lambda)) \in
\alg(\Gr_{\theta}(d;n))^{\oplus n}$ over the noncommutative
grassmannian which satisfy the equations
\begin{equation}
\label{eq2}
\sum_{\alpha=1}^{d+1}\, (-1)^{\alpha} \ \Big(\, \prod_{\beta=1}^d \, q_{ j_\alpha  j_\beta^\alpha } \, \Big) \ \Lambda^{J\setminus j_\alpha}\, f_{j_\alpha}(\Lambda) = 0 
\end{equation}
for every ordered $(d+1)$-multi-index $J=(j_1\cdots j_{d+1})$ with
$j_1<j_2<\cdots<j_{d+1}$, where the minors of order
$d$ obey the relations~(\ref{ncrelpr}).
The ``geometric'' interpretation of \eqref{eq2} is that we add the vector $f(\Lambda)=(f_1(\Lambda),\ldots ,f_n(\Lambda))$ to the $d\times n$ matrix $\Lambda$ and require that all the minors, expanded along the $(d+1)$-th row $f(\Lambda)$ via the noncommutative Laplace expansion \eqref{nlapkr}, are zero. This is equivalent to saying that the vector $f(\Lambda)$ belongs to the $d$-plane $\Lambda$, the condition which characterizes $\Scal_{\theta}$ inside the trivial bundle $\alg(\Gr_{\theta}(d;n))^{\oplus n}$. It is 
proven in \cite[Prop. 6.15]{CLSI} that the sheaf $\Scal_\theta$ is indeed locally free on $\Open(\Gr_\theta(d;n))$. 

We can use the Pl\"ucker map to regard the noncommutative minors $\Lambda^{{J\setminus j_\alpha}}$ as homogeneous coordinates in
$\mathbb{P}(\bigwedge_{\theta}^d V)$. In this case we have to consider the restriction of (\ref{eq2}) to those elements {$\Lambda^J$} which also satisfy the Young symmetry relations~(\ref{ncysr}). This gives the sheaf $\Scal_\theta$ the natural structure of a graded $\alg(\Gr_{\theta}(d;n))$-bimodule. 

\subsection{Twistor geometry}	
\label{twgeomsphere}

Consider the explicit algebraic relations for the noncommutative
Pl\"{u}cker embedding of the Klein quadric $\Gr_\theta(2;4)$ in
$\mathbb{CP}_{\Theta}^5$. The algebra $\alg(\Gr_{\theta}(2;4))$ has
six generators $\lam{ij}$, $1\leq i<j \leq 4$; we give to the multi-indices labelling the minors a lexicographic ordering such that $1=(1\,2)$, $2=(1\,3)$, $3=(1\,4)$, $4=(2\,3)$, $5=(2\,4)$, and $6=(3\,4)$. Then the expression (\ref{Theta}) for the skew-symmetric noncommutativity
matrix $\Theta$ in terms of entries of $\theta$ is given by
$$
\Theta=\begin{pmatrix}
\scriptstyle{0} & \scriptstyle{-\theta^{12}+\theta^{13}+\theta^{23}} &
\scriptstyle{-\theta^{12}+\theta^{14}+\theta^{24}} &
\scriptstyle{\theta^{12}+\theta^{13}+\theta^{23}} &
\scriptstyle{\theta^{12}+\theta^{14}+\theta^{24}} &
\scriptstyle{\theta^{13}+\theta^{14}+\theta^{23}+\theta^{24}} \\
  & \scriptstyle{0}                                  &
  \scriptstyle{-\theta^{13}+\theta^{14}+\theta^{34}} &
  \scriptstyle{\theta^{12}+\theta^{13}-\theta^{23}} &
  \scriptstyle{\theta^{12}+\theta^{14}-\theta^{23}+\theta^{34}} &
\scriptstyle{\theta^{13}+\theta^{14}+\theta^{34}} \\
  &                                      & \scriptstyle{0}
  &\scriptstyle{\theta^{12}+\theta^{13}-\theta^{24}-\theta^{34}}&
    \scriptstyle{\theta^{12}+\theta^{14}-\theta^{24}} &
\scriptstyle{\theta^{13}+\theta^{14}-\theta^{34}} \\
  &                                      &
  & \scriptstyle{0} &
  \scriptstyle{-\theta^{23}+\theta^{24}+\theta^{34}} &
  \scriptstyle{\theta^{23}+\theta^{24}+\theta^{34}} \\
  &                                      &
  &   & \scriptstyle{0} &
  \scriptstyle{\theta^{23}+\theta^{24}-\theta^{34}} \\
  &                                      &
  &   &   & \scriptstyle{0}
\end{pmatrix} \ .
$$
When considering noncommutative Pl\"ucker relations (\ref{ncysr}), various choices for the multi-indices $I$ and $J$ give two-term equations, which yield known noncommutative relations among minors. The only three-term Pl\"ucker equation, obtained from 
$I=(1\,2\,3)$ and $J=(4)$, is a noncommutative deformation of the well-known classical equation describing the Klein quadric $\Gr(2;4)\hookrightarrow\mathbb{CP}^5$. After rearranging all indices labelling the minors in increasing order using antisymmetry, and the minors
themselves using (\ref{ncrelpr}), we obtain
\begin{equation}
\label{pl24}
q_{31}\, q_{32}\, q_{34}\ \Lambda^{(12)}\,\Lambda^{(34)} -
q_{21}\, q_{23}\, q_{24}~\Lambda^{(13)}\,\Lambda^{(24)} +
q_{12}\, q_{13}\,q_{14}~\Lambda^{(23)}\,\Lambda^{(14)} = 0 \ .
\end{equation}

We will describe the homogeneous coordinate algebra
of the noncommutative partial flag variety $\Fl_\theta(1,2;4)$ using
the algebra projection~\cite{CLSI}
$$
\alg\big(\CP_\theta^3\big)\,\otimes_\theta\,
\alg\big(\Gr_\theta(2;4)\big)~\longrightarrow~
\alg\big(\Fl_\theta(1,2;4)\big) \ .
$$
The braided tensor product algebra $\alg(\CP_\theta^3)\,\otimes_\theta\,
\alg(\Gr_\theta(2;4))$, as an object in the category ${}^{\hil_\theta}\Module$, is generated by the homogeneous coordinate
elements $w_i$ with relations (\ref{ncrelpr}) for
$d=1$, i.e. $w_i\, w_j=q_{ij}^2\ w_j\, w_i$, for $i,j=1,2,3,4$, 
and by the noncommutative $2\times2$ minors
$\Lambda^{(j_1j_2)}$, $1\leq j_1<j_2\leq4$, obeying (\ref{ncrelpr}) for
$d=2$ and the quadric relation (\ref{pl24}). They also obey the structure equations (relations \eqref{ncmin} among minors of different order)
\beq
w_i\,\Lambda^{(j_1j_2)}=q_{12}^{-2}\,q_{i\,j_1}^2\,q_{i\,j_2}^2~\Lambda^{(j_1j_2)}\,w_i \ ,
\label{wLambda}\eeq
for all $i=1,2,3,4$ and $1\leq j_1<j_2\leq4$. The remaining
noncommutative Pl\"ucker equations come from the Young symmetry
relations \eqref{ncplfl} with $d=2$ and $d'=1$, and one finds
the additional relations
\begin{eqnarray}
q_{12}\, q_{13}\ \Lambda^{(23)}\, w_1 - q_{21}\, q_{23}\ \Lambda^{(13)}\, w_2 + q_{31}\, q_{32}\ \Lambda^{(12)}\, w_3&=&0 \ ,
\nonumber \\[4pt]
q_{12}\, q_{14}\ \Lambda^{(24)}\, w_1 - q_{21}\, q_{24}\ \Lambda^{(14)}\, w_2 + q_{41}\, q_{42}\ \Lambda^{(12)}\, w_4 &=&0 \ ,
\nonumber \\[4pt]
q_{13}\, q_{14}\ \Lambda^{(34)}\, w_1 - q_{31}\, q_{34}\ \Lambda^{(14)}\, w_3 + q_{41}\, q_{43}\ \Lambda^{(13)}\, w_4&=&0 \ ,
\nonumber \\[4pt]
q_{23}\, q_{24}\ \Lambda^{(34)}\, w_2 - q_{32}\, q_{34}\
\Lambda^{(24)}\, w_3 + q_{42}\, q_{43}\ \Lambda^{(23)}\, w_4&=&0 \ .
\label{flagpluck}\end{eqnarray}

Consider now the noncommutative correspondence diagram
\beq
\xymatrix{
 & \alg\big(\Fl_\theta(1,2;4)\big) & \\
\alg\big(\CP_\theta^3\big) \ar[ur]^{p_1} &  & 
\alg\big(\Gr_\theta(2;4)\big) \ar[ul]_{p_2}
}
\label{nccorr}\eeq
which is a noncommutative deformation of the usual Penrose twistor
correspondence, with $\alg(\CP_\theta^3)$ the ``noncommutative twistor
algebra''.  The morphism
$E\mapsto p_2{}^*\,p_{1*}(E)$ gives a map from sheaves in
$\coh(\CP_\theta^3)$ to sheaves in $\coh(\Gr_\theta(2;4))$, yielding a noncommutative deformation of the usual Penrose--Ward
twistor transform. Given a graded right
$\alg(\CP_\theta^3)$-module $M$ in $\gr(\alg(\CP_\theta^3))$, the push-forward 
\beq
M'=p_{1*}(M)=M\otimes_{\alg(\CP_\theta^3)}\alg\big(\Fl_{\theta}(1,2;4)\big) 
\label{p1forward}\eeq
is a bigraded right module over $\alg(\Fl_{\theta}(1,2;4))$,
where on the right-hand side we regard the algebra
$\alg(\Fl_{\theta}(1,2;4))$ as an $\alg(\CP_\theta^3)$-bimodule. The diagonal subspace of this module, in the sense of~\cite[\S8]{KKO}, induces the
push-forward functor
$p_{1*}:\coh(\CP_{\theta}^3)\to\coh(\Fl_{\theta}(1,2;4))$. Similarly,
one defines the push-forward functor
$p_{2*}:\coh(\Gr_{\theta}(2;4))
\to\coh(\Fl_{\theta}(1,2;4))$, which has a right adjoint functor 
$p_2{}^*:\coh(\Fl_{\theta}(1,2;4))\to\coh(\Gr_{\theta}(2;4))$.
In~\cite[Lem.~3.14]{CLSII} the noncommutative
twistor transforms of the elementary bundles
$\sheaf_{\CP_{\theta}^3}(k)$ for $k\in\zed$ are computed to be
$$
p_2{}^*\,p_{1*}\big(\sheaf_{\CP_{\theta}^3}(k)\big) = 
\left\{\begin{array}{rl}
{\rm Sym}^k_{\theta}(\Scal_{\theta}) \quad , & k\geq0 
\\ 0 \quad , & k<0  \end{array} \ , \right.
$$
where $\Scal_{\theta}$ is the tautological bundle on
$\Open(\Gr_{\theta}(2;4))$ and
${\rm Sym}^k_{\theta}(\Scal_{\theta})$ is the bundle
associated to the graded right $\alg\big(\Gr_{\theta}(2;4)\big)$-module
$$
\Gamma(\Scal_{\theta})^{\otimes k}\, \big/\, \big\langle s_1\otimes
s_2-\Psi_{\Gamma(\Scal_{\theta}), \Gamma(\Scal_{\theta})}^{\theta}(s_1\otimes
s_2)\big\rangle_{s_1,s_2\in\Gamma(\Scal_{\theta})} \ .
$$

\subsection{Real structures on the noncommutative Klein quadric}

In order to introduce a $*$-algebra structure on $\alg(\Gr_\theta(2;4))$ and a compatible twistor construction of instanton bundles later on, 
from now on we set 
\beq
q_{12}=q_{21}^{-1}=:q \ , \qquad q_{ij}=1 \quad \mbox{otherwise} \ . 
\nonumber\eeq
From the perspective of the Drinfel'd twist deformation, this choice
restricts the $\hil_\theta$-coaction to a coaction of
the subalgebra
$\complex_\theta(t_1,t_2)$ defined by setting $t_3=t_4=1$, i.e. the underlying
noncommutative complex torus $T_\theta=(\complex_\theta^\times)^4$ is reduced to
$(\complex_\theta^\times)^2$. This entails in particular that the generator $\lam{34}$ is central. We classify all possible real structures
on $\Gr_{\theta}(2;4)$ and determine the induced signature on the
Pl\"ucker relation \eqref{pl24}, which for the present choice becomes 
\begin{equation}
\label{pl24bis}
q \ \Lambda^{(12)}\,\Lambda^{(34)} - \Lambda^{(13)}\,\Lambda^{(24)} + \Lambda^{(14)} \, \Lambda^{(23)} = 0 \ .
\end{equation}
A real structure $\dagger : \alg(\Gr_{\theta}(2;4)) \rightarrow \alg(\Gr_{\theta}(2;4))$ is
defined on the generators $\lam{ij}$ and then extended to the whole of $\alg(\Gr_{\theta}(2;4))$ as a
conjugate linear anti-homomorphism; it must
therefore be compatible with the commutation relations (\ref{ncrelpr})
and the Pl\"ucker equation \eqref{pl24bis}. By imposing compatibility of
$\dagger$ with (\ref{ncrelpr}), we can easily determine how
$\dlam{ij}$ should transform under the coaction (\ref{grasstoric}). In
particular, $\dlam{34}$ must be coinvariant, i.e. it is central and
hence proportional to $\lam{34}$, while $\dlam{12}$ transforms with
weight $t_1\, t_2$ and hence is proportional to $\lam{12}$. We then
write
\begin{equation}
\label{eta}
\dlam{12} = \mu \, \lam{12} \ , \qquad \dlam{34} = \nu \, \lam{34}
\qquad \mbox{for} \quad \mu , \nu \in S^1=\{\zeta\in\complex \ | \
\zeta\, \bar{\zeta} = 1 \} \ .
\end{equation}   
Similarly, $\dlam{13}$ and $\dlam{14}$ must transform
either both with weight $t_1$ or both with weight $t_2$, and
$\dlam{23}$ and $\dlam{24}$ accordingly both with weight
$t_2$ or both with weight $t_1$.

Suppose first that $\dlam{13}$ and $\dlam{14}$ transform with
weight $t_1$; then $q\in S^1$ in order to ensure $(q\,
\lam{ij})^{\dagger} = q^{-1}\, \dlam{ij}$, and \emph{a priori} $\dlam{13}$
and $\dlam{14}$ are each linear combinations of both $\lam{13}$
and $\lam{14}$. Via a basis change automorphism of the
generators $(\lam{13},\lam{14})$ if necessary, we can assume that both
$\dlam{13}$ and $\dlam{14}$ are proportional to single generators
without loss of generality. The
$\GL(2)$ equivalence classes of real structures on
$\{\lam{13},\lam{14}\}$ are therefore exhausted by the representatives
\begin{equation}
\label{t1a}
\dlam{13} = \alpha \, \lam{13} \ , \qquad 
\dlam{14} = \beta \, \lam{14} \qquad \mbox{for} \quad \alpha,\beta\in S^1
\end{equation}
and
\begin{equation}
\label{t1b}
\dlam{13} = \alpha \, \lam{14} \ , \qquad
\dlam{14} = \bar{\alpha}^{-1} \, \lam{13} \qquad \mbox{for} \quad
\alpha\in\complex \ .
\end{equation}

Suppose now that $\dlam{13}$ and $\dlam{14}$ transform with weight
$t_2$; this requires $q\in\real$. A similar argument as in the
previous case shows that up to basis change automorphism one can
choose two classes of real structures with representatives
\begin{equation}
\label{t2a}
\dlam{13} = \alpha \, \lam{23} \ , \qquad \dlam{14}  =  \beta \,
\lam{24}  \qquad \mbox{for} \quad \alpha,\beta \in\complex
\end{equation}
and
\begin{equation}
\label{t2b}
\dlam{13}  = \alpha \, \lam{24} \ , \qquad \dlam{14} = \beta \,
\lam{23} \qquad \mbox{for} \quad \alpha,\beta\in \complex \ .
\end{equation}

We now impose compatibility with the Pl\"ucker equation
\eqref{pl24bis}. Since this is the only quadratic relation in
$\alg(\Gr_{\theta}(2;4))$, its image under $\dagger$ must be
proportional to itself; this requirement also determines in
\eqref{t1a}--\eqref{t1b} the real structure on the generators
$\lam{23}$ and $\lam{24}$. There are four cases in total, one each
coming from \eqref{t1a}--\eqref{t2b}. 

Let us first consider the choice of real structure \eqref{t1a}. Compatibility with the Pl\"ucker equation requires 
\begin{equation}
\nonumber
\dlam{23} = \gamma \, \lam{23} \ , \qquad \dlam{24} = \delta\,\lam{24}
\qquad \mbox{for} \quad \gamma,\delta\in S^1 \ .
\end{equation}
Remembering \eqref{eta} and that in this case $q\in S^1$, by applying
$\dagger$ to \eqref{pl24bis} we find the constraint
\beq
\alpha\, \delta =
\beta\, \gamma = \mu\, \nu \ .
\label{adbgmnrel}\eeq
To determine the signature of the quadratic form appearing in the
Pl\"ucker relation, we rewrite \eqref{pl24bis} as
\begin{multline*}
q \, \bar{\nu} \ \lam{12}\, \dlam{34} + q\, \bar{\mu} \ \lam{34}\,
\dlam{12} - \bar{\delta} \ \lam{13}\, \dlam{24} \nonumber \\  
-\, q^2\, \bar{\alpha} \
\lam{24}\, \dlam{13} + \bar{\gamma} \ \lam{14}\, \dlam{23} + q^2\,
\bar{\beta} \ \lam{23}\, \dlam{14} = 0 \ .
\end{multline*}
Requesting that the terms are pairwise real, together with (\ref{adbgmnrel}), amounts to setting
$$ \mu\, \nu = q^2 \ , \qquad \alpha= \pm\, 1 \ , \qquad \delta =
\pm\, q^2 \ , \qquad \beta=\pm\, 1 \ , \qquad \gamma = \pm\, q^2 \ . $$ 
A set of generators which diagonalizes the Pl\"ucker bilinear form is therefore
$$ 
X_{\pm}= q^{-1}\, \mu \, \lam{12} \pm \lam{34} \ , \qquad
V_{\pm}= \bar{\delta} \, \lam{13} \pm \lam{24} \  , \qquad
W_{\pm}= \bar{\gamma} \, \lam{14} \pm \lam{23}  
$$
giving
\begin{equation}
\nonumber
X_+\,X_+{}^{\dagger} - X_-\, X_-{}^{\dagger} - V_+\, V_+{}^{\dagger} +
V_-\, V_-{}^{\dagger} + W_+\, W_+{}^{\dagger} - W_-\, W_-{}^{\dagger}
= 0 \ ,
\end{equation}
which exhibits signature $(3,3)$. A completely analogous analysis
shows that the choices \eqref{t1b} and \eqref{t2a} also yield a
diagonal Pl\"ucker equation with signature $(3,3)$, with compatibility
conditions $\dlam{23}=- q^2\, \bar\alpha\, \lam{24}$ and $\mu=q^2\,
\bar\nu$ in the case \eqref{t1b}, while $\nu=\bar\mu$ and
$\beta=-\bar\alpha$ for \eqref{t2a}.

Let us finally consider the choice \eqref{t2b}. In this case $q$ is
real, and the image of the
Pl\"ucker equation under $\dagger$ imposes the constraint
$\bar{\alpha}^{-1}\, \alpha = \mu\, \nu = \bar{\beta}^{-1}\,
\beta$. We rewrite \eqref{pl24bis} in the form
\begin{multline*}
q\, \bar{\nu} \ \lam{12}\, \dlam{34} + q \, \bar{\mu} \ \dlam{12}\,
\lam{34} - \alpha^{-1} \ \lam{13}\, \dlam{13} \nonumber \\ 
 - \, \bar{\alpha} \
\dlam{24}\, \lam{24} + \bar{\beta} \ \dlam{23}\, \lam{23} + \beta^{-1}
\ \lam{14}\, \dlam{14} = 0 \ ,
\end{multline*}
so that asking for the terms to be pairwise real we get in addition $\nu=\bar{\mu}$ and $\alpha,\beta \in \mathbb{R}$.
We can diagonalize the first two terms by introducing the combinations
$ X_{\pm} = q\, \mu \, \lam{12} \pm \lam{34} $ giving $X_+\,
X_+{}^{\dagger} - X_-\, X_-{}^{\dagger}$; the remaining terms are
already in diagonal form. The global signature is therefore $(3,3)$ if
$\alpha\, \beta>0$, and $(5,1)$ if $\alpha\, \beta<0$.

\subsection{Noncommutative four-sphere\label{NCS4}}

We construct a noncommutative four-sphere by picking a compatible
$*$-involution on the algebra $\alg(\Gr_{\theta}(2;4))$ for which the
quadratic form in the Pl\"ucker equation (\ref{pl24bis}) has
signature $(5,1)$. As we have just shown this requires $q\in\real$ and there is a unique family of such
real structures parametrized as 
\begin{align}
& \dlam{12} =  \mu \,  \lam{12} \, , \quad \dlam{34} =  \bar{\mu} \, \lam{34}  \, , \nonumber \\
& \dlam{13} =  \alpha \, \lam{24}\, , \quad  \dlam{14} =  \beta \, \lam{23} 
\qquad \mbox{for} \quad \mu\in S^1 \ , \; \alpha,\beta \in \mathbb{R} \ , \;
\alpha\, \beta<0 \ .   
\label{stargrass}
\end{align}
The parameters $\mu,\alpha,\beta$ may be absorbed via a suitable
redefinition of the generators $\lam{ij}$, so that up to change of
basis automorphism in \eqref{stargrass} there is only one real structure with 
$\mu=\alpha=- \beta=1$. In what follows, however, it will be
convenient to make the choice
$$
\mu=1 \ , \qquad \alpha=q \ , \qquad \beta=-q^{-1}
$$
for the real structure, so that
\beq
\quad\quad\quad \dlam{12} =  \lam{12} \, , \quad \dlam{34} = \lam{34}  \, , \quad  \dlam{13} = q \, \lam{24} \, , \quad  
\dlam{14} = -q^{-1} \, \lam{23} \ .
\label{stargrass-bis}
\eeq
The real structure is not compatible with the whole toric symmetry
$\hil$. Compatibility forces us to restrict to a coaction coming from a real abelian  subgroup $G$ of $T$ endowed with a $*$-structure defined by $(t_1)^*=t_2$. This subgroup is a copy of 
$(\real^\times)^2$ whose unital $*$-algebra of coordinate functions we shall denote by $\galg=\alg(G)=\alg((\real^\times)^2)$. 
When restricted to $\galg$ the twist \eqref{Ftheta} becomes real (remembering that the deformation parameter $q$ is now real),
and everything is consistent within the framework of real Drinfel'd twists and associated deformations of $*$-algebras. 
Thus the algebra $\alg(\Gr_{\theta}(2;4))$ is a comodule $*$-algebra (only) for $\galg$. 

When substituted into the Klein quadric equation~\eqref{pl24bis} we obtain
\begin{equation}
\label{realkq}
q\ \Lambda^{(12)}\,\Lambda^{(34)} -
q^{-1}~\Lambda^{(13)}\,\Lambda^{(13)}\,^\dag -
q~\Lambda^{(14)}\,\Lambda^{(14)}\,^\dag = 0 \ . 
\end{equation}
In the classical case $q=1$, this relation would read as the equation of a four-sphere in a real slice of $\complex\P^5$. 
We thus interpret the $*$-algebra corresponding to \eqref{realkq} as
the coordinate algebra $\alg(S_\theta^4)$ of a noncommutative
four-sphere $S_\theta^4$ with a non-central ``radius''. For this, let
us redefine the generators by writing $ q \, (\Lambda^{(12)}-
\Lambda^{(34)}) =: 2 X$ and $ q \,(\Lambda^{(12)}+\Lambda^{(34)})=: 2 R$, where the hermitean elements $R$ and $X$ commute with each other but not with the remaining minors. Simple algebra then transforms the relation \eqref{realkq} into
\begin{equation}
\label{realkq-bis}
\Lambda^{(13)}\,\Lambda^{(13)}\,^\dag +
q^{2}~\Lambda^{(14)}\,\Lambda^{(14)}\,^\dag + X^2 = R^2 \ . 
\end{equation} 
This is a deformation of the equation of a four-sphere in homogeneous
coordinates in a real slice of $\complex\P^5$, with a 
non-central radius~$R$. As a consequence, one is not allowed to fix $R$ to some real number (typically $R=1$ in the classical case). Furthermore, the homogeneous element $R$ does not generate 
a right or left denominator set, as the corresponding fraction
elements do not close to an algebra.  
Thus it is not possible to apply Ore localization with respect to $R$, an operation that classically would correspond to rescaling the homogeneous coordinates to get an affine description of the sphere.
Hence for our noncommutative sphere there is no hope for any global affine description. However, one can still use localization to describe local patches for the sphere; using the two hermitean generators $\Lambda^{(12)}$ and $\Lambda^{(34)}$ we can define two
localizations to affine subvarieties of the grassmannian
$\Gr_{\theta}(2;4)$, whose ``real slices'' are interpreted as local
patches of the sphere $S_\theta^4$. The two generators
$\Lambda^{(12)}$ and $\Lambda^{(34)}$ are rather different in nature,
and hence so are the corresponding localizations, as $\Lambda^{(34)}$
is central while $\Lambda^{(12)}$ is not. 

The generators of the degree~zero subalgebra of
$\alg(\Gr_{\theta}(2;4))[\Lambda^{(34)}\,^{-1}]$ are 
\begin{eqnarray*}
\xi_1=-\Lambda^{(14)}\,\Lambda^{(34)}\,^{-1} \qquad &,& \qquad
\xi_2= -\Lambda^{(24)}\,\Lambda^{(34)}\,^{-1} \ , \\[4pt]
\bar\xi_1=\Lambda^{(23)}\,\Lambda^{(34)}\,^{-1} \qquad &,& \qquad
\bar\xi_2=\Lambda^{(13)}\,\Lambda^{(34)}\,^{-1} \ ,
\end{eqnarray*}
{together with $\rho=q\ \Lambda^{(12)}\,\Lambda^{(34)}\,^{-1}$}.
The Pl\"ucker relation (\ref{realkq}) becomes
$\rho=\xi_1\,\bar\xi_1-\bar\xi_2\,\xi_2$
so that the generator $\rho$ is redundant. The localized algebra $\alg(\Gr_{\theta}(2;4))[\Lambda^{(34)}\,^{-1}]_0$ is then isomorphic to the 
algebra generated by elements $\xi_i,\bar\xi_i$, $i=1,2$ with the relations
\begin{eqnarray*}
\xi_1\,\bar\xi_1=q^2~\bar\xi_1\,\xi_1 \qquad &,& \qquad
\xi_2\,\bar\xi_2=q^{-2}~\bar\xi_2\,\xi_2 \ , \\[4pt]
\xi_1\,\xi_2=q^2~\xi_2\,\xi_1 \qquad &,& \qquad
\bar\xi_1\,\bar\xi_2=q^{-2}~\bar\xi_2\,\bar\xi_1 \ , \\[4pt]
\xi_1\,\bar\xi_2=\bar\xi_2\,\xi_1 \qquad &,& \qquad
\xi_2\,\bar\xi_1=\bar\xi_1\,\xi_2 \ .
\end{eqnarray*}
This localized algebra may be thought of as dual to a noncommutative variety $\underline{\complex_\theta^4}$  regarded as an open affine subvariety of the noncommutative Klein quadric $\Gr_{\theta}(2;4)$. It will be
important later on to note that this deformation can be regarded as originating, via a Drinfel'd twist procedure, from the action of the torus $T=(\complex^\times)^2$ on $\complex^4$ giving a left coaction 
$
\Delta_{\alg(\complex_\theta^4)}:
\alg(\,\underline{\complex_\theta^4}\, ) \rightarrow \hil_\theta
\otimes \alg(\,\underline{\complex_\theta^4}\, )
$
with
\begin{eqnarray}
\Delta_{\alg(\,\underline{\complex_\theta^4}\,
  )}(\xi_1)=t_1\otimes\xi_1 \quad &,& \quad \Delta_{\alg(\,
  \underline{\complex_\theta^4}\, )}(\xi_2)=t_2\otimes\xi_2 \ , \nonumber \\[4pt]
\Delta_{\alg(\,\underline{\complex_\theta^4}\,
  )}(\bar\xi_1)=t_2\otimes\bar\xi_1 \quad &,& \quad \Delta_{\alg(\,
  \underline{\complex_\theta^4}\, )}(\bar\xi_2)=t_1\otimes\bar\xi_2 \ ,
\label{DeltaLR4}\end{eqnarray}
and extended as an algebra map; in particular
$\Delta_{\alg(\,\underline{\complex_\theta^4}\, )}(\rho)=(t_1\, t_2) \otimes \rho$.

The $*$-involution of \eqref{stargrass-bis} gives rise on the
generators $\xi_i,\bar\xi_i$, $i=1,2$ to the relations
\beq
\xi_1^\dag=q^{-1}~ \bar\xi_1 \ , \qquad \xi_2^\dag=-q^{-1}~ \bar\xi_2 \ ,  
\label{starxi}\eeq
from which it also follows that $\rho= q\ \xi_1\, \xi_1^\dag + q^3\ \xi_2\, \xi_2^\dag$. The corresponding 
$*$-algebra is dual to
a noncommutative real variety $\real_\theta^4$, regarded as an open
affine subvariety of the noncommutative four-sphere $S_\theta^4$. 
Arguing as we did after \eqref{stargrass-bis}, the coordinate algebra $\alg(\real_\theta^4)$ is a comodule $*$-algebra only for the coaction of $\galg=\alg((\real^\times)^2)$, not for the whole $\hil=\alg(T)$.

A second open affine subvariety of $S_\theta^4$ is obtained by
localization onto another affine subvariety of the grassmannian $\Gr_{\theta}(2;4)$ via the non-central hermitean minor 
$\Lambda^{(12)}$, which generates a left denominator set in $\alg(\Gr_{\theta}(2;4))$. The generators of the degree~zero subalgebra of
$[\Lambda^{(12)}\,^{-1}]\alg(\Gr_{\theta}(2;4))$ are 
\begin{eqnarray*}
\zeta_1=-\Lambda^{(12)}\,^{-1}\,\Lambda^{(14)} \qquad &,& \qquad
\zeta_2=-\Lambda^{(12)}\,^{-1}\,\Lambda^{(24)} \ , \\[4pt]
\bar{\zeta_1}=\Lambda^{(12)}\,^{-1}\,\Lambda^{(23)} \qquad &,& \qquad
\bar{\zeta_2}=\Lambda^{(12)}\,^{-1}\,\Lambda^{(13)} \ ,
\end{eqnarray*}
{together with $\tilde\rho=q\ \Lambda^{(12)}\,^{-1}\,\Lambda^{(34)}$}. 
The Pl\"ucker relation (\ref{realkq}) becomes
$
\tilde\rho= \bar\zeta_1 \,\zeta_1 - \zeta_2\, \bar\zeta_2 
$
showing that the generator $\tilde\rho$ is redundant. Using the multiplication rules for noncommutative Ore localization in $[\Lambda^{(12)}\,^{-1}]\alg(\Gr_{\theta}(2;4))_0$
one finds an algebra isomorphic to the 
algebra generated by elements $\zeta_i,\bar{\zeta_i}$, $i=1,2$ with the relations
\begin{eqnarray*}
\zeta_1\,\bar\zeta_1=q^{-2}~\bar\zeta_1\,\zeta_1 \qquad &,& \qquad
\zeta_2\,\bar\zeta_2=q^{2}~\bar\zeta_2\,\zeta_2 \ , \\[4pt]
\zeta_1\,\zeta_2=q^{-2}~\zeta_2\,\zeta_1 \qquad &,& \qquad
\bar\zeta_1\,\bar\zeta_2=q^{2}~\bar\zeta_2\,\bar\zeta_1 \ , \\[4pt]
\zeta_1\,\bar\zeta_2=\bar\zeta_2\,\zeta_1 \qquad &,& \qquad
\zeta_2\,\bar\zeta_1=\bar\zeta_1\,\zeta_2 \ .
\end{eqnarray*} 
The corresponding noncommutative complex variety is denoted $\underline{\widetilde\complex_\theta^4}$.
The counterpart of the coaction \eqref{DeltaLR4} is given 
by the dual left coaction 
$
\Delta_{\alg(\, \underline{\widetilde\complex_\theta^4}\, )}: \alg(\,
\underline{\widetilde\complex_\theta^4}\, ) \rightarrow\hil_\theta
\otimes \alg(\, \underline{\widetilde\complex_\theta^4}\, )
$
with
\begin{eqnarray}
\Delta_{\alg(\,\underline{\widetilde\complex_\theta^4}\,
  )}(\zeta_1)=t_2^{-1}\otimes\zeta_1 \quad &,& \quad \Delta_{\alg(\,
  \underline{\widetilde\complex_\theta^4}\, )}(\zeta_2)=t_1^{-1}\otimes\zeta_2 \ , \nonumber \\[4pt]
\Delta_{\alg(\, \underline{\widetilde\complex_\theta^4}\,
  )}(\bar\zeta_1)=t_1^{-1}\otimes \bar\zeta_1 \quad &,& \quad
\Delta_{\alg(\, \underline{\widetilde\complex_\theta^4}\, )}(\bar\zeta_2)=t_2^{-1}\otimes \bar\zeta_2 \ ,
\label{DeltaLR4tilde}\end{eqnarray}
and again extended as an algebra map; in particular
$\Delta_{\alg(\, \underline{\widetilde\complex_\theta^4}\, )}(\tilde\rho\, )= (t_1\,t_2)^{-1}\otimes\tilde\rho$.

The $*$-involution on the generators 
$\zeta_i,\bar\zeta_i$, $i=1,2$ induced from the one in (\ref{stargrass-bis}) reads 
\beq
\zeta_1^\dag=q~\bar \zeta_1 \ , \qquad 
\zeta_2^\dag=-q^{-3}~\bar \zeta_2 \ , 
\label{starzeta}\eeq
from which it also follows that $\tilde\rho= q\ \zeta_1 \,
\zeta_1^\dag + q^3\ \zeta_2 \, \zeta_2^\dag$. The corresponding noncommutative real
variety is denoted $\widetilde\real_\theta^4$. 
Also on this second patch of the noncommutative four-sphere 
$S_\theta^4$, only for the coaction of $\galg=\alg((\real^\times)^2)$
is there compatibility with the $*$-structure.

The intersection of the two
open affine subvarieties $\real_\theta^4$ and $\widetilde\real_\theta^4$ is
described by adjoining the element $\tilde\rho$ to
$\alg(\real_\theta^4)$ and $\rho$ to
$\alg(\widetilde\real_\theta^4)$. The corresponding
gluing automorphism is a $*$-isomorphism
$G:\alg(\real_\theta^4)[\tilde\rho\, ]\to
[\rho]\alg(\widetilde\real_\theta^4)$ in the category
${}^{\galg_{\theta}}\Module$ by~\cite[Prop.~3.25]{CLSII}. This morphism has a natural ``geometric''
interpretation. In the overlap of the two patches there are two sets
of generators that describe ``points'', and $G$ describes how to pass
from the affine coordinates $\xi_j$ to the affine coordinates
$\zeta_j$; it is indeed just the identity map in terms of the
homogeneous coordinates $\Lambda^J$ on the grassmannian, i.e. we do
not ``move'' points, we simply describe how the coordinates of the two
patches are related. Note that only the projection
$\alg(\Gr_{\theta}(2;4))\to \alg(\real_\theta^4)$ corresponding to the central generator
$\Lambda^{(34)}$ is an algebra homomorphism, and it would be
interesting to find a better interpretation of the two affine
subvarieties as defining a ``cover'' in the noncommutative case $\theta\neq0$.

\subsection{Twistor bundle on $S_\theta^4$}

The {noncommutative twistor algebra} $\alg^{\rm
  tw}=\alg(\CP_{\theta}^3)$ is a particular instance of the
noncommutative projective spaces described in \S\ref{NCprojvar}; it is the
homogeneous coordinate algebra generated by $w_i$,
$i=1,2,3,4$ and the relations
\bea
w_i\,w_k&=&w_k\,w_i \ , \qquad i=1,2,3,4~,~k=3,4 \ , \nonumber\\[4pt]
w_1\,w_2&=&q^2~w_2\,w_1 \ .
\label{CP3rels}\eea
It is dual to the (complex) twistor space of the noncommutative sphere~$S_\theta^4$.  
{With} $q\in\real$,
there is a natural real structure on $\alg^{\rm tw}$ such that
\beq
w_1^\dag=w_2 \ , \qquad w_2^\dag=w_1 \ , \qquad w_3^\dag=w_4 \ ,
\qquad w_4^\dag=w_3 \ .
\label{starCP3}\eeq

At the level of the noncommutative
correspondence algebra $\alg(\Fl_{\theta}(1,2;4))$, regarded as
an $\alg(\Gr_{\theta}(2;4))$-bimodule, the Pl\"ucker
equations (\ref{flagpluck}) in the localized coordinate algebra
$\alg(\Fl_{\theta}(1,2;4))[\Lambda^{(34)}\,^{-1}]_0$ now read
\beq
w_1=-w_3\,\xi_1-w_4\,\bar\xi_2 \ , \qquad
w_2=-w_3\,\xi_2-w_4\,\bar\xi_1 \ .
\label{w1w2triv}\eeq
Using the commutation relations (\ref{wLambda}) together with the
multiplication rules of noncommutative Ore
localization, one easily checks that the generators
$w_3,w_4$ commute not only among themselves but also with
$\xi_i,\bar\xi_i$, $i=1,2$, and it follows that
\beq
\alg\big(\Fl_{\theta}(1,2;4)\big)
\big[\Lambda^{(34)}\,^{-1}\big]_0
\cong\alg\big(\real_\theta^4\big)\otimes \alg\big(\CP^1\big) \ ,
\label{flagloc34}\eeq
with $\alg(\CP^1)=\complex[w_3,w_4]$ the homogeneous coordinate algebra of a commutative projective line $\CP^1$. This isomorphism implies that, in the image of the restriction functor $
j^\bullet: \coh(\Gr_{\theta}(2;4))\rightarrow
\coh(\real_\theta^4)$ induced by localization on the real
grassmannian, the tautological bundle $\Scal_{\theta}$ obtained
through the twistor transform restricts to the free right
$\alg(\real_\theta^4)$-module of rank~two, spanned by $w_3$ and $w_4$.

The situation is somewhat different for the noncommutative localization with respect to the minor $\Lambda^{(12)}$. The Pl\"ucker equations (\ref{flagpluck}) in
${}_0[\Lambda^{(12)}\,^{-1}]\alg(\Fl_{\theta}(1,2;4))$ are
\beq
w_3=-q~ \bar\zeta_1\, w_1 + q^{-1}\ \bar\zeta_2\,w_2 \ , \qquad
w_4=q~ \zeta_2\,w_1 - q^{-1}\ \zeta_1\,w_2 \ .
\label{w3w4triv}\eeq
Now the generators $w_1,w_2$ do not commute with
$\zeta_i,\bar\zeta_i$, $i=1,2$ in general: one finds that the localized coordinate algebra
has the structure of the braided tensor product algebra
$$
{}_0\big[\Lambda^{(12)}\,^{-1}\big]\alg\big(\Fl_{\theta}(1,2;4)\big)
\cong
\alg\big(\CP_\theta^1\big)\otimes_\theta\,
\alg\big(\widetilde\real_\theta^4\big) \ ,
$$
where the noncommutative projective line $\CP_\theta^1$ has
homogeneous coordinates $w_1,w_2$ subject to the relations in
(\ref{CP3rels}). Coherent sheaves on $\CP_\theta^1$ can be
functorially identified with coherent sheaves on a commutative line $\CP^1$~\cite[Prop.~4.2]{CLSII}, and hence, in the image of the restriction functor $\tilde j\,^\bullet:\coh(\Gr_{\theta}(2;4)) \to
\coh(\widetilde\real_\theta^4)$ induced by the localization,
the tautological bundle $\Scal_\theta$ restricts to the free right
$\alg(\widetilde\real_\theta^4)$-module of rank~two.

The free modules
$\alg(\real_\theta^4)^{\oplus2}$ and
$\alg(\widetilde\real_\theta^4)^{\oplus2}$ carry natural
$*$-involutions induced by (\ref{starxi}), (\ref{starzeta}) and
(\ref{starCP3}), and there is naturally a $*$-isomorphism
$
G_2: \alg(\real_\theta^4)[\tilde\rho\,]^{\oplus2}
\rightarrow[\rho]\alg(\widetilde\real_\theta^4)^{\oplus2}
$
in the category ${}^{\galg_\theta}\Module$ which is compatible with
the module structures. This describes
the twistor fibration over the noncommutative sphere $S_\theta^4$.

\subsection{Braided ADHM parametrization of instanton moduli}\label{ncADHM}

We shall now study framed sheaves on $\Open(\CP_\theta^2)$ and relate
  their moduli to a noncommutative deformation of the standard ADHM data.

\subsubsection*{Instanton moduli spaces}

With $q=\exp(\frac\ii2\,\theta)$ and $\theta\in\complex$, 
consider the homogeneous coordinate algebra
$\alg=\alg(\CP_\theta^2)$ whose generators have relations 
\beq
w_1\,w_2=q^2~w_2\,w_1 \ , \qquad w_1\,w_3=w_3\,w_1 \ , \qquad
w_2\,w_3=w_3\,w_2  \ .
\label{CP2homcoordrels}\eeq
{Let $\alg_\infty:=\alg/ \langle w_3 \rangle$;  
we identify
$\alg_\infty=\alg(\CP_\theta^1)$ as the homogeneous coordinate algebra
dual to a {noncommutative projective line} $\CP_\theta^1$. 
The algebra projection $p:\alg\to\alg_\infty$ is dual to a closed
embedding $\CP_\theta^1\hookrightarrow\CP_\theta^2$ of noncommutative
projective varieties.
Let $i:\alg_\infty\hookrightarrow \alg$ be the algebra inclusion. 
Then the inclusion functor $i^\bullet$ gives a map 
from $\alg$-modules to $\alg_\infty$-modules, 
$M\mapsto M_\infty:=M/ (M\cdot\langle w_3\rangle)$. }
There is a natural equivalence of abelian categories
$\coh(\CP_\theta^1)\cong\coh(\CP^1)$ for any $\theta\in\complex$
\cite[Prop.~4.2]{CLSII}, which enables us to exploit the known cohomology of sheaves on the commutative projective line $\CP^1$.

Let us fix complex vector spaces $V$ and $W$
of dimensions $k$ and $r$, respectively. A {framed sheaf} on
$\CP_\theta^2$ is a
coherent torsion free sheaf $E$ on $\Open(\CP_\theta^2)$ together with
isomorphisms $H^1(\CP_\theta^2,E(-1))\cong V$ and
$\varphi:E_\infty:=i^\bullet(E)\to W\otimes\sheaf_{\CP_\theta^1}$; the map
$\varphi$ is called a
{framing of $E$ at infinity}. A {morphism of framed sheaves} $(E,\varphi)$ and $(E',\varphi'\,)$ is a
homomorphism of $\alg$-modules $\xi:E\to E'$ which
preserves the framing isomorphisms, i.e. $\varphi'\circ i^\bullet(\xi) = \varphi$.
The \emph{instanton moduli space $\Mcal_\theta(r,k)$} is the set of isomorphism
classes $[(E,\varphi)]$ of framed sheaves $(E,\varphi)$. 
In \cite[Cor. 4.8]{CLSII} it is shown that an isomorphism class $[(E,\varphi)]\in\Mcal_\theta(r,k)$
has invariants
$$
{\rm rank}(E)=r \ , \qquad \chi(E)=r-k \ ,
$$
where $\chi(E):=\sum_{p\geq0}\, (-1)^p\, \dim_\complex H^p(\CP^2_\theta,E)$ is the noncommutative
Euler characteristic of the framed sheaf $E$.

{Our goal now} is to reduce the study of moduli of framed sheaves on $\Open(\CP_\theta^2)$ to a problem of \emph{linear} algebra, 
defining a noncommutative deformation of the standard ADHM data. For
this, we consider quadruples 
\bea
\qquad (B_1,B_2,I,J)  \ \in \ \Xcal(W,V) := \End_\complex(V)^{\oplus2}
~\oplus~ \Hom_\complex(W,
V)~ 
\oplus~\Hom_\complex(V,W) \ .
\label{BIJsubvar}\eea
When we consider $\hil_\theta$-coequivariant
sheaves, there are natural coactions of the cotwisted Hopf algebra $\hil_\theta$
induced by the torus $T=(\complex^\times)^2$ on the moduli
spaces. The finite-dimensional vector spaces $V$ and $W$ are then
$\hil_\theta$-comodules, i.e. objects of the tensor category
${}^{\hil_{\theta}}\Module$, and
so are the vector spaces $\End_\complex(V)\cong V^*\otimes V$ 
and $\Hom_\complex(W,V)\cong W^*\otimes V$. The explicit coaction on the quadruples
(\ref{BIJsubvar}) is given by \cite[\S4.2]{CLSII}
\beq
\Delta_{\Xcal(W,V)}(B_1,B_2,I,J)=\big(t_1\otimes B_1\,,\,t_2\otimes
B_2\,,\,(t_1\,t_2)\otimes I\,,\, 1\otimes J \big) \ .
\label{DeltaLBIJ}\eeq
Here we only regard the parameter subspace $\End_\complex(V)$ as a
vector space object of ${}^{\hil_{\theta}}\Module$, hence we do not
deform the product on the endomorphism algebra where the matrices $B_i$ live. 
The variety $\Mcal_\theta^{{\tiny \rm ADHM}}(r,k)$ of \emph{noncommutative
  complex ADHM data} is the locally closed subvariety of quadruples
(\ref{BIJsubvar}) subject to the following two conditions: 
\begin{itemize}
\item[{\tt (I.1)}] {\tt Noncommutative complex ADHM equation:}
\beq
[B_1,B_2]_{-\theta} +I\, J=0
\label{NCADHMeqs}\eeq
in $\End_\complex(V)$, where the {braided commutator} 
$$ 
[B_1,B_2]_\theta:=
B_1\,B_2-q^{2}~B_2\,B_1
$$
is induced by the functorial isomorphism (\ref{PsithetaVW}) of the
category ${}^{\hil_{\theta}}\Module$; indeed using (\ref{brF}) we have
$\Psi^{\theta}(B_1\otimes B_2) = \Rcal_{\theta}(t_2\otimes t_1) \, B_2\otimes B_1 = q^2\, B_2\otimes B_1 $. \\
\item[{\tt (I.2)}] {\tt Stability:} \ There are no proper
  $B_i$-invariant subspaces of $V$ for $i=1,2$ which contain the
  image of $I$. 
\end{itemize}
The general linear group $\GL(k)$ of basis change automorphisms of the vector space $V$ acts naturally on the variety $\Mcal_\theta^{{\tiny \rm ADHM}}(r,k)$ as 
\beq
g\triangleright(B_1,B_2,I,J)=\big(g\, B_1\, g^{-1}\,,\, g\, B_2\,
g^{-1}\,,\,
g\, I\,,\,J\, g^{-1}\big) \ , \qquad g\in\GL(k) \ .
\label{GLVaction}\eeq
This action is free and proper \cite[Lem.~4.20]{CLSII}, and so
the quasi-projective variety of closed $\GL(k)$-orbits on the space $\Mcal_\theta^{{\tiny \rm ADHM}}(r,k)$ is given by the geometric invariant theory quotient
\beq
\widehat\Mcal_\theta^{{\tiny \rm ADHM}}(r,k):=\Mcal_\theta^{{\tiny \rm ADHM}}(r,k)\,
\big/\,\GL(k) \ .
\label{MhatADHM}
\eeq

There is a natural (set theoretic) bijection between the moduli space
$\widehat\Mcal_\theta^{{\tiny \rm ADHM}}(r,k)$ of braided linear
algebraic ADHM data and the moduli space $\Mcal_\theta(r,k)$
of framed sheaves on $\Open(\CP_\theta^2)$. 
The proof of this result is quite technical, see \cite[\S 4.3]{CLSII};
here we only briefly highlight the two essential steps that we need in the
following:
\begin{itemize}
\item[{\tt 1.}]
Realise framed sheaves on $\Open(\CP_\theta^2)$ as the middle cohomology of a linear monad complex
$$
\underline{\calg}\,_\bullet\,:\, 0~\longrightarrow~
V\otimes\sheaf_{\CP_\theta^2}(-1)~
\xrightarrow{ \ \sigma \ }~V_0 \otimes\sheaf_{\CP_\theta^2} 
~ \xrightarrow{ \ \tau \ }~V\otimes\sheaf_{\CP_\theta^2}(1)~
\longrightarrow~0
$$
with $V_0=V\oplus V\oplus W$. Conversely, essentially any linear monad
$\underline{\calg}\,_\bullet$ on $\CP_\theta^2$ of this form
defines an isomorphism class in $\Mcal_\theta(r,k)$,
see~\cite[Thm. 4.25]{CLSII}. \\
\item[{\tt 2.}]
Show that the ADHM moduli space \eqref{MhatADHM} has a similar monadic
description. Given a quadruple of ADHM data $(B_1, B_2,I,J)\in \Mcal_\theta^{{\tiny \rm ADHM}}(r,k)$, we define canonical sheaf morphisms
\bea
\sigma = \begin{pmatrix}
B_1\, w_3+q\, w_1 \\ B_2\, w_3+q^{-1}\,
w_2 \\  J\, w_3 \end{pmatrix}
\, : \, V\otimes\sheaf_{\CP_\theta^2}(-1)~
\longrightarrow~ V_0
\otimes\sheaf_{\CP_\theta^2}
\label{sigmaBIJ}\eea
and
\bea\label{tauBIJ}
\qquad \tau = \begin{pmatrix} B_1\, w_3+ q^{-1}\, w_1 & B_2\, w_3+q\,
w_2 & I\, w_3 \end{pmatrix} \, : \, V_0
\otimes\sheaf_{\CP_\theta^2}~
\longrightarrow~ V
\otimes\sheaf_{\CP_\theta^2}(1) \ .
\eea
These maps determine a chain $\underline{\calg}\,_\bullet$ of morphisms of coherent sheaves on
$\Open(\CP_\theta^2)$, which by~\cite[Thm. 4.32]{CLSII} is a monad of
the type considered above. Every framed torsion free sheaf on $\CP_\theta^2$
arises from this construction.
\end{itemize}
In the next section we will describe this correspondence as an
isomorphism between smooth algebraic varieties.

\subsubsection*{Instanton charge one}

In the case $k=1$ we have $B_i=b_i\in\complex$ for $i=1,2$, and we can regard the morphisms $I$ and
$J$ as vectors 
$$ I=(i_1,\dots,i_r)\ \in \ W^* \ , \qquad J=\begin{pmatrix} \, j_1 \,
  \\ \vdots \\
  \, j_r \, \end{pmatrix}\ \in \ W \ .
$$
The braided ADHM equation (\ref{NCADHMeqs}) then
defines a quadric in $\complex^{2r+2}$ given by
\beq
\big(1-q^{-2}\big)\  b_1\,  b_2+\sum_{l=1}^r\, i_l\, j_l =0 \ .
\label{ADHMquadric}\eeq
Stability is equivalent to $i_l\neq0$ for at least one index $l=1,\dots,r$, showing that the moduli space 
$\widehat\Mcal_{\theta}^{{\tiny \rm ADHM}}(r,1)$ is quasi-projective. An element $t\in\GL(1)=\complex^\times$ 
acts trivially on $ b_1, b_2$, and as multiplication by $t$ on $i_l$ and by $t^{-1}$ on $j_l$ for each $l=1,\dots,r$. 
In the patch $i_r\neq 0$ we can use this scaling symmetry to set $i_r=1$, and then use (\ref{ADHMquadric}) to eliminate
$j_r\in\complex$. This coordinatizes  $\widehat\Mcal_{\theta}^{{\tiny \rm ADHM}}(r,1)$ as a patch
$\complex^2\times \complex^{r-1} \times \complex^{r-1}$. The last factor is the 
cotangent space at the point $[I]\in \P(W^*) \cong \CP^{r-1}$: with
$i_r=1$ and $j_r = - (1-q^{-2}) \, b_1\, b_2$, the quadric \eqref{ADHMquadric} reads 
\beq
\sum_{l=1}^{r-1}\, i_l\, j_l =0 \ , 
\label{ADHMquadricbis}\eeq
which is an equation for $(j_1,\dots,j_{r-1}) \in T_{[I]}^*\CP^{r-1}$. This gives the charge~$1$ instanton moduli space
\beq
\Mcal_{\theta}(r,1) \ \cong \ \complex^2 \times T^*\CP^{r-1} 
\label{Mcalthetar1}\eeq
as a complex variety for all $\theta\in\complex$ and $r\geq1$.

\subsubsection*{Self-conjugate instanton modules}

We will now provide another characterization of the moduli space
$\Mcal_\theta(r,k)$ in terms of ``real'' linear algebraic data. For
this, we fix hermitean inner products on the complex vector spaces $V$
and $W$. 
{Then, with $q\in\real$, the space of quadruples $\Xcal(W,V)$ in 
(\ref{BIJsubvar})
acquires a natural quaternionic structure given by 
the conjugate linear anti-involution
\begin{equation}\label{q-s}
\Jscr\,:\,\Xcal(W,V)~\longrightarrow~ \Xcal(W,V) \ , \qquad
\Jscr(B_1,B_2,I,J)=\big(-B_2^\dag\,,\,B_1^\dag\,,\, -J^\dag\,,\,I^\dag\,\big)
\end{equation}
extended as an anti-algebra map. The variety $\Mcal_\theta^\real(r,k)$ of \emph{noncommutative real
  ADHM data} is the subspace of $\Mcal_\theta^{{\tiny \rm ADHM}}(r,k)$
consisting of quadruples (\ref{BIJsubvar}) which satisfy, in addition to
conditions~{\tt (I.1)} and~{\tt (I.2)} of the complex data, the real
ADHM equation
\beq\label{realADHMeq}
\big[B_1\,,\,B_1^\dag\,\big]_{-\theta} + q^{-2}~
\big[B_2\,,\,B_2^\dag\,\big]_\theta +I\, I^\dag-J^\dag\,
J=0
\eeq
in $\End_\complex(V)$.
The natural $\GL(k)$-action on $\Mcal_\theta^{{\tiny \rm ADHM}}(r,k)$ reduces 
on $\Mcal_\theta^\real(r,k)$ to an action of the group $\U(k)$ of
unitary basis change automorphisms of the vector space $V$. The corresponding
space of stable orbits is denoted $\widehat\Mcal_\theta^{\rm
  tw}(r,k)$. This quotient is related to a particular class of framed
torsion free sheaves on $\CP_\theta^2$, as we explain momentarily.

Let us first make two remarks concerning the map defined by
(\ref{q-s}): 
\begin{itemize}
\item We can rewrite the real ADHM equation (\ref{realADHMeq}) as
$$ \big[\Jscr(B_1)\,,\,B_2\big]_{-\theta} + \big[B_1\,,\,\Jscr(B_2)
\big]_{-\theta} + \Jscr(I) \, J + I \, \Jscr(J) = 0 \ . $$
This is precisely the equation that would arise if we let the
quaternionic structure $\Jscr$ act on the complex ADHM equation
(\ref{NCADHMeqs}) as a derivation, rather than as an anti-algebra map;
hence if we interpret $\Jscr$ as a symmetry generator in this way then
the real ADHM equation is automatically satisfied. \\
\item 
Both the complex ADHM equation \eqref{NCADHMeqs}
and the real ADHM equation \eqref{realADHMeq} are invariant for the
full coaction associated to $T$. On the other hand, requiring that the quaternionic structure in \eqref{q-s} be coequivariant, i.e.
$(\Id\otimes\Jscr)\circ \Delta_{\Xcal(W,V)}=\Delta_{\Xcal(W,V)}\circ\Jscr$ so that it is 
a morphism in the category ${}^{\hil_\theta}\Module$ of left $\hil_\theta$-comodules, 
restricts the toric symmetry to a coaction associated with the real subgroup $G\subset T$ 
with $(t_1)^*=t_2$. Here we take $B_i^\dag$ to transform with weight
$t_i^*$ for $i=1,2$, $J^\dag$ with weight $(t_1\,t_2)^*$, and
$I^\dag$ to be coinvariant.
\end{itemize}

We will now realize a correspondence between $\widehat\Mcal_\theta^\real(r,k)$ and linear monads on $\CP_{\theta}^3$. We endow the noncommutative twistor algebra $\alg^{\rm tw}=\alg(\CP_{\theta}^3)$ with 
a conjugate linear
anti-involution $\Jscr:\alg^{\rm tw} \to (\alg^{\rm tw})^{\rm op}$
acting on generators as the quaternionic structure
\beq
\Jscr(w_1,w_2,w_3,w_4)=(w_2,-w_1,w_4,-w_3) \ .
\label{Jscrw}\eeq
There is a natural embedding of the noncommutative projective plane $\CP_\theta^2$ into the noncommutative twistor space $\CP_{\theta}^3$. The homogeneous coordinate algebra of the original $\CP_\theta^2$ is recovered through $\alg\cong\alg^{\rm tw}/\langle w_4\rangle$. Let $\iota:\alg\hookrightarrow\alg^{\rm tw}$ be the natural algebra inclusion. We again denote by
$i:\alg_\infty\hookrightarrow\alg^{\rm tw}$ the algebra inclusion of the noncommutative projective line $\CP_\theta^1$ with
$\alg_\infty\cong \alg^{\rm tw}/\langle w_3,w_4\rangle$. 
Consider a linear monad on $\CP_{\theta}^3$ of the form
$$
\underline{\calg}\,^{\rm tw}_\bullet\,:\, 0~\longrightarrow~
V\otimes\sheaf_{\CP_{\theta}^3}(-1)~
\xrightarrow{ \ \sigma \ }~V_0\otimes\sheaf_{\CP_{\theta}^3}~
\xrightarrow{ \ \tau \ }~V\otimes\sheaf_{\CP_{\theta}^3}(1)~
\longrightarrow~0 \ .
$$
Its restriction
$i^\bullet(\,\underline{\calg}\,^{\rm tw}_\bullet)$ is again a monad on
$\CP_\theta^1$ which is quasi-isomorphic to
$W\otimes\sheaf_{\CP_\theta^1}$. The anti-homomorphism $\Jscr$ induces
a functor $\Jscr^\bullet$ between categories of sheaves
which, when extended to the derived category of
$\coh(\CP_{\theta}^3)$ and applied to a monad
$\underline{\calg}\,^{\rm tw}_\bullet$, gives a monad
$$
\underline{\calg}\,^{\rm tw}_\bullet\,^\dag\,:\, 0~\longrightarrow~
\overline{V}\,^*\otimes\sheaf_{\CP_{\theta}^3}(-1)~
\xrightarrow{ \ \sigma^\dag \ }~\overline{V_0}\,^*
\otimes\sheaf_{\CP_{\theta}^3}~
\xrightarrow{ \ \tau^\dag \ }~\overline{V}\,^*
\otimes\sheaf_{\CP_{\theta}^3}(1)~\longrightarrow~0 \ .
$$
Here the bars denote complex conjugation while
$\sigma^\dag:=\Jscr^\bullet(\sigma)^*$ and $\tau^\dag:=\Jscr^\bullet(\tau)^*$. We say that a monad
$\underline{\calg}\,^{\rm
  tw}_\bullet$ is {self-conjugate} if there is an isomorphism
$\underline{\calg}\,^{\rm tw}_\bullet\,^\dag \cong
\underline{\calg}\,^{\rm tw}_\bullet$ of complexes.

Then there is a natural (set-theoretic) bijective correspondence between isomorphism
classes of self-conjugate linear monad complexes $\underline{\calg}\,^{\rm
  tw}_\bullet$ on $\CP_{\theta}^3$ and isomorphism
classes in the moduli space~$\widehat\Mcal_\theta^\real(r,k)$ of
braided real ADHM data; the proof follows a similar argument to that
sketched above, once we take into account the real structure (see~\cite[Thm.~4.40]{CLSII} for details). The restriction $\iota^\bullet(\,\underline{\calg}\,^\real_\bullet)$ of a self-conjugate monad on $\CP_{\theta}^3$ is a monad on $\CP_\theta^2$ which we know defines an isomorphism
class in $\Mcal_\theta(r,k)$. This gives a map of moduli spaces
$\widehat\Mcal_\theta^\real(r,k)\to \Mcal_\theta(r,k)$. At the
level of noncommutative ADHM data, this map is just the natural
inclusion of varieties $\Mcal_\theta^\real(r,k)\hookrightarrow
\Mcal_\theta^{{\tiny \rm ADHM}}(r,k)$. 

\subsection{Construction of noncommutative instantons}	
\label{constrinst}

We will now demonstrate that
the smaller class of torsion free sheaves on $\Open(\CP_\theta^2)$
parametrized by the moduli space $\widehat\Mcal_\theta^\real(r,k)$
corresponds directly to a class of anti-selfdual connections on a
canonically associated ``instanton bundle'' {over the noncommutative sphere $S^4_\theta$}.

\subsubsection*{Instanton modules}

We use the monadic description of noncommutative
real ADHM data on the twistor space $\CP_{\theta}^3$ to construct
canonical bundles, called ``instanton bundles'', on the
noncommutative sphere $S_\theta^4$. This is achieved via the twistor
transform of real framed torsion free sheaves with $E^\dag\cong E$ on $\Open(\CP_{\theta}^3)$ via the noncommutative correspondence diagram \eqref{nccorr}.
We can apply the derived functor
of the twistor transform $p_2{}^*\,p_{1*}$ to a self-conjugate
linear monad complex $\underline{\calg}\,^{\rm tw}_\bullet$ on
$\CP_{\theta}^3$ to get the complex
\beq \qquad
p_2{}^*\,p_{1*}\big(\,\underline{\calg}\,^{\rm tw}_\bullet\big)
\,:\, 0~\longrightarrow~V_0\otimes \sheaf_{\Gr_{\theta}(2;4)}~
\xrightarrow{(\Id_V\otimes\hat\eta)\circ\tau}~V\otimes
\Scal_{\theta}~\longrightarrow~0 \ ,
\label{twisttransfmonad}\eeq
where $\hat\eta$ is the rank~two projector which defines the tautological bundle over the Grassmann variety as $\Scal_{\theta}\cong
\hat\eta\big(\alg(\Gr_{\theta}(2;4))^{\oplus4}\big)$ (via the
noncommutative Euler sequence, see~\cite[eq.~(3.7)]{CLSII}). It follows that
the image of a cohomology sheaf $E=H^0\big( \,\underline{\calg}\,^{\rm
  tw}_\bullet\big)$ under the twistor transform is the sheaf on $\Open(\Gr_{\theta}(2;4))$
given by
$$
E'=p_2{}^*\,p_{1*}(E)=\ker\big((\Id\otimes\hat\eta)\circ\tau\big)
\ .
$$

It follows from (\ref{flagloc34}) that the restriction of the complex (\ref{twisttransfmonad}) to the patch 
$\real_\theta^4$ of the noncommutative sphere $S^4_\theta$ yields the ``Dirac operator''
\bea
&& \quad 
\Dcal:=j^\bullet\,(\Id\otimes\hat\eta)\circ\tau \,:\, V_0 \otimes\alg\big(\real_\theta^4\big) ~\longrightarrow~
(V\oplus V)\otimes\alg\big(\real_\theta^4\big) \ ,
\label{Diracop}\eea
which is written explicitly as
\begin{eqnarray}
\Dcal = \begin{pmatrix}
B_1-q^{-1}\, \xi_1 &
B_2-q\,\xi_2 & I
\\ -B_2^\dag-q^{-1}\,\bar\xi_2 & B_1^\dag-q\,
\bar\xi_1 & -J^\dag
\end{pmatrix} \ .
\label{Diracopexpl}\end{eqnarray}
The operator $\Dcal$ is a surjective morphism of free right
$\alg(\real_\theta^4)$-modules and $\Dcal^\dag$ is injective; here the $\dag$-involution is the tensor product of the real structure given by (\ref{starxi}) and those of the chosen hermitean structures on the vector spaces $V$ and $W$. The ``Laplace operator''
\beq
\triangle:=\Dcal\circ\Dcal^\dag \,:\, (V\oplus
V)\otimes\alg\big(\real_\theta^4\big) ~\longrightarrow~ (V\oplus
V)\otimes\alg\big(\real_\theta^4\big)
\label{Laplaceop}
\eeq
is an isomorphism and there is a decomposition
\beq
V_0\otimes\alg\big(\real_\theta^4\big) =\im\big(\Dcal^\dag\,\big)
\oplus \ker(\Dcal)
\label{imkerdecomp}\eeq
of $\alg(\real_\theta^4)$-modules~\cite[Prop.~5.5]{CLSII}. 
Moreover, the operators $\Dcal$, $\Dcal^\dag$ and $\triangle$ are all
$\galg_\theta$-coequivariant morphisms. The right $\alg(\real_\theta^4)$-module
\beq
\bun :=\ker(\Dcal) =\ker\big(
j^\bullet\,(\Id\otimes\hat\eta)\circ\tau \big)
\label{instbun}\eeq
is projective by (\ref{imkerdecomp}). It is also finitely generated,
and has rank $\dim_\complex(W)=r$ since $\Dcal$ is surjective. The
corresponding projection can be given as
\beq
P:= \Id - \Dcal^\dag\circ\triangle^{-1}\circ \Dcal\, :\,
V_0\otimes\alg\big(\real_\theta^4\big) ~\longrightarrow~ \bun \ ,
\label{instbunproj}\eeq
with $P^2=P=P^\dag$ and trace $\Tr
P=\dim_\complex(V_0)-\dim_\complex(V\oplus V)= r$. The module (\ref{instbun}) is called an \emph{instanton bundle} over $\real_\theta^4$; it defines an object of the category ${}^{\galg_{\theta}}\Module$. One
can show that the isomorphism class of $\bun$ depends only
on the class of the noncommutative ADHM data $(B_1,B_2,I,J)$ in the moduli
variety $\Mcal_\theta^\real(r,k)$, and that the reality condition
$\bun^\dag\cong \bun$ is satisfied. 
One can repeat these arguments to
construct an instanton bundle $\widetilde\bun$ on the second patch
$\widetilde{\real}_\theta^4$ of $S_\theta^4$, and then use the gluing
automorphism $G$ between the two patches to induce an isomorphism
between the modules $\bun[\tilde\rho\,]:=\bun\otimes_{\alg(\real_\theta^4)}
\alg(\real_\theta^4)[\tilde\rho\,]$ and $[\rho]\tilde\bun:=
[\rho]\alg(\widetilde\real_\theta^4)\otimes_{\alg(\widetilde\real_\theta^4)}
\tilde\bun$ which is compatible
with the $\dag$-involutions, thus defining a $\galg_\theta$-coequivariant instanton
bundle on the noncommutative sphere $S_\theta^4$.

\subsubsection*{Instanton gauge fields}

We will now construct canonical connections on the instanton
bundle. For this, let $\Omega^\bullet_{\real^4}=\bigwedge^\bullet\Omega_{\real^4}^1$ be the usual classical differential calculus on the coordinate
algebra $\alg(\real^4)$. It is generated as a differential graded
algebra by degree zero elements $\xi_i,\bar\xi_i$ and degree one
elements $\dd\xi_i,\dd\bar\xi_i$ satisfying the skew-commutation relations
$$
\dd\xi_i\wedge\dd\xi_j=-\dd\xi_j\wedge\dd\xi_i \ , \qquad
\dd\xi_i\wedge\dd\bar\xi_j=-\dd\bar\xi_j\wedge\dd\xi_i \ , \qquad
\dd\bar\xi_i\wedge\dd\bar\xi_j=-\dd\bar\xi_j\wedge\dd\bar\xi_i
$$
and the symmetric $\alg(\real^4)$-bimodule structure
$$
\xi_i~\dd\xi_j=\dd\xi_j~\xi_i \ , \qquad
\xi_i~\dd\bar\xi_j=\dd\bar\xi_j~\xi_i \ , \qquad
\bar\xi_i~\dd\bar\xi_j=\dd\bar\xi_j~\bar\xi_i
$$
for $i,j=1,2$. The differential
$\dd:\Omega^0_{\real^4}:=\alg(\real^4)\to\Omega^1_{\real^4}$ is
defined by $\xi_i\mapsto\dd\xi_i$, $\bar\xi_i\mapsto\dd\bar\xi_i$ and 
extended uniquely to a map of degree one, $\dd:\Omega^n_{\real^4}\to
\Omega^{n+1}_{\real^4}$, with $\dd^2=0$, using $\complex$-linearity and the graded
Leibniz rule
$$
\dd(\omega\wedge\omega'\,)=\dd\omega\wedge\omega'+(-1)^{{\rm
    deg}(\omega)}\, \omega\wedge\dd\omega'
$$
where the product is taken over $\alg(\real^4)$ using the
bimodule structure of $\Omega^\bullet_{\real^4}$. We will demand that the differential calculus on
$\alg(\real^4)$ is \emph{coequivariant} for the real torus coaction
$\Delta_L$ given by (\ref{DeltaLR4}). Then we can extend this coaction to a left coaction $\Delta_L:\Omega^\bullet_{\real^4}\to
\galg \otimes\Omega^\bullet_{\real^4}$ such that  $\Delta_L$ is an $\alg(\real^4)$-bimodule morphism and
$\dd$ is a left $\galg$-comodule morphism .
We next use the Drinfel'd twist $F_\theta$ of \eqref{Ftheta} to deform
$\Omega^\bullet_{\real^4}$, following the general prescription for
$\galg$-comodule algebras. This defines the braided exterior product
$\wedge_\theta$ as the projection of the tensor product on the quotient of the tensor algebra
$$
T_\theta\big(\Omega_{\real^4}^1\big)=\alg\big(\real_\theta^4\big) \ \oplus \ \bigoplus_{n\geq1}\,
\big(\Omega_{\real^4}^1\big)^{\otimes_{\alg(\real_\theta^4)}\, n}
$$ 
by the ideal generated by the braided skew-commutation relations (see (\ref{qest})). The undeformed differential $\dd$ is still a derivation (of degree one) of the deformed product $\wedge_\theta$, as one can easily check. The above construction defines a canonical differential graded algebra $\Omega_{\real_\theta^4}^\bullet= \bigwedge^\bullet_\theta\Omega_{\real^4}^1$ 
subject to the braided skew-commutation relations
\begin{eqnarray}
\dd\xi_i\wedge \dd\xi_j\= -q_{ij}^2~\dd\xi_j\wedge\dd\xi_i \quad &,& \quad
\dd\bar\xi_i\wedge \dd\bar\xi_j\= -q_{i+1\,j+1}^2~\dd\bar\xi_j\wedge\dd\bar\xi_i \ , \nonumber \\[4pt] 
\dd\xi_i\wedge \dd\bar\xi_j &=& -q_{i\,j+1}^2~\dd\bar\xi_j\wedge\dd\xi_i
\label{R4skewcomm}\end{eqnarray}
and the braided symmetric $\alg(\real_\theta^4)$-bimodule structure
\begin{eqnarray*}
\xi_i~\dd\xi_j\= q_{ij}^2~\dd\xi_j~\xi_i \quad &,& \quad
\bar\xi_i~\dd\bar\xi_j\= q_{i+1\,j+1}^2~\dd\bar\xi_j~\bar\xi_i \ , \\[4pt]
\xi_i~\dd\bar\xi_j\= q_{i\,j+1}^2~\dd\bar\xi_j~\xi_i \quad &,& \quad
\bar\xi_i~\dd\xi_j\= q_{i+1\,j}^2~\dd\xi_j~\bar\xi_i \ .
\end{eqnarray*}
Here we drop the explicit deformation symbols from the products. The
real structure (\ref{starxi}) extends to
$\Omega_{\real_\theta^4}^\bullet$ by graded extension of the morphism
$\xi_i\mapsto\xi_i^\dag$ {and $(\dd \xi_i)^\dag=\dd (\xi_i^\dag)$}.

We are ready to give a canonical connection on the right $\alg(\real_\theta^4)$-module $\bun$: a linear map
$$
\nabla\,:\, \bun ~\longrightarrow~ \bun\otimes_{\alg(\real_\theta^4)} \Omega^1_{\real_\theta^4}
$$
satisfying the Leibniz rule
$$
\nabla(\sigma\triangleleft f)=(\nabla\sigma)\triangleleft f+ \sigma\otimes \dd f
$$
for $f\in\alg(\real_\theta^4)$ and $\sigma\in\bun$. 
Let $\iota:\bun\to
V_0\otimes\alg(\real_\theta^4)$ denote the natural inclusion, and
$\dd:\alg(\real_\theta^4)\to \Omega^1_{\real_\theta^4}$ the
differential introduced above. We then take $\nabla$ to be the
Grassmann connection associated to the projection $P$, which is
defined by the composition
$$
\nabla = P\circ\dd\circ\iota \ .
$$
It is easily seen to be compatible
with the $\dag$-involution, and its curvature is given by
$$
F_\nabla=\nabla^2 = P~(\dd P)^2 \ \in\ \Hom_{\alg(\real_\theta^4)} 
\big( \bun\,,\, \bun
\otimes_{\alg(\real_\theta^4)}\Omega^2_{\real_\theta^4}\big) \ .
$$
Gauge transformations $g\in\Aut_{\alg(\real_\theta^4)}(\bun)$ act
naturally through $\nabla\mapsto g\,\nabla\, g^{-1}$, from which it
follows that
$F_\nabla\mapsto g\, F_{\nabla}\,g^{-1}$; the gauge equivalence class of the instanton connection $\nabla=P\circ\dd\circ\iota$ depends only on the class of the noncommutative ADHM data $(B_1,B_2,I,J)$ in the moduli variety $\Mcal_\theta^\real(r,k)$~\cite[Lem.~5.13]{CLSII}.
Similarly one builds the instanton connection
$\widetilde\nabla$ on the left
$\alg(\widetilde\real_\theta^4)$-module $\tilde\bun$, 
with curvature $F_{\widetilde\nabla}=\widetilde P~(\dd\widetilde
P)^2\in \Hom_{\alg(\widetilde\real_\theta^4)}\big(\tilde\bun\,,\, \Omega^2_{\widetilde\real_\theta^4} \otimes_{\alg(\widetilde\real_\theta^4)}
\tilde\bun\big)$.

The instanton connections
$\nabla_{\tilde\rho}$ and ${}_\rho\widetilde\nabla$ on the adjoinment
bundles $\bun[\tilde\rho\, ]$ and $[\rho]\tilde\bun$, respectively,
are related by a gauge transformation, which thereby defines the instanton connection on the noncommutative sphere~$S_\theta^4$.

\subsubsection*{BPS equations}

We will now demonstrate that the Grassmann connections constructed above define noncommutative instantons in analogy with
the classical case, i.e. that they satisfy some form of anti-selfduality equations with respect to a suitable Hodge duality operator acting on two-forms. The classical Hodge duality operator on two-forms $*:\Omega^2_{\real^4}\to\Omega^2_{\real^4}$, induced by the euclidean metric on $\real^4$, is given by
\begin{eqnarray}
*(\dd\xi_1\wedge\dd\bar\xi_2) = -\dd\xi_1\wedge\dd\bar\xi_2 \quad &,&
\quad *(\dd\xi_1\wedge\dd\xi_2) = \dd\xi_1\wedge\dd\xi_2 \ ,
\nonumber \\[4pt] 
*(\dd\xi_1\wedge\dd\bar\xi_1) = -\dd\xi_2\wedge\dd\bar\xi_2 \quad &,&
\quad *(\dd\xi_2\wedge\dd\bar\xi_2) = -\dd\xi_1\wedge\dd\bar\xi_1 \ ,
\nonumber \\[4pt]
*(\dd\bar\xi_1\wedge\dd\bar\xi_2) = \dd\bar\xi_1\wedge\dd\bar\xi_2
\quad &,& \quad *(\dd\xi_2\wedge\dd\bar\xi_1) = -\dd\xi_2\wedge\dd\bar\xi_1 \ .
\label{Hodgeclass}\end{eqnarray}
In contrast to the case of isospectral deformations \cite{BLvS}, the
coaction (\ref{DeltaLR4}) of the real torus algebra $\galg =\alg((\real^\times)^2)$ on $\alg(\real^4)$
is \emph{not} isometric. However, it does coact by conformal
transformations on $\alg(\real^4)$, and hence preserves the Hodge
operator. This is easily checked using (\ref{Hodgeclass}) and
(\ref{DeltaLR4}) which shows that the operator
$*:\Omega^2_{\real^4}\to\Omega^2_{\real^4}$ is coequivariant,
$$
\Delta_L(*\omega)=(\Id\otimes *)\Delta_L(\omega) \qquad \mbox{for} \quad
\omega\in\Omega^2_{\real^4} \ ,
$$
and hence defines a morphism of the category ${}^\galg\Module$. Since
the vector space $\Omega_{\real_\theta^4}^2$ coincides with its
classical counterpart $\Omega^2_{\real^4}$, and the quantization
functor $\mathscr{F}_\theta:
{}^{\hil}\Module\to {}^{\hil_\theta}\Module$ acts as the identity
  on objects and morphisms of ${}^{\hil}\Module$, we can define a
  Hodge duality operator $*_\theta:\Omega_{\real_\theta^4}^2\to
  \Omega_{\real_\theta^4}^2$ by the same formula (\ref{Hodgeclass}),
  which by construction is a morphism in the category
  ${}^{\galg_\theta}\Module$. In particular, it satisfies
  $*_\theta^2=\Id$ and the $\galg_\theta$-coequivariant decomposition
$$
\Omega_{\real_\theta^4}^2 =\Omega_{\real_\theta^4}^{2,+} \oplus
\Omega_{\real_\theta^4}^{2,-} \ ,
$$
of the right $\alg(\real_\theta^4)$-module $\Omega_{\real_\theta^4}^2$
into submodules corresponding to eigenvalues $\pm\,1$ of $*_\theta$, 
is identical as a vector space to that of the classical
case. Hence the eigenmodules of selfdual and anti-selfdual
two-forms are given respectively by
\begin{align}
\Omega_{\real_\theta^4}^{2,+} & =
\alg\big(\real_\theta^4\big)\big\langle
\dd\xi_1\wedge\dd\xi_2\,,\, \dd\bar\xi_1\wedge\dd\bar\xi_2\,,\,
\dd\xi_1\wedge\dd\bar\xi_1- \dd\xi_2\wedge\dd\bar\xi_2 \big\rangle \ ,
\nonumber \\[4pt]
\Omega_{\real_\theta^4}^{2,-} & =
\alg\big(\real_\theta^4\big)\big\langle \dd\xi_1\wedge\dd\bar\xi_2\,,\,
\dd\xi_2\wedge\dd\bar\xi_1\,,\, \dd\xi_1\wedge\dd\bar\xi_1 +
\dd\xi_2\wedge\dd\bar\xi_2 \big\rangle \ .
\label{OmegapmR4}
\end{align}

Then the curvature $F_\nabla$ of the instanton
connection is anti-selfdual, i.e. as a two-form it obeys the anti-selfduality equation 
{(with the notational simplification $\Id \otimes *_\theta \simeq *_\theta$)}
$$
*_\theta F_\nabla=-F_\nabla \ .
$$
The proof can be found in~\cite[Prop.~5.21]{CLSII}, and it consists in showing that $F_\nabla$ is proportional to $\dd\Dcal^\dag\wedge\dd \Dcal$ as a two-form. Once the exterior derivative of the Dirac operator (\ref{Diracopexpl}) and its adjoint are computed,
using the commutation relations (\ref{R4skewcomm}) one finds
$$
\dd\Dcal^\dag\wedge \dd\Dcal = \begin{pmatrix} \
  -q^{-1}\,\big(\dd\xi_1\wedge\dd\bar\xi_1+
  \dd\xi_2\wedge\dd\bar\xi_2\big) & -\big(q+q^{-1}\big)\,
  \dd\xi_2\wedge\dd\bar \xi_1 & 0 \ \\[6pt] \big(q+q^{-1}\big)\,
  \dd\xi_1\wedge\dd\bar \xi_2 & q^3\,\big(\dd\xi_1\wedge\dd\bar
  \xi_1+\dd\xi_2\wedge\dd\bar\xi_2\big) & 0 \ \\[6pt] 0 & 0 & 0
  \ \end{pmatrix}  \ ,
$$
and comparing with (\ref{OmegapmR4}) we see that each entry of $F_\nabla$ belongs to the submodule $\Omega_{\real_\theta^4}^{2,-}$.
The anti-selfduality equation for the curvature
$F_{\widetilde\nabla}$ follows in a completely analogous way:
by using the real $\galg$-coaction
(\ref{DeltaLR4tilde}) one constructs a morphism $\tilde *_\theta:\Omega_{\widetilde\real_\theta^4}^2\to
  \Omega_{\widetilde\real_\theta^4}^2$ in the category
  ${}^{\galg_\theta}\Module$ with the same formula (\ref{Hodgeclass})
  but with affine coordinates $\xi_i$ substituted with $\zeta_i$, and
   overall changes of sign reflecting the change of
  ``orientation''.
The consistency condition for the corresponding morphisms
on the adjoinment bimodules defines the $\galg_\theta$-coequivariant
Hodge operator, and hence the instanton equations,
on the noncommutative sphere~$S_\theta^4$.

\newsection{Noncommutative gauge theory partition functions\label{gaugepartfn}}

In this section we construct smooth moduli spaces for the instanton
counting problem on the noncommutative projective plane
$\CP_\theta^2$ and natural bundles over them. We then evaluate various
equivariant K-theory partition functions for supersymmetric gauge
theories on the complex Moyal plane $\complex_\theta^2$, and compare
them with the known results in the commutative case $\theta=0$. In
particular, we extend the study of pure gauge theories considered
in~\cite{CLSII} to gauge theories with both adjoint and fundamental
matter fields. Although the technical details of the
derivations differ, our noncommutative instanton counting functions coincide with
those of the classical case, as desired.

\subsection{Instanton moduli spaces}

To construct our instanton moduli spaces, we look not for a set of
objects but rather for a space which parametrizes those objects. For
this, we consider functors from the category $\Alg$ of commutative,
unital noetherian $\complex$-algebras to the category $\Set$ of
sets. Let $A$ be a commutative, unital noetherian $\complex$-algebra. We write $\CP_{\theta,A}^2$ for the noncommutative variety dual to the algebra $\alg(\CP_{\theta,A}^2):= A\otimes\alg(\CP_\theta^2)$. 

We consider such \emph{families} of algebras $\alg(\CP_{\theta,A}^2)$ in order to study the \emph{global} structure of our moduli spaces, and endow them with geometric data.
A {family of framed sheaves parametrized by the algebra $A$} is an
$A$-flat torsion free module $\bun$ of rank $r$ on
{$A\otimes\alg(\CP_\theta^2)$} together with families of isomorphisms $H^1\big(\CP_{\theta,A}^2 \,,\, \bun(-1)\big)\cong A\otimes V$ and
$\varphi_A:\bun_\infty:= A\otimes i^\bullet(E)\to W\otimes A\otimes
\sheaf_{\CP_{\theta}^1}$. Two families of framed sheaves
$(\bun,\varphi_A)$ and $(\bun', \varphi_A')$ are
  \emph{isomorphic} if they are isomorphic as framed modules
  on $A\otimes\alg(\CP_\theta^2)$.
For $A=\complex$, i.e. for a family parametrized by a one-point space,
this definition reduces to our earlier notion of framed sheaf.
The \emph{instanton moduli functor} is the covariant functor
$
\Module_{\rm inst}(r,k):\Alg  \rightarrow
\Set
$
that associates to every algebra $A$ the set
$\Module_{\rm inst}(r,k)(A)$ of isomorphism classes of
families $(\bun,\varphi_A)$ of framed sheaves parametrized by $A$, and to every algebra morphism
$f:A\to B$ associates the pushforward $\Module_{\rm inst}(r,k)(f)$ sending families
parametrized by $A$ to families parametrized by $B$. A \emph{fine instanton moduli space} is a variety dual to a universal object
representing the instanton moduli functor, i.e. a triple $(\hat
A,\hat\bun, \hat\varphi_{\hat A})$,
where $\hat A$ is an object of the category $\Alg$ and $(\hat\bun,
\hat\varphi_{\hat A}) \in
\Module_{\rm inst}(r,k)(\hat A)$, such that for any triple
$(A,\bun,\varphi_A)$ with $A$ an object of $\Alg$ and
$(\bun, \varphi_A)\in\Module_{\rm inst}(r,k)(A)$ there exists a
unique morphism $\alpha\in \Hom_{\Alg}(\hat A,A)$ with
$\Module_{\rm inst}(r,k)(\alpha)(\hat\bun,\hat\varphi_{\hat A})\cong
(\bun,\varphi_A)$.
By Yoneda's lemma, a fine instanton moduli space given by $(\hat
A,\hat\bun,\hat\varphi_{\hat A})$ corresponds bijectively to a representation of
the instanton moduli functor via an isomorphism of functors
$
\Homfun_{\hat A} := \Hom_\Alg(\hat A,-) \rightarrow
\Module_{\rm inst}(r,k)
$,
with universal framed sheaf represented by
$(\hat\bun,\hat\varphi_{\hat A}) =\Homfun_{\hat
  A}(\Id)$.

The constructions of the previous section have the following
properties:
\begin{itemize}
\item For any $r\geq1$, $k\geq0$, the quotient variety
$\widehat\Mcal_\theta^{{\tiny \rm ADHM}}(r,k)$ is a fine instanton moduli
space for framed sheaves on $\Open(\CP_\theta^2)$ of rank $r$ and Euler characteristic
$\chi=r-k$. \\
\item For $r\geq1,k\geq0$ coprime integers, the instanton moduli
space $\Mcal_\theta(r,k)$, when non-empty, is a smooth
quasi-projective variety of dimension
$$
\dim_\complex\big(\Mcal_\theta(r,k)\big)=2\, r\,k \ .
$$

\item The tangent space to $\Mcal_\theta(r,k)$ at a point $[(E,\varphi)]$ is canonically
the vector space
$$
T_{[(E,\varphi)]}\Mcal_\theta(r,k)=\Ext^1\big(E\,,\, E(-1) \big) \ .
$$
\end{itemize}
The proof of these statements can be found in~\cite[\S6.1]{CLSII} and it makes use of the moduli space constructions of \cite{NS,DNVdB}. 
In fact, the noncommutative toric variety $\CP_\theta^2$ occurs
in the universal flat family $\alg=\alg(\CP_\theta^2)$ parametrized
by the commutative unital algebra $\alg(\complex^\times)=\complex(t)$ over $\complex$ dual to the smooth irreducible curve
$\bigwedge^2T\cong \complex^\times$. This family
includes the commutative polynomial algebra
$\alg(\CP^2):=\complex[w_1,w_2,w_3]$ (for $\theta=0$). The
moduli spaces constructed by Nevins and Stafford in~\cite{NS} all behave well in this family, and are $\complex$-schemes in the usual sense: they are constructed as geometric invariant theory quotients of subvarieties of products of grassmannians.

A much more powerful description of the local geometry of the
instanton moduli spaces is provided by the \emph{instanton deformation
  complex}. By~\cite[Thm.~6.14]{CLSII}, the K-theory class of the
tangent space $T_{[(E, \varphi)]}\Mcal_\theta(r,k)$ to the instanton moduli
space at a closed point $[(E,\varphi)]= [(B_1,B_2,I,J)]$ is isomorphic to the first
cohomology group of the complex
\beq
0\ \longrightarrow\ 
\End_\complex(V)~
\xrightarrow{ \ \dd\phi \ }~\begin{matrix}
  \End_\complex(V)^{\oplus2}
  \\[4pt] \oplus \\[4pt] \Hom_\complex(W,V)\\[4pt] \oplus \\[4pt] \Hom_\complex(V,W) \end{matrix}
~ \xrightarrow{ \ \dd\mu \ }~\End_\complex(V) \ \longrightarrow\ 0 \ ,
\label{instdefcomplex}\eeq
where the differential $\dd\mu$ is the linearization of the braided
ADHM equation (\ref{NCADHMeqs}) while the differential $\dd\phi$ 
is the linearization of the action (\ref{GLVaction}) of the gauge group $\GL(k)$ on~$\Xcal(W,V)$.

\subsection{Deformations of Hilbert schemes\label{Rank1}}

Instantons of rank one will play a prominent role in our ensuing
considerations. A torsion free sheaf $E\in\coh(\CP_\theta^2)$ has rank
one if and only
if $M=\Gamma(E)\in\gr(\alg)$ is isomorphic to a shift $\Ical(m)$ of a
right ideal $\Ical\subset \alg$~\cite[\S4.3]{CLSI}. In this case one
may identify the instanton moduli space $\Mcal_\theta(1,k)$ with an
open subset of the
\emph{commutative} deformation $\Hilb_\theta^k(\CP_\theta^2)$ for $\theta\in\complex$,
constructed in~\cite[Cor.~6.6~(1)]{NS}, of the Hilbert scheme of points
$\Hilb^k(\CP^2)$ parametrizing zero-dimensional subschemes of degree
$k$ in $\CP^2$. This is a smooth, projective fine moduli space of dimension $2k$ for torsion free 
{$\alg(\CP^2_\theta)$}-modules in
$\coh(\CP_\theta^2)$ with rank $r=1$ and
Euler characteristic $\chi=1-k$ which
also behaves well in the family $\alg=\alg(\CP^2_\theta)$, in the
sense described above. In particular, it is irreducible, hence connected, and is non-empty
for all $k\geq0$.
De~Naeghel and Van~den~Bergh~\cite{DNVdB} describe the deformation
$\Hilb_\theta^k(\CP_\theta^2)$ as the scheme parametrizing torsion free
graded $\alg$-modules $\Ical=\bigoplus_{n\geq0}\, \Ical_n$ of projective
dimension one such that
\beq
\dim_\complex(\alg_n)-\dim_\complex(\Ical_n)=k \qquad \mbox{for} \quad
n\gg0 \ .
\label{DNVdBHilb}\eeq
In particular, the module $\Ical$ corresponds to an \emph{ideal sheaf} in the
sense of~\cite[\S4.3]{CLSI}, and hence corresponds to a closed
subscheme of $\CP_\theta^2$ by~\cite[Thm.~4.10]{CLSI}. 

In this {rank one} case there is a bijection between
the set of ideals of codimension $k$ in the coordinate algebra $\alg(\complex_\theta^2)$
of the complex Moyal plane and the set of triples
$(B_1,B_2,I)\in \End_\complex(V)^{\oplus
  2}\oplus\Hom_\complex(\complex,V)$ satisfying the braided ADHM equation
\beq
B_1\, B_2-q^{-2}\ B_2\, B_1 = 0 \ , 
\label{B1B2qcomm}
\eeq
such that no proper $B_i$-invariant subspaces of $V$ contain $I(1)$ for $i=1,2$. We may regard the scheme $\Hilb_\theta^k(\complex_\theta^2)$ 
as a commutative deformation of the resolution of the $k$-th symmetric
product singularity $\Sym^k(\complex^2):=(\complex^2)^k/S_k$ provided by the Hilbert--Chow morphism
$\Hilb^k(\complex^2)\to \Sym^k(\complex^2)$, which sends an ideal to its
support. For this, we note that $\Sym_\theta^k(\complex_\theta^2):= \widehat\Mcal_\theta^{\rm
  ADHM}(0,k)$ for $\theta\in\complex$ parametrizes isomorphism classes
of $k$-dimensional irreducible representations
(\ref{B1B2qcomm}) of the complex Moyal plane $\complex_\theta^2$. In
particular, it represents the moduli functor
$\Alg\to\Set$ which sends a $\complex$-algebra $A$ to the set of
simple $A\otimes \alg(\complex_\theta^2)$-module structures on
$A^{\oplus k}$; for $A=\complex$ this set consists of isomorphism
classes of irreducible
representations of $\alg(\complex_\theta^2)$ on the vector space
$V= \complex^k$. Note that since the plane is
rigid against commutative deformations, there are no commutative deformations of the Hilbert scheme
$\Hilb^1(\complex^2)=\complex^2$, and so in the charge one case $k=1$ one has
$$
\Hilb_\theta^1\big(\complex_\theta^2\big)\ \cong \ \complex^2
$$
as a complex variety for all $\theta\in\complex$, in agreement with (\ref{Mcalthetar1}).

\subsection{Equivariant K-theory partition functions\label{subsec:Tfixedpoints}} 

In the classical case, instanton partition functions of topologically twisted supersymmetric
gauge theories are given by integrating suitable equivariant
characteristic classes over the instanton moduli
spaces~\cite{nekrasov}; in the following we will treat all of our
partition functions within the framework of K-theory, as this is more
natural from the point of view of the functorial quantization taken in
this paper. The
equivariant partition function is then the generating function for the K-theory
integrals
$$
Z_{\rm inst}(r;Q, z)=\sum_{k=0}^\infty\, Q^k\
\int_{\Mcal_\theta(r,k)^{\widetilde T}}\, \omega(z) \ ,
$$
where $Q$ is a formal variable and $\omega(z)$ is an equivariant K-theory class depending
on the canonical generators $z$ of the coordinate algebra of the torus
$\widetilde T=T \times(\complex^\times)^{r-1}$ for $r\geq1$, with
$T= (\complex^\times)^2$ the ``geometrical'' torus used for the
deformation of $\CP^2$ and of the instanton moduli. 
The integral is evaluated formally by applying the localization
theorem in equivariant K-theory, hence
$\int_{\Mcal_\theta(r,k)^{\widetilde T}}\, \omega(z)$ is a rational
function in the coordinate algebra $\alg(\widetilde T\, )$, i.e. an
element of the quotient field of the representation ring of $\widetilde{T}$.
A combinatorial
formula for the instanton counting functions can be computed
explicitly by classifying the $\widetilde T$-fixed points in the instanton
moduli space $\Mcal_\theta(r,k)$, and computing the Euler
characteristic classes of the equivariant normal bundles to the
fixed loci {of the torus action}.

We first describe the
natural action of the torus $\widetilde T$ and its fixed points on the
instanton moduli space.
We denote by $z= (t_1,t_2,\rho_1,\dots,\rho_{r})$ the canonical generators of the coordinate algebra of 
$\widetilde T$, where we identify $(\complex^\times)^{r-1}$ with the maximal torus of ${\rm SL}(r)$ 
given as the hypersurface
$
\rho_1\cdots \rho_r=1
$
in $(\complex^\times)^r$. Using this presentation of  
$(\complex^\times)^{r-1}$, we denote its characters by
$m=(m_1,\dots,m_r)\in\zed^r$. For any $[(E,\varphi)]\in\Mcal_\theta(r,k)$, 
a natural coaction of the Hopf algebra
$\hil^{(r)}:=\complex(\rho_1,\dots,\rho_r)$ on the framing module
$E_\infty \cong \big( \sheaf_{\CP_\theta^1} \big)^{\oplus r}$ is defined by
$$
\Delta_{E_\infty}^{(r)}\, :\, E_\infty \ \longrightarrow \ \hil^{(r)}\otimes E_\infty
 \ , \qquad \Delta_{E_\infty}^{(r)}(f)= \sum_{l=1}^r\, \rho_l\otimes f_l
$$
for $f=(f_1,\dots,f_r)\in \big(\sheaf_{\CP_\theta^1}\big)^{\oplus r}$. 
The coaction of the Hopf algebra $\hil_\theta$ on the moduli space
$\widehat\Mcal_\theta^{{\tiny \rm ADHM}}(r,k)$ is given by
(\ref{DeltaLBIJ}). To describe the coaction of $\hil^{(r)}$, we represent
the linear maps $I$ and $J$ by complex matrices $I=(I_{a,l})$ and
$J=(J_{l,a})$ of sizes $k\times r$ and $r\times k$, respectively. Then the left $\hil^{(r)}$-coaction on the noncommutative ADHM data $(B_1,B_2,I,J)$ is given by
\beq
\Delta_{\Xcal(W,V)}^{(r)}(B_1,B_2,I,J) = \Big(1\otimes B_1\,,\,
1\otimes B_2 \,,\, \big(\rho_l^{-1}\otimes I_{a,l} \big) \,,\, \big(
\rho_l\otimes J_{l,a} \big) \Big) \ .
\label{Deltainstr}\eeq
This makes the vector spaces $V,W$ objects and
$(B_1,B_2,I,J)$ morphisms in the category ${}^{\hil^{(r)}}\Module$.
By construction, the isomorphism $\Mcal_\theta(r,k)\rightarrow \widehat\Mcal_\theta^{{\tiny \rm ADHM}}(r,k)$ is $\widetilde T$-coequivariant.

A fixed point $[(E,\varphi)]\in\Mcal_\theta(r,k)^{\widetilde T}$ is an isomorphism
class of a framed coequivariant sheaf $(E,\varphi)$ (see~\cite[\S 2.5]{CLSII})
which is equiped with a coaction of the Hopf algebra $\widetilde \hil_\theta:=\hil_\theta\otimes\hil^{(r)}$; hence $E$ decomposes into a finite direct sum of torsion free {$\alg(\CP_\theta^2)$}-modules graded by the character lattice of the torus $\widetilde T$ as
\beq
E=\bigoplus_{p\in \zed^2}\ \bigoplus_{m\in\zed^r}\, E(p,m) \ .
\label{Edecomp}\eeq
The left $\widetilde\hil_\theta$-coactions $\widetilde\Delta_{E(p,m)} :E(p,m)\rightarrow\widetilde\hil_\theta\otimes
E(p,m)$
are given for $f\in E(p,m)$ by
$\widetilde\Delta_{E(p,m)}(f)=t^p\otimes\rho^m\otimes f$. The set of
$\widetilde{T}$-fixed points in $\Mcal_{\theta}(r,k)$, i.e. the $\widetilde\hil_\theta$-coinvariants, admits a
characterization in terms of Young diagrams. This follows from writing
analogous character decompositions for the $\widetilde \hil_\theta
$-comodules $V$ and $W$, and using the braided ADHM equation
(\ref{NCADHMeqs}) together with the stability condition to find a
combinatorial characterization of the finite lattice of characters
$p=(p_1,p_2)\in\nat^2$ defined by the non-trivial isotopical
components $V(p,m)\subset V$ and the non-trivial components of the
linear maps $B_i(p,m)$ for $i=1,2$. From the discussion in
\S\ref{Rank1},  it follows that fixed
points $[(B_1,B_2,I,J)] \in\widehat\Mcal_\theta^{{\tiny \rm ADHM}}(r,k)^{\widetilde T}$ correspond
bijectively to ideals of codimension $k$ in the
affine coordinate algebra $\alg(\complex_\theta^2)$. In particular, the
character decomposition (\ref{Edecomp}) truncates to
$$
E=\bigoplus_{l=1}^r\, \Ical_l \ ,
$$
where $\Ical_l$ for each
$l=1,\dots,r$ is an $\hil_\theta$-coequivariant torsion free sheaf of
rank one on $\Open(\CP_\theta^2)$. In this way it is
proven in~\cite[Prop.~7.8]{CLSII} that  $\Mcal_\theta(r,k)^{\widetilde
  T}$ is a finite set in bijective correspondence with length $r$
sequences $\vec\lambda=(\lambda^1,\dots,\lambda^r)$ of Young diagrams
$\lambda^l$ of size $|\,\vec\lambda\, |:= \sum_l \, |\lambda^l|=k$,
where $|\lambda^l|$ is the total number of points in $\lambda^l$. This
fixed point classification coincides with that of the classical case
$\theta=0$~\cite[Prop.~2.9]{NakYosh}. It can be understood in terms of the noncommutative toric geometry as follows. By~\cite[Prop.~4.15]{CLSI}, $\hil_\theta$-coequivariant ideal sheaves on $\Open(\CP_\theta^2)$ are in bijective correspondence with
$\zed^2$-graded subschemes of $\CP_\theta^2$, with $\zed^2$ the
character lattice of $T=(\complex^\times)^2$; in particular,
irreducible subschemes correspond to prime ideals in the spectrum of
the homogeneous coordinate algebra {$\alg(\CP_\theta^2)$}. Moreover, the
$\hil_\theta$-coinvariant ideals $\Ical \subset {\alg(\CP_\theta^2)}$ are
\emph{monomial} ideals. If $\Ical$ obeys the condition
(\ref{DNVdBHilb}), then it determines a finite partition
$\lambda_{\Ical}(k)$ of $k$ by considering lattice points which correspond to monomials \emph{not} contained in $\Ical$.

The equivariant characters of the $\widetilde\hil_\theta$-comodules
$V=V_{\vec\lambda}$ and $W=W_{\vec\lambda}$ corresponding to a fixed point parametrized by $\vec\lambda$ are given by the elements
\beq
\ch_{\widetilde T}(V_{\vec\lambda})=\sum_{l=1}^r\
\sum_{p\in\lambda^l}\,\rho_l\ t_1^{1-p_1}\, t_2^{1-p_2} \ , \qquad
\ch_{\widetilde T}(W_{\vec\lambda})=\sum_{l=1}^r\, \rho_l
\label{VWchars}\eeq
in the representation ring of $\widetilde T$, as in the classical case~\cite{NakYosh}. The restriction
of the instanton deformation complex (\ref{instdefcomplex}) to a $\widetilde T$-fixed point $\vec\lambda$ is a complex in the category ${}^{\widetilde\hil_\theta}\Module$. The computation of the $\widetilde T$-equivariant character of the tangent bundle over the instanton moduli space thus proceeds exactly as in the classical case
(see~\cite{FlumePog} and~\cite[Thm.~2.11]{NakYosh}). At a fixed point
parametrized by a length $r$ sequence
$\vec\lambda=(\lambda^1,\dots,\lambda^r)$ of Young diagrams with
$|\,\vec\lambda\,|=k$, one has, in the representation ring of
$\widetilde T$, the character
\bea
\ch_{\widetilde T}\big(T_{\vec\lambda}\Mcal_\theta(r,k) \big) =
\sum_{l,l'=1}^r\, \rho_l^{-1}\, \rho_{l'}\, \Big(\,
\sum_{p\in\lambda^l}\, t_1^{p_1-(\lambda^{l'})^t_{p_2}}\,
t_2^{\lambda_{p_1}^l-p_2+1} +
\sum_{p'\in\lambda^{l'}}\, t_1^{(\lambda^l)_{p_2'}^t-p_1'+1}\,
t_2^{p_2'-\lambda_{p_1'}^{l'}}\, \Big) \nonumber \\
\label{tildeTchar}\eea
where $\lambda_i$ and $\lambda_j^t$ denote respectively the number of
points in the $i$-th column and $j$-th row of
the Young diagram $\lambda$. In particular, the corresponding top form
is the
equivariant Euler class of the normal bundle to the fixed point; it is
given by
\bea
\mbox{$\bigwedge_{-1}$}T_{\vec\lambda}^*\Mcal_\theta(r,k) &=& 
\prod_{l,l'=1}^r\ \prod_{p\in\lambda^l}\, \Big( 1-\rho_l\, \rho_{l'}^{-1}\, t_1^{(\lambda^{l'})^t_{p_2}-p_1}\,
t_2^{p_2-\lambda_{p_1}^l-1} \Big) \nonumber\\ && \qquad \qquad  \times\
\prod_{p'\in\lambda^{l'}}\, \Big( 1-\rho_l\, \rho_{l'}^{-1}\, t_1^{p_1'-(\lambda^l)_{p_2'}^t-1}\,
t_2^{\lambda_{p_1'}^{l'}-p_2'} \Big) \ . \nonumber
\eea

With these ingredients one can now write down equivariant
instanton partition functions for noncommutative supersymmetric gauge theories. The equivariant K-theory class $\omega(z)$ is, by
the localization theorem, given by the pullback of a class
$\tilde\omega$ on $\Mcal_\theta(r,k)$ which descends from a $\widetilde{T}$-equivariant class, evaluated at the fixed point
$\vec\lambda$, and divided by the Euler character
$\bigwedge_{-1}T_{\vec\lambda}^*\Mcal_\theta(r,k)$. The generic form
of the equivariant K-theory partition function is thus
\beq
Z_{\rm inst}(r;Q, z)=\sum_{k=0}^\infty\, Q^k\ \sum_{\vec\lambda\, :\,
  |\,\vec\lambda\,|=k} \ \frac{\tilde\omega(\vec\lambda\,)}{\mbox{$\bigwedge_{-1}$}
    T_{\vec\lambda}^*\Mcal_\theta(r,k)} \ .
\label{Zinstgen}\eeq
In the following we look at some explicit instances and extensions of
this K-theory counting function.

\subsection{$\Ncal=4$ gauge theory}

The simplest example is obtained by taking $\tilde\omega$ to be
the Euler class of the tangent bundle over $\Mcal_\theta(r,k)$,
i.e. the class of the virtual bundle
$$
\tilde\omega = \sum_{j=0}^{2\, r\,k}\, (-1)^j \ \mbox{$\bigwedge^j$}
    T^*\Mcal_\theta(r,k) \ .
$$
In
this case $\omega(z)=1$ (independently of the equivariant parameters~$z$) and the localization
integral simply counts the fixed points of the $\widetilde T$-action on
the instanton moduli spaces. Then (\ref{Zinstgen}) results in the Vafa--Witten partition function~\cite{VafaWitten}
for $\Ncal=4$ gauge theory given by
\beq
Z_{\rm inst}^{\Ncal=4}(r;Q)=\sum_{\vec\lambda}\, Q^{|\,\vec\lambda\,|}
= \prod_{n=1}^\infty\, \frac1{\big(1-Q^n\big)^r} \ ,
\label{VWpartfn}\eeq
which computes the Euler characteristic of the instanton moduli space.

\subsection{$\Ncal=2$ gauge theory}

For
$\tilde\omega=1$, the expression (\ref{Zinstgen}) reproduces the K-theory version of Nekrasov's partition
function~\cite{nekrasov} for pure $\Ncal=2$ gauge theory given by
\beq
Z_{\rm inst}^{\Ncal =2}(r;Q,z)= \sum_{k=0}^\infty\
\sum_{\vec\lambda\, :\, |\,\vec\lambda\,|=k}\
\frac{Q^{|\,\vec\lambda\,|}}{\mbox{$\bigwedge_{-1}$}
    T_{\vec\lambda}^*\Mcal_\theta(r,k)} \ .
\label{Nekpartfn}\eeq

\subsection{Adjoint matter fields\label{adjmatter}}

By twisting $T\Mcal_\theta(r,k)$ with a torus character $m$, the Euler
class computes
the Chern polynomial of the tangent bundle; the resulting
noncommutative gauge theory
is called the $\Ncal=2^*$ theory and it
is a deformation of the $\Ncal=4$ theory obtained by adding a section
of the tangent bundle $T\Mcal_\theta(r,k)$, i.e. an adjoint matter
field, of mass $m$. We set $\tilde\omega=c(T\Mcal_\theta(r,k),m)$,
where the
Chern polynomial of a vector bundle $\Vcal\to\Mcal_\theta(r,k)$ of
rank $n$ is defined by the formula
$$
c(\Vcal,u)= \sum_{i=0}^n\, c_i(\Vcal)\ u^{n-i}
$$
with $c_i(\Vcal)$ the $i$-th Chern class of $\Vcal$.
This class descends from the equivariant class
$$
c_{\widetilde{T}}\big(T\Mcal_\theta(r,k)\,,\,m \big) = \mbox{$\bigwedge_{-1}$}\big(
    T\Mcal_\theta(r,k)\otimes M\big)_{\widetilde{T}\times T_m} \ ,
$$
where $M=\complex$ is the natural irreducible module over the complex torus
$T_m=\complex^\times$ parametrized by $\mu:= \e^{-m}$; the torus
$\widetilde{T}$ (resp. $T_m$) acts trivially on $M$ (resp. $T\Mcal_\theta(r,k)$). The
$\widetilde{T}\times T_m$-equivariant character at a fixed point
parametrized by $\vec\lambda$ is a simple
modification of (\ref{tildeTchar}) given by
$$
\ch_{\widetilde{T}\times T_m}\big(T_{\vec\lambda}\Mcal_\theta(r,k)\otimes M \big)
= \mu\ \ch_{\widetilde T}\big(T_{\vec\lambda}\Mcal_\theta(r,k)\big) \
,
$$
and hence the noncommutative $\Ncal=2^*$ gauge theory partition function obtained
from (\ref{Zinstgen}) is
$$
Z_{\rm inst}^{\Ncal=2^*}(r;Q, z; \mu) = \sum_{k=0}^\infty\, Q^k\ \sum_{\vec\lambda\, :\,
  |\,\vec\lambda\,|=k} \ \frac{\mbox{$\bigwedge_{-1}$} \big(
    T_{\vec\lambda}^*\Mcal_\theta(r,k)\otimes M\big)}{\mbox{$\bigwedge_{-1}$}
    T_{\vec\lambda}^*\Mcal_\theta(r,k)}
$$
where
\bea
\mbox{$\bigwedge_{-1}$} \big(
    T_{\vec\lambda}^*\Mcal_\theta(r,k)\otimes M\big) &=&
    \sum_{j=0}^{2\, r\,k}\, (-1)^j\, \mu^j\ \ch_{\widetilde{T}}\big(\mbox{$\bigwedge^j$}
    T^*\Mcal_\theta(r,k)\big) \nonumber\\[4pt]
&=& 
\prod_{l,l'=1}^r\ \prod_{p\in\lambda^l}\, \Big( 1-\mu\, \rho_l\, \rho_{l'}^{-1}\, t_1^{(\lambda^{l'})^t_{p_2}-p_1}\,
t_2^{p_2-\lambda_{p_1}^l-1} \Big) \nonumber\\ && \qquad \qquad  \times\
\prod_{p'\in\lambda^{l'}}\, \Big( 1-\mu\, \rho_l\, \rho_{l'}^{-1}\, t_1^{p_1'-(\lambda^l)_{p_2'}^t-1}\,
t_2^{\lambda_{p_1'}^{l'}-p_2'} \Big) \ . \nonumber
\eea
This partition function can be regarded as a ``homotopy'' between (\ref{VWpartfn}) and
(\ref{Nekpartfn}) in the sense that it has the limits
$$
Z_{\rm inst}^{\Ncal=2^*}(r;Q, z; \mu=0) = Z_{\rm inst}^{\Ncal
  =2}(r;Q,z)
 \ , \qquad 
Z_{\rm inst}^{\Ncal=2^*}(r;Q, z; \mu=1) = Z_{\rm inst}^{\Ncal=4}(r;Q) \ .
$$

\subsection{Dirac bundles and fundamental matter fields\label{fundmatter}}

Because of the existence of a universal framed module
$(\hat\bun,\hat\varphi_{\hat A})$ over
$\alg(\Mcal_\theta(r,k))\otimes \alg(\CP_\theta^2)$, we can consider the sheaf $\bun$ on
$\Mcal_\theta(r,k)\times\Mcal_\theta(r,k'\,)$ with fibre over a pair
of framed torsion free sheaves $((E,\varphi),(E',\varphi'\, ))$ on $\Open(\CP_\theta^2)$ given by
\beq
\bun\big|_{([(E,\varphi)],[(E',\varphi'\,)])} = \Ext^1\big(E'\,,\,E(-1)\big) \ .
\label{bunExtfibre}\eeq
Since the cohomological dimension of the category $\coh(\CP_\theta^2)$
is~two, by Serre duality it follows that
$\Ext^p(E',E(-1))=0$ for all $p>2$~\cite[\S6.2]{CLSI}; this is consistent with the fact
that the Koszul algebra $\alg(\complex_\theta^2)$ has projective
dimension~two, so that its Hochschild cohomology
$HH^p(\alg(\complex_\theta^2),M)$ vanishes in degrees $p>2$ for arbitrary
$\theta\in\complex$ and any $\alg(\complex_\theta^2)$-bimodule
$M$. Moreover, 
it is straightforward to modify the proofs of~\cite[Lem.~7.14]{NS}
and~\cite[Thm.~6.4]{CLSII} (which consider the limiting case $k=k'$,
$E=E'$) to
show that $\Hom(E',E(-1))=\Hom(E',E(-k))=0$ for $k>0$ due to the
framing conditions, and that $\Ext^2(E',E(-1))=0$ for the same reason
preceded by (twisted) Serre duality; for $E'=\sheaf_{\CP_\theta^2}$
these vanishing results are the content of~\cite[Prop.~4.7]{CLSII}. This implies that the vector
spaces (\ref{bunExtfibre}) all have the same dimension as $([(E,\varphi)],[(E',\varphi'\,)])$
varies over $\Mcal_\theta(r,k)\times\Mcal_\theta(r,k'\,)$. It follows that $\bun$ is a
vector bundle over $\Mcal_\theta(r,k)\times\Mcal_\theta(r,k'\,)$ whose
rank may be computed by using~\cite[Cor.~6.2]{NS} to get
\bea
\dim_\complex\Ext^1\big(E'\,,\, E(-1)\big) = \rank(E'\,)\,
\big[\rank\big(E(-1)\big) -\chi\big(E(-1)\big)\big] 
- \chi(E'\,)\, \rank\big(E(-1)\big) \nonumber
\eea
with $\chi(E(-1))=-k$ (see~\cite[Thm.~6.4]{CLSII}); whence the rank of $\bun$ is
$$
\rank_\complex(\bun) = r\, (k+k'\,) \ .
$$
These are precisely the same arguments which show
that the instanton moduli space $\Mcal_\theta(r,k)$ is smooth of
dimension $2\,r\,k$.

By suitably twisting $\bun$ with torus characters, we can 
compute characteristic classes associated with bifundamental matter
fields and use them to build noncommutative linear $\Ncal=2$ quiver gauge theory
partition functions. In the limiting case
$k=k'$, there are two particular restrictions of the vector
bundle $\bun$
that we are interested in. Firstly, the restriction of $\bun$ to the
diagonal subspace of $\Mcal_\theta(r,k)\times\Mcal_\theta(r,k)$ is the
tangent bundle $T\Mcal_\theta(r,k)$ of rank $2\,r\,k$ whose sections
were studied above. Secondly, the
push-forward of $\bun$ by the projection of
$\Mcal_\theta(r,k)\times\Mcal_\theta(r,k)$ onto the first factor is a
vector bundle $\Vcal\to\Mcal_\theta(r,k)$ of rank $k$ called the
\emph{Dirac bundle}; its fibre over a framed torsion free sheaf $(E,\varphi)$ is
$H^1(\CP_\theta^2,E(-1))\cong V$. Sections of this bundle, suitably
twisted with torus characters, are called
fundamental matter fields; we shall now evaluate the corresponding
noncommutative gauge theory
partition functions~(\ref{Zinstgen}).

For this, we fix a positive integer $n_f$ and let $T_M\cong(\complex^\times)^{n_f}$ be the
maximal torus of the flavour group $\GL(M)\cong\GL(n_f)$ acting on its
fundamental representation $M=\complex^{n_f}$, with parametrization
$\mu_a:=\e^{-m_a}$ for $a=1,\dots,n_f$. The Euler class
$\bigwedge_{-1}(\Vcal\otimes M)$ integrates over $\Mcal_\theta(r,k)$
to a (virtual) class of degree $(2r-n_f)\,k$. The
$\Ncal=2$ gauge
theory with fundamental matter fields is said to be ``conformal'' when
$n_f=2r$ and ``asymptotically free'' when $n_f<2r$; henceforth we
assume that $n_f\leq2r$ in order to avoid the use of virtual
classes and virtual localization techniques. We take $\tilde\omega$ to descend from the equivariant class
$$
\mbox{$\bigwedge_{-1}$}(\Vcal\otimes M)_{\widetilde{T}\times T_M} =
\prod_{a=1}^{n_f}\, c(\Vcal,m_a) \ ,
$$
where the torus $\widetilde{T}$ (resp. $T_M$) acts trivially on $M$
(resp. $\Vcal$). The $\widetilde{T}\times T_M$-equivariant character
at a fixed point parametrized by $\vec\lambda$ is a simple
modification of the $\widetilde{T}$-equivariant character
$\ch_{\widetilde{T}}(V_{\vec\lambda})$ from (\ref{VWchars}) given by
$$
\ch_{\widetilde{T}\times T_M}(\Vcal_{\vec\lambda}\otimes
M)= \sum_{a=1}^{n_f}\, \mu_a \ \ch_{\widetilde{T}}(V_{\vec\lambda}) \ ,
$$
and whence the noncommutative $\Ncal=2$ gauge theory partition function
(\ref{Zinstgen}) with $n_f\leq2r$ fundamental matter hypermultiplets is
given by
$$
Z_{\rm inst}^{\Ncal=2,n_f}(r;Q, z; \mu) = \sum_{k=0}^\infty\, Q^k\ \sum_{\vec\lambda\, :\,
  |\,\vec\lambda\,|=k} \ \frac{\mbox{$\bigwedge_{-1}$}
  \big(\Vcal_{\vec\lambda}^* \otimes M\big)}{\mbox{$\bigwedge_{-1}$}
    T_{\vec\lambda}^*\Mcal_\theta(r,k)}
$$
where
\bea
\mbox{$\bigwedge_{-1}$} \big(\Vcal_{\vec\lambda}^*\otimes M\big) =
\prod_{a=1}^{n_f} \ \prod_{l=1}^r \ \prod_{p\in\lambda^l}\, \Big(
1-\mu_a \, \rho_l^{-1}\, t_1^{p_1-1}\,
t_2^{p_2-1} \Big) \ . \nonumber
\eea
For $\mu_a=0$ this reduces to the partition function for
the pure $\Ncal=2$ gauge theory (\ref{Nekpartfn}).

\newsection{Noncommutative vortices}

One of the most tantalizing extensions of our constructions would be
to formulate the counting problem for generalized instantons on
noncommutative toric threefolds, and to investigate its connection with a
noncommutative version of Donaldson--Thomas theory. However,
at present such a higher-dimensional extension is not understood, and
we anticipate that much more sophisticated moduli space techniques,
such as perfect obstruction theory, will be required; see~\cite{CS}
for a detailed comparison in this context between the instanton counting problems in
four and six dimensional topologically twisted supersymmetric gauge theories. On the
other hand, non-abelian vortices play the roles of
instantons in two dimensions. We will construct
noncommutative vortices within our framework via certain \emph{reductions} of our instanton
moduli constructions, inspired by the works~\cite{HT,BTZ} which show
via D-brane constructions that
the moduli spaces of non-abelian vortices in two-dimensional
$\Ncal=(2,2)$ supersymmetric gauge theories, with adjoint and fundamental matter
fields, can be described as holomorphic submanifolds of the moduli
spaces of instantons in four dimensions. We shall also compute the corresponding
noncommutative vortex partition functions via localization in
equivariant K-theory.

\subsection{Vortices on the projective line}

Let $\hil_\zeta=\complex(\zeta)$ be the Hopf algebra associated to a
complex torus $T_\zeta=\complex^\times$ which {we make to} 
coact on the homogeneous coordinate algebra of the noncommutative projective plane
$\alg(\CP_\theta^2)$ through
$$
\Delta_{\zeta}(w_1) = 1\otimes w_1 \ , \qquad
\Delta_{\zeta}(w_2) = \zeta\otimes w_2 \ , \qquad
\Delta_{\zeta}(w_3) = 1\otimes w_3 \ .
$$
{The subalgebra of coinvariants for this $\hil_\zeta$-coaction, generated by $w_1$ and $w_3$, is the algebra
$\alg(\CP^1)$ which is dual to a commutative projective line $\CP^1$.}
Note that such
a reduction to a commutative variety is unavoidable, as generic
deformations of holomorphic curves are always commutative (see~\cite[\S4.1]{CLSII}). 
As a toric variety, in this reduction we restrict to 
{the real subalgebra $\galg_t$ of the Hopf algebra $\hil$ with $t:=t_1$ and $t_2=t^*$ (as described after 
\eqref{stargrass-bis})}, so that the induced coaction of
$\galg_t$ on $\alg(\CP^1)$ is given by
$$
\Delta_t(w_1)= t\otimes w_1 \ , \qquad \Delta_t(w_3) = 1\otimes w_3 \ .
$$
{The need to restrict the toric symmetry to $\galg_t$ will be made
  apparent below when we deal with 
the real noncommutative plane.} 

We study the moduli of sheaves over $\alg(\CP^1)$ which are
pullbacks of framed torsion free sheaves on $\CP_\theta^2$ by the algebra 
{inclusion $\alg(\CP^1) \hookrightarrow \alg(\CP_\theta^2)$}, which induces a functor
$\coh(\CP_\theta^2)\to \coh(\CP^1)$.
The $\hil_\zeta$-coaction on $\alg(\CP_\theta^2)$ has a natural lift
to the instanton moduli space $\Mcal_\theta(r,k)$,
defined on the braided ADHM data (\ref{BIJsubvar}) by the coaction
\beq
\Delta_\zeta(B_1,B_2,I,J) = \big(1\otimes B_1\,,\,\zeta\otimes B_2
\,,\,1\otimes I \,,\,
\zeta\otimes J \big) \ .
\label{zetacoactBIJ}\eeq
The subvariety of coinvariants consists of quadruples $(B_1=B,0,I,0)$
and is identified as
$$
\Xcal_{\rm vort}(W,V) = \End_\complex(V) \ \oplus \ \Hom_\complex(W,V)
\ .
$$
The coaction (\ref{zetacoactBIJ}) preserves the noncommutative complex
ADHM equation (\ref{NCADHMeqs}) and the action of the gauge group
(\ref{GLVaction}). The pullback of the equation
(\ref{NCADHMeqs}) to the subvariety of coinvariants is a trivial identity: $0=0$. The $\GL(k)$-orbits of the data
$(B,I)$ define the \emph{vortex moduli space}
$$
\Mcal_{\rm vort}(r,k) := \widehat\Mcal_\theta^{\rm
  ADHM}(r,k)^{T_\zeta} = \Xcal_{\rm vort}(W,V)\,
\big/\,\GL(k) \ ,
$$
where the quotient is defined by restricting to \emph{stable} pairs
$(B,I)$ such that there are no proper $B$-invariant subspaces of $V$
which contain the image of $I$; this restriction removes the pairs at
which the action of the gauge group
$$
g\triangleright(B,I)=\big(g\, B\, g^{-1}\,,\, g\, I \big) \ , \qquad
g\in\GL(k)
$$
is not free. As shown
by~\cite{HT,EINOS,BTZ}, this subvariety of
the instanton moduli space $\Mcal_\theta(r,k)=
\widehat\Mcal_\theta^{\rm ADHM}(r,k)$ is a complex manifold. The tangent space
$T_{[(B,I)]}\Mcal_{\rm vort}(r,k)$ is described in K-theory by the
cohomology of the corresponding pullback of the instanton deformation
complex (\ref{instdefcomplex}) which is given by
\beq\label{vortdefcomplex}
0\ \longrightarrow\ 
\End_\complex(V)~
\xrightarrow{ \ \dd\phi \ }~\begin{matrix}
  \End_\complex(V)
  \\[4pt] \oplus \\[4pt] \Hom_\complex(W,V) \end{matrix}
\ \longrightarrow\ 0 \ .
\eeq
In particular, the moduli space of vortices has dimension
$$
\dim_\complex\big(\Mcal_{\rm vort}(r,k)\big)= r\,k \ .
$$
In this way the vortices on $\CP^1$ correspond to instantons on
$\CP_\theta^2$ which are invariant under the action of the complex
torus $T_\zeta$. Moreover, the $\galg_t$-coaction on $\alg(\CP^1)$ also
naturally lifts to the vortex moduli space to give the coaction
\beq
\Delta_t(B,I) = \big(t\otimes B\,,\,1\otimes I \big) \ .
\label{DeltatBI}\eeq

For example, in the rank one case $r=1$ the vortex moduli space $\Mcal_{\rm vort}(1,k)$ can be identified with the Hilbert
scheme $\Hilb^k(\complex)$ which is the $k$-th symmetric product orbifold
$\Sym^k(\complex)=\complex^k/S_k = \complex^k$. This space is non-singular and one has
an isomorphism
$$
\Mcal_{\rm vort}(1,k) \ \cong \ \complex^k
$$
of smooth complex varieties. 

On the other hand, for vorticity $k=1$ we have $B=b\in\complex$ and $I=(i_1,\dots,i_r)\in W^*$,
with the stability condition implying $i_l\neq0$ 
{for at least one index $l=1,\dots,r$}. The action of the gauge group $\GL(1)=\complex^\times$
{leaves $b$ alone and}
can be used to rescale $i_r=1$, describing a patch
{$\complex \times \complex^{r-1}$} of the vortex moduli space, which is therefore given globally by
\beq
\Mcal_{\rm vort}(r,1) \ \cong \ {\complex \times \CP^{r-1} }
\label{Mvortr1}\eeq
as a smooth complex variety. For $r,k>1$, the moduli space $\Mcal_{\rm vort}(r,k)$ is a resolution of the symmetric product
$\Sym^k(\complex \times \CP^{r-1})$. 

\subsection{Vortices on the noncommutative plane}

Despite their origin within the moduli space of noncommutative
instantons on $\CP_\theta^2$, at the level of holomorphic varieties
the vortex moduli space essentially coincides with its classical
limit; the moduli space $\Mcal_{\rm vort}(r,k)$ is constructed
directly in~\cite{EINOS} from a field theory analysis.
What differs is its geometry as an embedded submanifold of the instanton moduli space. In
the classical case it can be regarded
as a symplectic quotient with respect to the natural (hyper-K\"ahler) symplectic
structure on the instanton moduli space~\cite{HT,BTZ}, while
in the noncommutative case the instanton moduli space
$\Mcal_\theta(r,k)$ is not even a symplectic manifold (at least in the
sense of standard hamiltonian reduction);
see~\cite[\S6.6]{CLSII} for a braided version of symplectic reduction
in this context
which may illuminate how the noncommutative deformation affects the
induced geometry of the smooth variety $\Mcal_{\rm
  vort}(r,k)$. Alternatively, we may explore
the effects of the quantization by examining the reductions of the instanton
moduli on $S_\theta^4$, which requires introducing suitable
real structures. The situation is reminescent of the manner in which
dimensional reductions of invariant instantons over the quantum two-sphere yield commutative
$q$-deformations of the usual non-abelian vortices~\cite{LS}.

The coaction of the Hopf algebra $\hil_\zeta$ above has a natural extension
to the algebra {$\alg(\real_\theta^4)$} from \S\ref{NCS4} defined by
$$
\Delta_\zeta(\xi_1)=1\otimes\xi_1 \ , \qquad \Delta_\zeta(\xi_2)=\zeta
\otimes\xi_2 \ , \qquad \Delta_\zeta(\bar\xi_1)=1\otimes \bar\xi_1 \ ,
\qquad 
{\Delta_\zeta(\bar\xi_2)=\zeta^* \otimes\bar\xi_2}
$$
for a suitable $*$-structure on $\hil_\zeta$ which is compatible with
the coaction on $\alg(\real_\theta^4)$ below defining the desired coinvariants.
The subalgebra of coinvariants
$
\alg(\real_\theta^2)= \alg(\real_\theta^4)^{\hil_\zeta}
$
is the algebra dual to a noncommutative real plane $\real_\theta^2$, 
generated by elements $\xi,\bar\xi$ with commutation relations
$$
\xi\,\bar\xi=q^2 \ \bar\xi \, \xi \ , 
$$
and $*$-structure from \eqref{starxi}, 
$$
\xi^\dag = q^{-1} \ \bar\xi \ .
$$
The induced toric coaction $\Delta_t:\alg(\real_\theta^2)\to \galg_t\otimes \alg(\real_\theta^2)$
is given by 
$$
\Delta_t(\xi)= t\otimes\xi \ , \qquad \Delta_t(\xi^\dag\, ) = t^* \otimes \xi^\dag \ , 
$$
for the real limit $t=t_1$ and $t_2=t^*$. 
The algebra inclusion
 $\alg(\real_\theta^2) \hookrightarrow \alg(\real_\theta^4)$ induces a functor
$\module\big(\alg(\real_\theta^4) \big) \to \module\big(
\alg(\real_\theta^2) \big)$, giving a map from
bundles on $\real_\theta^4$ to bundles on~$\real_\theta^2$.
The $\hil_\zeta$-coaction lifts to the real  ADHM data by \eqref{zetacoactBIJ}, together with
$$
\Delta_\zeta\big(B_1^\dag\,,\,B_2^\dag\,,\,I^\dag\,,\,J^\dag \big) = \big(1\otimes B_1^\dag\,,\, 
\zeta^* \otimes B_2^\dag \,,\,1\otimes I^\dag \,,\,
\zeta^* \otimes J^\dag \big) \ ,
$$
and the $\hil_\zeta$-coaction does not preserves the braided real ADHM equation
(\ref{realADHMeq}) unless we restrict to the coinvariants for this coaction.
The subvariety $\Mcal_{\rm
  vort}^\real(r,k)_\theta \subset \Mcal_{\rm vort}(r,k)$
consists of pairs $(B,I)$ which additionally
satisfy the corresponding pullback of
(\ref{realADHMeq}) given by
\beq\label{Mvortrealdef}
\big[B\,,\,B^\dag\,\big]_{-\theta} +I\, I^\dag =0
\eeq
in $\End_\complex(V)$, with $q\in\real$ and together with the pullback of the natural
$\GL(k)$-action on $\Mcal_{\rm vort}(r,k)$ to an action of the unitary gauge
group $\U(k)$. We identify $\Mcal_{\rm
  vort}^\real(r,k)_\theta$ as a moduli space of (local) vortices on the
noncommutative plane $\real_\theta^2$. It can be regarded as the phase
space of an abstract quantum integrable system; for $r=1$ this
integrable system is a
$q$-deformation of the Calogero--Moser model.

As an example, consider the vorticity one case. The real ADHM equation (\ref{Mvortrealdef}) in this instance reads
$$
\big(1-q^{-2}\big) \ b\,\bar b+\sum_{l=1}^r\, i_l\, \bar i_l=0 \ ,
$$
which has solutions only when
$q\in(0,1)$; in this case the real vortex moduli space $\Mcal_{\rm
  vort}^\real(r,1)_\theta$ is a (resolved) hyperspherical cone in the space
(\ref{Mvortr1}). The set $\Mcal_{\rm
  vort}^\real(r,1)_\theta$ is empty in the classical case $\theta=0$,
in contrast to the corresponding instanton moduli space $\widehat\Mcal_\theta^\real(r,1)$ which is
non-empty for all {$q\in\real$}.

\subsection{Noncommutative vortex counting functions}

The $\Ncal=(2,2)$ gauge theory in two dimensions with $r$ fundamental chiral multiplets
can be obtained as a dimensional reduction of four-dimensional
$\Ncal=1$ supersymmetric Yang--Mills theory. For completeness, we will
now work out the various combinatorial vortex partition functions in
parallel to what we did in \S\ref{gaugepartfn}; similar calculations
appear for the classical case
in~\cite{Yoshida,BTZ}. Following~\cite{BTZ}, we can interpret these
partition functions in terms of surface operators in the original four-dimensional
noncommutative $\Ncal=2$ gauge theory, with the two-dimensional gauge
theory living on the defect curve $\CP^1\subset\CP_\theta^2$.

As previously, we consider the left $\hil^{(r)}$-coaction on the
vortex moduli $(B,I)$ given by
\beq\label{Deltavortr}
\Delta_{\Xcal_{\rm vort}(W,V)}^{(r)}(B,I) = \Big(1\otimes B \,,\, \big(\rho_l^{-1}\otimes I_{a,l} \big) \Big) \ .
\eeq
Note that both torus coactions (\ref{DeltaLBIJ}) and (\ref{Deltainstr})
are coequivariant for the $\hil_\zeta$-coaction
(\ref{zetacoactBIJ}), i.e. one has
$$
\big(\Id\otimes\Delta_{\Xcal(W,V)}\big)\circ\Delta_\zeta = \Delta_\zeta\circ
\Delta_{\Xcal(W,V)} \ , \qquad \big(\Id\otimes
\Delta_{\Xcal(W,V)}^{(r)} \big) \circ \Delta_\zeta = \Delta_\zeta\circ
\Delta_{\Xcal(W,V)}^{(r)} \ .
$$
It follows that the set of coinvariants $\Mcal_{\rm
  vort}(r,k)^{\widetilde{T}}$ for the coaction of
$\widetilde{\hil}_t:={\galg_t} \otimes\hil^{(r)}$ can be obtained from the
subset of anti-diagonal $\hil_\zeta$-coinvariants in $\Mcal_\theta(r,k)^{\widetilde
  T}$ for $t=t_1={t_2^{*}}$. In particular, the vector space $V$ has the
isotopical decomposition
$$
V=\bigoplus_{p\in\zed} \ \bigoplus_{m\in\zed^r} \, V(p,m) \ ,
$$
where here
$\zed$ is the character lattice of the torus $T_t=\complex^\times$ with
parameter $t$. Now only the components of the linear map $B=B_1$ are
involved with $B(p,m):V(p,m)\to V(p-1,m)$ for $p\in\zed$ and
$m\in\zed^r$. It follows that the parametrization of the fixed points in the vortex moduli
space are the same as in the instanton moduli space~\cite[\S7.2]{CLSII} with a
reduction of one of the directions of the character lattice of
$T=(\complex^\times)^2$; they are thus parametrized by $r$-vectors of column Young
diagrams, i.e. $r$-component partitions $\vec k=(k_1,\dots,k_r)$ of the vorticity
$k=|\,\vec k\,|:= k_1+\dots+ k_r$, where
$k_l=\dim_\complex\big(\bigoplus_{p\in\zed}\, V(p,l) \big)$ for an
$\widetilde{\hil}_t$-comodule $V=V_{\vec k}$ corresponding to a fixed
point parametrized by $\vec k$ with
$\dim_\complex\big(V(p,l)\big)=0,1$.

The equivariant characters of the $\widetilde\hil_t$-comodules
$V=V_{\vec k}$ and $W=W_{\vec k}$ corresponding to a fixed point
parametrized by $\vec k$ are given from (\ref{VWchars}) by the elements
\beq
\ch_{\widetilde{T}_t}(V_{\vec k})=\sum_{l=1}^r\
\sum_{i=1}^{k_l} \,\rho_l\ t^{1-i} = \sum_{l=1}^r\, \rho_l \ \frac{1-t^{-k_l}}{1-t^{-1}} \ , \qquad
\ch_{\widetilde{T}_t}(W_{\vec k})=\sum_{l=1}^r\, \rho_l
\label{VWcharsvort}\eeq
in the representation ring of
$\widetilde{T}_t=T_t\times(\complex^\times)^{r-1}$; the dual modules
$V_{\vec k}^*$ and $W_{\vec k}^*$ transform with inverse weights
$\rho_l^{-1}$ and $t^{-1}$. From the cohomology of the deformation complex
(\ref{vortdefcomplex}) one easily computes the
$\widetilde{T}_t$-equivariant character of the tangent space at a
fixed point parametrized by $\vec k$ as
\bea
\ch_{\widetilde{T}_t}\big(T_{\vec k}\Mcal_{\rm vort}(r,k) \big) &=&
\ch_{\widetilde{T}_t}(V_{\vec k})\,
\big(\ch_{\widetilde{T}_t}(W_{\vec k}^*)+
(t-1) \, \ch_{\widetilde{T}_t}(V_{\vec k}^*)\big) \nonumber \\[4pt]
&=& \sum_{l,l'=1}^r\, \rho_l \, \rho_{l'}^{-1} \,
\frac{1-t^{-k_{l}}}{1-t^{-1}}\, \Big(1+
(t-1)\, \frac{1-t^{k_{l'}}}{1-t} \, \Big)  \nonumber \\[4pt] &=& \sum_{l,l'=1}^r \,
\rho_l^{-1} \, \rho_{l'} \, \frac{t^{k_l}-t^{k_l-k_{l'}}}{1-t^{-1}} \ , \nonumber
\eea
and hence the corresponding equivariant Euler class is given by
$$
\mbox{$\bigwedge_{-1}$}T_{\vec k}^*\Mcal_{\rm vort}(r,k) =
\prod_{l,l'=1}^r \, \Big(\, 1-\rho_l\,
\rho_{l'}^{-1}\, \frac{t^{-k_l}-t^{k_{l'}-k_l}}{1-t}\, \Big) \ .
$$

Proceeding as in \S\ref{gaugepartfn} we can now compute various
equivariant K-theory partition functions for vortex counting in the noncommutative
gauge theory which by the localization theorem are of the generic form
\bea
&& Z_{\rm vort}(r;Q, z) = \sum_{k=0}^\infty\, Q^k\
\int_{\Mcal_{\rm vort}(r,k)^{\widetilde{T}_t}}\, \omega(z) = \sum_{k=0}^\infty\, Q^k\ \sum_{\vec k \, :\,
  |\,\vec k \,|=k} \ \frac{\tilde\omega(\vec k \,)}{\mbox{$\bigwedge_{-1}$}
    T_{\vec k}^*\Mcal_{\rm vort}(r,k)}  \ ,
\label{Zvortgen}\eea
where $\omega(z)$ is an equivariant K-theory class depending
on $z=(t,\rho_1,\dots,\rho_r)$ which is given by the pullback of a class
$\tilde\omega$ on $\Mcal_{\rm vort}(r,k)$; in our examples $\tilde\omega$
will be regarded as the pullback to the smooth equivariant embedding $\Mcal_{\rm
  vort}(r,k)\subset \Mcal_\theta(r,k)$ of the corresponding instanton
counting class. Let us now look at some
explicit examples of this partition function.

\subsubsection*{$\Ncal=(4,4)$ gauge theory}

By taking $\tilde\omega$ to be the Euler class of the tangent bundle
$T\Mcal_{\rm vort}(r,k)$, we recover the enhanced noncommutative
$\Ncal=(4,4)$ supersymmetric gauge theory with $r$ fundamental
hypermultiplets obtained by dimensional reduction of $\Ncal=4$ gauge
theory in four dimensions~\cite{BTZ}. In this case the partition function (\ref{Zvortgen}) computes the Euler
characteristic of the vortex moduli space which is given by
\beq
Z_{\rm vort}^{\Ncal=(4,4)}(r;Q) = \sum_{\vec k}\, Q^{|\,\vec k\,|}
= \frac1{(1-Q)^r} \ .
\label{44partfn}\eeq

\subsubsection*{$\Ncal=(2,2)$ gauge theory}

Setting $\tilde\omega=1$ in (\ref{Zvortgen}) we obtain the K-theory
version of the partition function for the $\Ncal=(2,2)$ gauge theory
with $r$ fundamental hypermultiplets~\cite{Yoshida,BTZ} given by
\beq
Z_{\rm vort}^{\Ncal =(2,2)}(r;Q,z)= \sum_{k=0}^\infty \, Q^k \
\sum_{\vec k \, :\, |\,\vec k \,|=k}\ \prod_{l,l'=1}^r\  \frac1{1-\rho_l\,
\rho_{l'}^{-1}\,\frac{t^{-k_l}-t^{k_{l'}-k_l}}{1-t} } \ .
\label{22partfn}\eeq

\subsubsection*{Adjoint matter fields}

The dimensional reduction of $\Ncal=2^*$ gauge theory in four
dimensions likewise leads to an analogous field theory in two dimensions~\cite{BTZ}.
By adding an adjoint matter field of twisted mass $m$ to the $\Ncal=(4,4)$
vector multiplet analogously to \S\ref{adjmatter}, i.e. taking $\tilde\omega$ to
descend from the equivariant class $\bigwedge_{-1}(T\Mcal_{\rm
  vort}(r,k)\otimes M)_{\widetilde{T}_t\times T_m}$ with $M=\complex$, we obtain from
(\ref{Zvortgen}) the
partition function for the noncommutative $\Ncal=(2,2)^*$ gauge theory
given by
$$
Z_{\rm vort}^{\Ncal =(2,2)^*}(r;Q,z;\mu)= \sum_{k=0}^\infty \, Q^k \
\sum_{\vec k \, :\, |\,\vec k \,|=k}\
\prod_{l,l'=1}^r\  \frac{1-\mu\, \rho_l\,
\rho_{l'}^{-1}\,\frac{t^{-k_l}-t^{k_{l'}-k_l}}{1-t} }{ 1-\rho_l\,
\rho_{l'}^{-1}\,\frac{t^{-k_l}-t^{k_{l'}-k_l}}{1-t} } \ .
$$
As for instanton counting, this partition function continuously
interpolates between the partition functions (\ref{22partfn}) at
$\mu=0$ and (\ref{44partfn}) at $\mu=1$.

\subsubsection*{Anti-fundamental matter fields}

The $\Ncal=(2,2)$ gauge theory in two dimensions is already endowed
with $r$ fundamental matter fields; the coordinates
$(\rho_1,\dots,\rho_r)$ of $(\complex^\times)^{r-1}$ give twisted mass
parameters $a_l=-\log\rho_l$ to the fundamental multiplets, with $a_1+\dots + a_r=0$. We can consider an
additional $n_f$ anti-fundamental matter fields by pulling back the
rank $k$
Dirac bundle $\Vcal\to\Mcal_\theta(r,k)$ from \S\ref{fundmatter} to
the smooth equivariant embedding $\iota:\Mcal_{\rm vort}(r,k)\hookrightarrow \Mcal_\theta(r,k)$ of the
vortex moduli space; the corresponding Euler class
$\bigwedge_{-1}(\iota^*\Vcal\otimes M)$ with $M=\complex^{n_f}$ now
integrates to a stable class provided $n_f\leq r$. The fibre of the
bundle $\iota^*\Vcal$ over a closed point $[(B,I)]\in\Mcal_{\rm
  vort}(r,k)$ is the vector space $V$ with equivariant
character at a fixed point parametrized by a partition $\vec k$ given in (\ref{VWcharsvort}). With
$\tilde\omega$ the descendent of the equivariant class
$\bigwedge_{-1}(\iota^*\Vcal\otimes M)_{\widetilde{T}_t\times T_M}$,
  the $\Ncal=(2,2)$ gauge theory partition function (\ref{Zvortgen})
  with $n_f\leq r$ anti-fundamental matter multiplets is given by
$$
Z_{\rm vort}^{\Ncal =(2,2),n_f}(r;Q,z;\mu)= \sum_{k=0}^\infty \, Q^k \
\sum_{\vec k \, :\, |\,\vec k \,|=k}\
\frac{\displaystyle{\prod_{a=1}^{n_f} \ \prod_{l=1}^r \, \Big(\, 1-\mu_a\, \rho_l^{-1}\,\frac{1-t^{k_l}}{1-t}\, \Big)}}{\displaystyle{\prod_{l,l'=1}^r\,\Big(\, 1-\rho_l\,
\rho_{l'}^{-1}\,\frac{t^{-k_l}-t^{k_{l'}-k_l}}{1-t} \, \Big) }} \ .
$$

\end{document}